\documentclass[pdftex,twocolumn,epjc3]{svjour3}
\pdfoutput=1
\usepackage{lineno,graphicx,amsmath,amssymb,slashed,verbatim,color,cite,url}

\journalname{MCNET-16-30, CP3-16-40, LPSC16158}

\newcommand{\nn}{\nonumber}
\newcommand{\perc}{\%}
\newcommand\sss{\scriptscriptstyle}
\newcommand{\tth}{t\bar t H}

\begin{document}

\title{\emph{tWH} associated production at the LHC}

\author{
 Federico Demartin\thanksref{addr1,e1}
 \and
 Benedikt Maier\thanksref{addr2,addr3}
 \and
 Fabio Maltoni\thanksref{addr1}
 \and
 Kentarou Mawatari\thanksref{addr4,addr5}
 \and
 Marco Zaro\thanksref{addr6}
}

\authorrunning{F.~Demartin et al.}

\thankstext{e1}{e-mail: federico.demartin@uclouvain.be}

\institute{%
 Centre for Cosmology, Particle Physics and Phenomenology (CP3),
 Universit\'e catholique de Louvain,
 B-1348 Louvain-la-Neuve, Belgium\label{addr1}
 \and
 Institut f{\"u}r Experimentelle Kernphysik, 
 Karlsruher Institut f{\"u}r Technologie (KIT), 
 D-76131 Karls\-ruhe, Germany\label{addr2}
 \and
 Laboratory for Nuclear Science, 
 Massachusetts Institute of Technology, 
 Cambridge, MA 02139, U.S.A.\label{addr3}
 \and
 Laboratoire de Physique Subatomique et de Cosmologie, 
 Universit\'e Grenoble-Alpes,
 CNRS/IN2P3, Avenue des Martyrs 53, F-38026 Grenoble, France\label{addr4}
 \and
 Theoretische Natuurkunde and IIHE/ELEM, Vrije Universiteit Brussel\\
 \emph{and} International Solvay Institutes, 
 Pleinlaan 2, B-1050 Brussels, Belgium\label{addr5}
 \and
 Sorbonne Universit\'es, UPMC Univ. Paris 06, UMR 7589, LPTHE, 
 F-75005, Paris, France\\
 \emph{and}
 CNRS, UMR 7589, LPTHE, F-75005, Paris, France\label{addr6}
}

\date{}

\maketitle

\abstract{
We study Higgs boson production in association with a top quark and a $W$ boson at the LHC.
At NLO in QCD, $tWH$ interferes with $t\bar t H$ and a procedure to meaningfully separate 
the two processes needs to be employed. 
In order to define $tWH$ production for both total rates and differential distributions, 
we consider the diagram removal and diagram subtraction techniques that have been previously 
proposed for treating intermediate resonances at NLO, in particular in the context of $tW$ production.
These techniques feature approximations that need to be carefully taken into account 
when theoretical predictions are compared to experimental measurements.
To this aim, we first critically revisit the $tW$ process, for which an extensive literature exists 
and where an analogous interference with $t \bar t$ production takes place. 
We then provide robust results for total and differential cross sections for $tW$ and $tWH$ at 13~TeV, 
also matching short-distance events to a parton shower. 
We formulate a reliable prescription to estimate the theoretical uncertainties, 
including those associated to the very definition of the process at NLO. 
Finally, we study the sensitivity to a non-Standard-Model relative phase between the 
Higgs couplings to the top quark and to the $W$ boson in $tWH$ production.
}



\section{Introduction}
\label{sec:intro}

The study of the Higgs boson is one of the main pillars of the physics programme 
of the current and future LHC runs. Accurate measurements of the Higgs boson properties 
are crucial both to validate the Standard Model (SM) as well as to possibly discover 
new physics through the detection of deviations from the SM predictions. 
Another main pillar of the LHC research programme of the coming years is the study of the top quark.  
Being the heaviest quark, the top quark also plays a main role in Higgs boson phenomenology.  
In particular, the main production channel for the Higgs boson at the LHC entails a top-quark loop,
while very soon Run II will be sensitive to on-shell top--anti-top pair production in association with the Higgs boson, 
a process that will bring key information on the strength of the top-quark Yukawa interaction.

Exactly as when no Higgs is present in the final state, top quark and Higgs boson associated production 
can proceed either via a top pair production mediated by QCD interactions, 
or as a single top (anti-)quark process mediated by electroweak interactions. 
The latter case, despite being characterised by much smaller cross sections with respect to the QCD production, 
displays a richness and peculiarities that make it phenomenologically very interesting. 
For example, it is sensitive to the relative phase between the Higgs coupling to the top quark and 
to the $W$ boson. 
Single-top production (in association with a Higgs boson) can be conveniently classified in three main channels:
$t$-channel, $s$-channel (depending on the virtuality of the intermediate $W$ boson) and $tW(H)$ associated production. 
For the first two channels, this classification is unambiguous only up to next-to-leading order (NLO) accuracy 
if a five-flavour scheme (5FS) is used. 
Beyond NLO, the two processes interfere and cannot be uniquely separated. 
The associated $tW(H)$ production, on the other hand, can be easily defined only at leading order (LO) accuracy 
and in the 5FS, i.e. through the partonic process $g b \to tW (H)$. 
At NLO, real corrections of the type $g g \to tWb(H)$ arise that can feature a resonant $\bar t$ 
in the intermediate state and therefore overlap with $ g g \to t \bar t (H)$, i.e. with $t \bar t (H)$ production at LO.
This fact would not be necessarily a problem per se, were it not for the fact that the cross section 
of $t \bar t (H)$ is one order of magnitude larger than $tW(H)$, and its subtraction
-- which can only be achieved within some approximation --
leads to ambiguities that have to be carefully estimated and entails both conceptual issues and 
practical complications.

A fully consistent and theoretically satisfying treatment of resonant contributions can be achieved by starting from the
complete final state $WbWb(H)$ in the four-flavour scheme (4FS), including all contributions, i.e. doubly, singly and 
non-resonant diagrams. 
Employing the complex mass scheme~\cite{Denner:1999gp,Denner:2005fg} to deal with the finite width of the top quark 
guarantees the gauge invariance of the amplitude and the possibility of consistently going to NLO accuracy in QCD. 
This approach has been followed already for $WbWb$ and other processes calculations 
at NLO~\cite{Denner:2010jp,Bevilacqua:2010qb,Denner:2012yc,Frederix:2013gra,Cascioli:2013wga,Heinrich:2013qaa}. 
Recent advances have also proven that these calculations can be consistently matched to 
parton showers (PS)~\cite{Jezo:2015aia,Frederix:2016rdc,Jezo:2016ujg}. 
However, from the practical point of view, such calculations are computationally very expensive 
and would entail the generation of large samples including resonant and non-resonant contributions 
as well as their interference. 
This approach does not allow to distinguish between top-pair and single-top production in the event generation.
One would then need to generate signal and background together in the same sample
(a procedure that would entail complications from the experimental point of view, for example in data-driven analyses) 
and communicate experimental results and their comparison with theory only via fiducial cross sections measurements. 
In any case, results for $WbWbH$ are currently available at NLO
accuracy only with massless $b$ quarks~\cite{Denner:2015yca}, and therefore cannot be used for studying $tWH$. 
 
A more pragmatic solution is to adopt a 5FS, define final states in terms of on-shell top quarks, and
remove overlapping contributions by controlling the ambiguities to a level such that the NLO accuracy 
of the computation is not spoiled, and total cross section as well as differential distributions 
can be meaningfully defined.
To this aim, several techniques have been developed with a different degree of flexibility, some being suitable only
to evaluate total cross sections, others being employable in event generators. 
They have been applied to $tW$ production and to the production of particles in SUSY or in other extensions of the SM, 
where the problem of resonances appearing in higher-order corrections is recurrent. 
Two main classes of such techniques exist for event generation, and they are generally dubbed as diagram removal (DR) 
and diagram subtraction (DS). 
Unavoidably, all these approaches have their own shortcomings, some of them of more theoretical nature, 
such as possible violation of gauge invariance (which, however, turns out not to be worrisome), 
or ambiguities in the far off-shell regions which need to be kept into account and studied 
on a process-by-process basis. 
As it will be reminded in the following, DR and DS actually feature complementary virtues and vices.
An important point of the 5FS approach is that the combination of the separate $t \bar t(H)$ and $tW(H)$ 
results ought not to depend on the technical details used to define the $tW(H)$ contribution, 
in the limit where overlapping is correctly removed and possible theoretical ambiguities are under control.
In practice, the most common approach is to organise the perturbative expansion in poles of the top propagator,
where $t \bar t(H)$ production is computed with on-shell top quarks
(this approach can also be used in the 4FS \cite{Denner:2010jp,Bevilacqua:2010qb,Denner:2012yc,Cascioli:2013wga}).
In this case, the complementary $tW(H)$ contribution should encompass all the remaining effects,
e.g. including the missing interference with $t \bar t(H)$ if that is not negligible.
We are interested in finding a practical and reliable procedure to generate $tW(H)$ events under this scenario.

As already mentioned above, Higgs and top-quark associated processes can provide further information 
on the top--Higgs interaction. 
While at the Run I the LHC experiments have not claimed observation yet for these processes, 
setting only limits on the signal strength~\cite{Chatrchyan:2013yea,Khachatryan:2014qaa,Aad:2014lma,Khachatryan:2015ila,Aad:2015gra,Aad:2015iha,Khachatryan:2015ota}, 
$\tth$ is expected to be soon observed at the Run II, allowing a first direct measurement of the top-quark Yukawa coupling $y_t$.
Indeed, unlike the dominant Higgs production mode via gluon fusion, where the extraction of $y_t$ is indirect,
in the case of $\tth$ such an extraction is (rather) model-independent. 
In addition, $\tth$ production is known to be sensitive to the Higgs CP properties~\cite{Frederix:2011zi,Artoisenet:2012st,Ellis:2013yxa,Demartin:2014fia,Khatibi:2014bsa,He:2014xla,Boudjema:2015nda,Kolodziej:2015qsa,Buckley:2015vsa,Li:2015kaa,Mileo:2016mxg,Gritsan:2016hjl}. 
On the other hand, Higgs production in association with a single top quark ($tH$ and $tWH$), though rare, 
is very sensitive to departures from the SM, since the total rate can increase by more than 
an order of magnitude~\cite{Farina:2012xp,Demartin:2015uha} due to constructive
interference effects, becoming comparable to or even larger than $\tth$. 
In particular, Higgs plus single top allows to access the phase of $y_t$, which remains unconstrained 
in gluon fusion and $\tth$;
a preliminary, yet not enough sensitive exploration has been carried out already at Run I~\cite{Khachatryan:2015ota}. 
At variance with $t$-channel and $s$-channel processes, predictions for $tWH$ cross sections are only available at LO. 
Accurate predictions for $tWH$ are not only important for the measurement of $tWH$ itself, 
but also as a possible background to $tH$ production, and in view of the observation of $t\bar t H$ 
and of the consequent extraction of Higgs couplings.  

The main aim of this paper is to present the first predictions at NLO accuracy for $tWH$ cross sections at the LHC. 
In order to do that, we first review the different techniques that can be used to remove resonant contributions 
from NLO corrections and also make a proposal for an improved  DS scheme. 
We then study the $tW$ process in detail, and compare our findings with the results already available in the literature.
Finally, we apply these techniques to get novel results for $tWH$ production.

At this point, we stress that even though it is not really the original motivation of this work, 
a critical analysis of $tW$ is certainly welcome. 
The relevance of which approach ought to be used to describe $tW$ production is far from being 
only of academic interest: 
already during the Run I, single-top production has been measured by both ATLAS and CMS 
in the $t$-channel~\cite{Chatrchyan:2011vp,Chatrchyan:2012ep,Aad:2012ux,Khachatryan:2014iya}, 
$s$-channel~\cite{Aad:2015upn,Khachatryan:2016ewo} 
and $tW$~\cite{Chatrchyan:2012zca,Aad:2015eto,Chatrchyan:2014tua} modes. 
In particular, in $tW$ analyses the difference between the two aforementioned methods, 
DR and DS (without including the $t\bar t$--$tW$ interference), has been added to the theoretical uncertainties. 
In view of the more precise measurements at the Run II, a better understanding of the $t\bar t$--$tW$ overlap 
is desirable, in order to avoid any mismodelling of the process and incorrect estimates 
of the associated theoretical uncertainties, both in the total cross section and in the shape of distributions. 
Furthermore, given the large amount of data expected at Run II and beyond, 
a measurement aimed at studying the details of the $t\bar t$--$tW$ interference may become feasible, 
and this gives a further motivation to study the best modelling strategy. 
Finally, a sound understanding of $tW$ production will also be beneficial for the numerous 
analyses which involve $t\bar t$ production as a signal or as background. 
This is particularly true in analyses looking for a large number of jets in the final state,
which typically employ Monte Carlo samples based on 
NLO merged~\cite{Frederix:2012ps,Lonnblad:2012ix,Hoeche:2014qda} events,
where stable top quarks are produced together with extra jets ($t \bar t + nj$). 
In this case, all kinds of non-top-pair contributions, like $tW$, need to be generated separately.  
While these effects are expected to be subdominant, their importance has still to be assessed 
and may become relevant after specific cuts, given also the plethora of analyses; 
an example can be the background modelling in $\tth$ or $tH$ searches.
Note that results for $WbWb$ plus one jet have been recently published~\cite{Bevilacqua:2015qha,Bevilacqua:2016jfk},
but the inclusion of extra radiation in merged samples is much more demanding if one starts from 
the $WbWb$ final state, and thus may be impractical.
Last but not least, a reliable 5FS description of $tW$ is desirable
in order to assess residual flavour-scheme dependence between the 4FS ($WbWb$) and
the 5FS ($t \bar t + tW$) modelling of this process.
Such a comparison can offer insights on the relevance of initial-state logarithms resummed
in the bottom-quark PDF, which are an important source of theoretical uncertainty.

The paper is organised as follows: in section~\ref{sec:DRandDS} we review the definitions  
of the DR and DS techniques, and we also include a proposal for an improved DS scheme.  
In section~\ref{sec:nlosetup} we describe our setup for NLO computations, also matched to parton shower.
In section~\ref{sec:tW_nlo} we review the results from these 
techniques in the well-studied case of $tW$ production, performing
a thorough study of their possible shortcomings,
considering the impact of interference effects between 
top-pair and single-top processes, and investigating what happens after
typical cuts are imposed to define a fiducial region for the $tW$ process. 
In section~\ref{sec:tWH_nlo} we repeat a similar study for the SM $tWH$
process at NLO.
We also include the study of the $tWH$ process
going beyond the SM Higgs boson, investigating results from a 
generic CP-mixed Yukawa interaction between the Higgs and the top quark.
Our study is complemented in \ref{app:twb_lo_4FS}
by a quantitative assessment of the $tWb$ and $tWbH$ channels, 
studied as standalone processes in the 4FS and at the partonic level. 
In section~\ref{sec:summary} we summarise our findings and propose an 
updated method to estimate the impact of theoretical systematics 
in the definition of $tW$ and $tWH$ at NLO in the 5FS.

\section{Subtraction of the top quark pair contribution}
\label{sec:DRandDS}

\begin{figure}
\center 
\includegraphics[width=0.56\columnwidth]{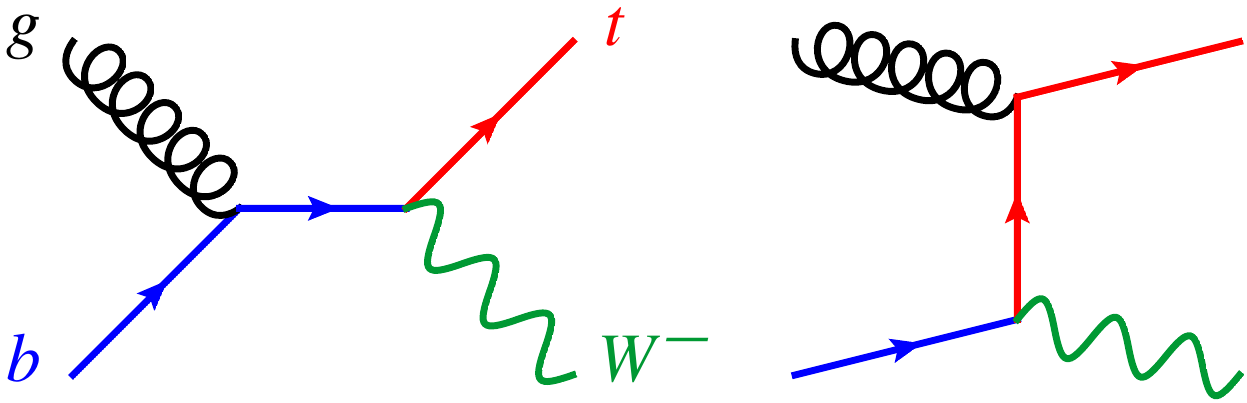}
\caption{LO Feynman diagrams for $tW^{-}$ production in the 5FS.}
\label{fig:tw_5FS_diagrams}
\end{figure} 

\begin{figure}
\center 
\includegraphics[width=0.96\columnwidth]{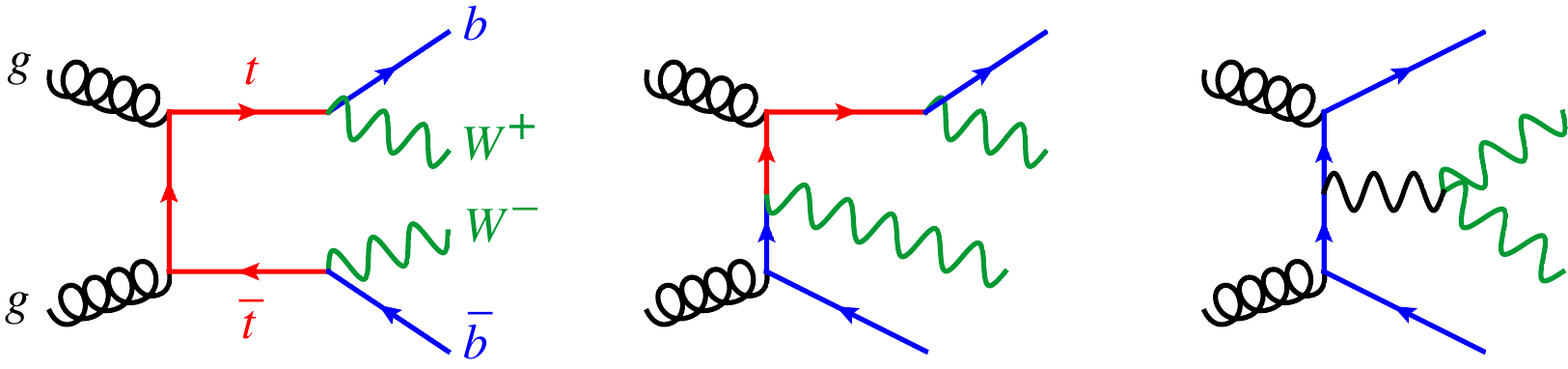}\\
\hspace*{1em} ($\mathcal{A}_{2t}$) \hspace*{6.2em} 
($\mathcal{A}_{1t}$) \hspace*{6.2em} ($\mathcal{A}_{0t}$)
\caption{Examples of doubly resonant (left), 
singly resonant (center) and 
non-resonant (right) diagrams contributing 
to $WbWb$ production. 
The first two diagrams on the left (with the $t$ line cut) 
describe the NLO real-emission contribution to the $tW^{-}$ process. 
} 
\label{fig:twb_4FS_diagrams}
\end{figure} 

As discussed in the introduction, the computation of higher-order corrections to $tW(H)$ 
requires the isolation of the $t\bar t(H)$ process, and its consequent subtraction. 
In this section we review the techniques to remove such a resonant contribution which appears 
in the NLO real emissions of the $tW(H)$ process. 

In the case of fixed-order calculations, and in particular when only the total cross section is computed, 
a {\it global subtraction} (GS) of the on-shell top quark can be employed,
which just amounts to the subtraction of the total cross section for $t\bar t (H)$ production times
the $t\to bW$ branching ratio~\cite{Tait:1999cf,Zhu:2002uj}:
\begin{align}
& \sigma_{\rm NLO}({tW(H)})_{\rm{GS}} =  \nn\\
& \lim_{\Gamma_t \to 0}  \left[
\sigma_{\rm NLO}({tW(H)}) - \sigma_{\rm LO}(t\bar t(H)) \frac{\Gamma({t \to W b})}{\Gamma_t} \right]\,,
\end{align}
where $\Gamma({t \to W b})$ is the physical width, while $\Gamma_t$ is introduced in the resonant top-quark 
propagator as a regulator, and gauge invariance is ensured in the $\Gamma_t\to 0$ limit. 
A conceptually equivalent version, that can be applied {\it locally} in the virtuality 
of the resonant particle and in an analytic form,%
\footnote{It differs only by tiny boundary effects, see~\cite{Hollik:2012rc}.} 
has been employed in the NLO computations 
for pair production of supersymmetric particles~\cite{Beenakker:1996ch,Hollik:2012rc} 
and for charged Higgs boson production~\cite{Berger:2003sm,Dao:2010nu}. 

On the other hand, NLO+PS simulations require a subtraction which is {\it fully local} in the phase space. 
In order to achieve such a local subtraction, two main schemes have been developed, 
known as {\it diagram removal} (DR) and {\it diagram subtraction} (DS)~\cite{Frixione:2008yi}. 
These subtraction schemes have been studied in detail for $tW$ production matched to parton shower 
in {\sc MC@NLO}~\cite{Frixione:2008yi,White:2009yt} and in {\sc Powheg}~\cite{Re:2010bp},
as well as in the case of $tH^-$~\cite{Weydert:2009vr}
and for supersymmetric particle pair production~\cite{Binoth:2011xi,GoncalvesNetto:2012yt,Gavin:2013kga,Gavin:2014yga}.

To keep the discussion as compact as possible, we focus on $tW$ production 
(see fig.~\ref{fig:tw_5FS_diagrams} for the LO diagrams) and
consider the specific case of the $tW^-\bar b$ real emission and of its overlap with
$t\bar t$ production. The extension to the process with an extra Higgs boson is straightforward.
Strictly speaking, one should 
consider $t\bar t$ and $tW^-\bar b$ ($\bar tW^+b$) processes as doubly resonant and singly resonant 
contributions to $WbWb$ production, which also contains the set of non-resonant diagrams as shown 
in fig.~\ref{fig:twb_4FS_diagrams}.
However, as discussed in detail in \ref{app:twb_lo_4FS}, the contribution from non-resonant $WbWb$ production
and off-shell effects for the final-state top quark are tiny,
as well as possible gauge-dependent effects due to the introduction of a finite top width. 
Therefore, we will treat one top quark as a final-state particle with zero width, so that
the only intermediate resonance appears in top-pair amplitudes.
The squared matrix element for producing a $tW^-\bar b$ final state can be written as
\begin{align}
\label{eq:M_tWb}
|\mathcal{A}_{tWb}|^2 
 & = | \mathcal{A}_{1t} + \mathcal{A}_{2t} |^2 \nn\\
 & = | \mathcal{A}_{1t} |^2 
 + 2 \mathrm{Re} (\mathcal{A}_{1t}^{}\mathcal{A}_{2t}^*) 
 + | \mathcal{A}_{2t} |^2 \,,
\end{align}
where $\mathcal{A}_{1t}$ denotes the single-top amplitudes, 
considered as the real-emission corrections to the $tW$ process,
while $\mathcal{A}_{2t}$ represents the resonant top-pair amplitudes describing $t\bar t$ production, 
where the intermediate $\bar t$ can go on-shell.
The corresponding representative Feynman diagrams are shown in fig.~\ref{fig:twb_4FS_diagrams}.
In the following, we will discuss the DR and DS techniques in detail. 

\paragraph{DR (diagram removal):}
Two different version of DR have been proposed in the literature:

\begin{itemize}

\item {\bf DR1 (without interference):}
This was firstly proposed in~\cite{Frixione:2008yi} 
for $tW$ production and its implementation in \textsc{MC@NLO}. 
One simply sets $\mathcal{A}_{2t}=0$\,, 
removing not only $| \mathcal{A}_{2t} |^2$, which can be identified 
with $t \bar t$ production, but also the interference term 
$2 \mathrm{Re} (\mathcal{A}_{1t}^{}\mathcal{A}_{2t}^*)$\,, 
so that the only contribution left is
\begin{align}
 \label{eq:DR1}
 |\mathcal{A}_{tWb}|^2_{\mathrm{DR1}} = | \mathcal{A}_{1t} |^2 \,. 
\end{align}
This technique is the simplest from the implementation point of view and, 
since diagrams with intermediate top quarks are completely removed from the calculation,
it does not need the introduction of any regulator. \\ 

\item {\bf DR2 (with interference):}
This second version of DR was firstly proposed 
in~\cite{Hollik:2012rc} for squark-pair production.
In this case, one removes only $| \mathcal{A}_{2t} |^2$, 
keeping the contribution of the interference between singly and doubly resonant diagrams
\begin{align}
 \label{eq:DR2}
 |\mathcal{A}_{tWb}|^2_{\mathrm{DR2}}
 = | \mathcal{A}_{1t} |^2 + 2 \mathrm{Re} (\mathcal{A}_{1t}^{}\mathcal{A}_{2t}^*) \,.
\end{align}
Note that the DR2 matrix element is not positive-definite, at variance with DR1.
In this case, while the integral is finite even with $\Gamma_t \to 0$, in practice one has to introduce
    a finite $\Gamma_t$ in the amplitude $\mathcal{A}_{2t}$ in order to improve the numerical stability of the phase-space integration. 
\end{itemize}

\noindent 
An important remark concerning the DR schemes is that, as they are based on removing contributions all over 
the phase space, they are not gauge invariant. 
However, for $tW$ the issue was investigated in detail in~\cite{Frixione:2008yi}, 
and effects due to gauge dependence have been found to be negligible. 
We have confirmed this finding for both $tW$ and $tWH$ in a different way,
and we discuss the details in \ref{app:twb_lo_4FS},
where we show that gauge dependence is not an issue if one uses a covariant gauge, 
such as the Feynman gauge implemented in \textsc{MadGraph5\_aMC@NLO}.

\paragraph{DS (diagram subtraction):}
DS methods, firstly proposed for the \textsc{MC@NLO} $tW$
implementation, have been developed explicitly to avoid the problem of 
gauge dependence which, at least in principle, affects DR techniques. 
The DS matrix element is written as
\begin{align}
 \label{eq:DS}
 |\mathcal{A}_{tWb}|^2_{\mathrm{DS}} = | \mathcal{A}_{1t} + 
 \mathcal{A}_{2t} |^2 - \mathcal{C}_{2t}\,,
\end{align}
where the local subtraction term $\mathcal{C}_{2t}$, 
by definition, must \cite{Frixione:2008yi,Re:2010bp}: 

\begin{enumerate}
\item cancel exactly the resonant matrix element $ | \mathcal{A}_{2t}|^2 $ when the 
kinematics is exactly on top of the resonant pole;
\item be gauge invariant;
\item decrease quickly away from the resonant region.
\end{enumerate}

\begin{figure*}
\center 
\includegraphics[width=0.45\textwidth]{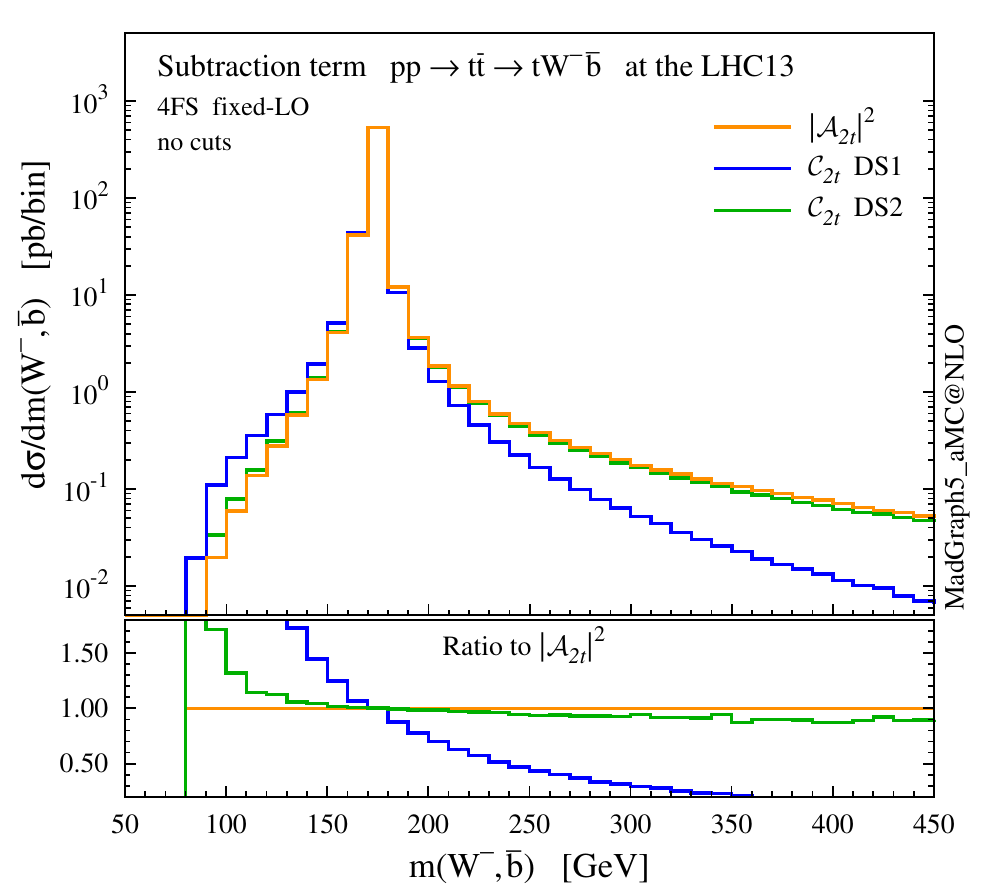}
\includegraphics[width=0.45\textwidth]{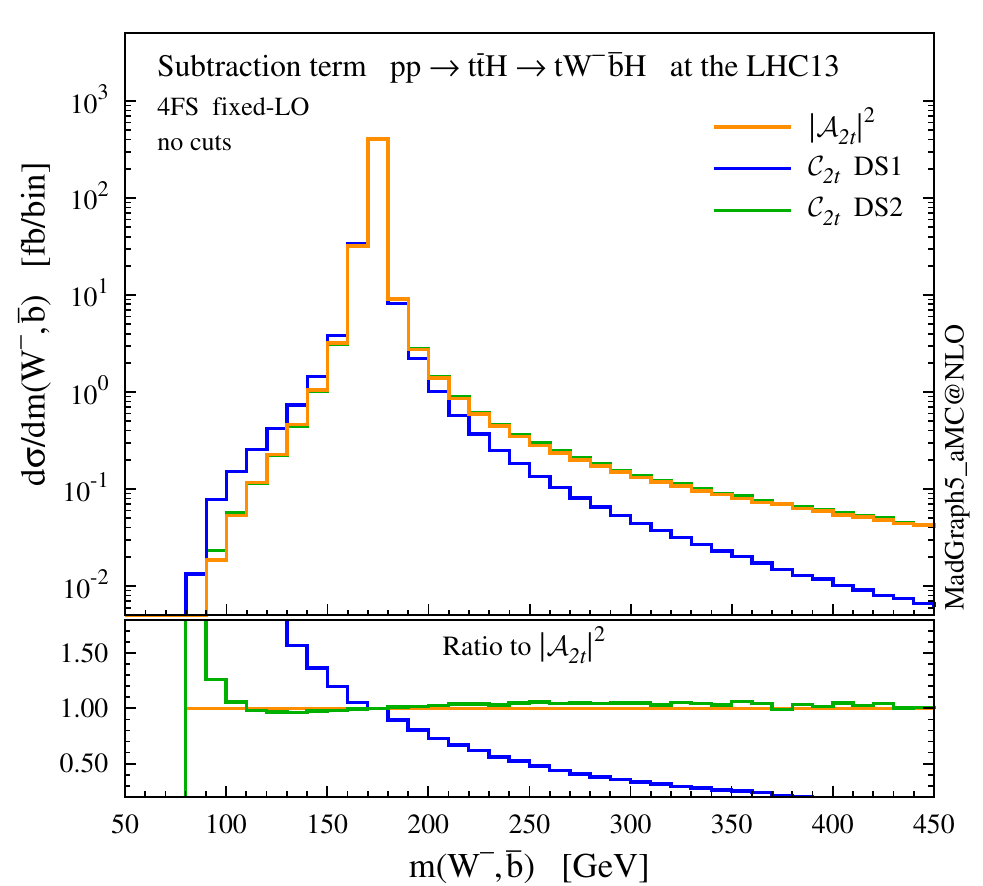}
\caption{Invariant mass $m(W^-, \bar b)$ distributions in the $pp\to tW^- \bar b$ process (left) 
 and in the $pp\to tW^- \bar bH$ process (right), for comparison between $|{\cal A}_{2t}|^2$ 
 and $\mathcal{C}_{2t}$ with two different Breit--Wigner forms, DS1 and DS2.}
\label{fig:mwb}
\end{figure*}

Given the above conditions, a subtraction term can be written as
\begin{align}
\label{eq:DS_counterterm}
\mathcal{C}_{2t}(\{p_i\}) = 
f(p_{Wb}^2)\,
\big| \mathcal{A}_{2t}(\{q_i\}) \big|^2 \,,
\end{align}
where $p_{Wb}=(p_W+p_b)$, and 
$\{p_i\}$ is the set of momenta of the external particles 
(i.e. the phase-space point),
while $\{q_i\}$ are the external momenta after a reshuffling 
that puts the internal anti-top quark on mass-shell, i.e.
\begin{align}
\{q_i\} \ : \
 q_{Wb}^2 \equiv (q_W+q_b)^2 = m_t^2 \,.
\end{align}
Such a reshuffling is needed in order to satisfy gauge invariance 
of $\mathcal{C}_{2t}$, which in turn implies gauge invariance
of the DS matrix element of eq.~\eqref{eq:DS} in the $\Gamma_t \to 0$ limit.
There is freedom to choose the prefactor $f(p_{Wb}^2)$, and the Breit--Wigner profile 
is a natural option to satisfy the third condition.
Here, we consider two slightly different Breit--Wigner distributions:

\begin{itemize}

\item {\bf DS1:}
\begin{align}
\label{ds1}
 f_1(s) = 
\frac{(m_t\Gamma_t)^2}{( s - m_t^2 )^2 + (m_t\Gamma_t)^2} \,,
\end{align}
which is just the ratio between the two Breit--Wigner functions for the top quark
computed before and after the momenta reshuffling, as implemented in \textsc{MC@NLO}
and \textsc{POWHEG} for $tW$~\cite{Frixione:2008yi,Re:2010bp}.\\

\item {\bf DS2:}
\begin{eqnarray}
\label{ds2}
 f_2(s) = 
\frac{(\sqrt{s}\,\Gamma_t)^2}{( s - m_t^2 )^2 + (\sqrt{s}\,\Gamma_t)^2} \,.
\label{eq:ds2}
\end{eqnarray}
This off-shell profile of the resonance differs from DS1 
by the replacement $m_t\Gamma_t \to \sqrt{s}\,\Gamma_t$~\cite{Olive:2016xmw,Gigg:2008yc}.
The exact shape of a resonance may be process-dependent, and
in the specific case of $tW(H)$ we find that this profile is in better agreement than DS1 with 
the off-shell lineshape of the amplitudes $|{\cal A}_{2t}|^2$ (away from $Wb$ threshold), 
as can be seen in fig.~\ref{fig:mwb}.
In particular, we have checked that the agreement between the $|{\cal A}_{2t}|^2$ profile
and the $\mathcal{C}_{2t}$ subtraction term in DS2 holds for the separate $q \bar q$ and $gg$ channels;
at least in the $q \bar q$ channel there is no gauge-related issue, 
off-shell effects in top-pair production are correctly described by $|{\cal A}_{2t}|^2$,
and DS2 captures these effects better.
As it will be shown later, this modification in the resonance profile leads to appreciable differences 
between the two DS methods at the level of total cross sections as well as differential distributions. 
\end{itemize}

\noindent
Apart from the different resonance lineshapes, another important remark on DS is about the reshuffling of the momenta.
Such a reshuffling is not a Lorentz transformation, since it changes the mass 
of the $Wb$ system, therefore different momenta transformations 
could result in different subtraction terms. 
Actually, there is an intrinsic arbitrariness in defining the on-shell 
reshuffling, potentially leading to different counterterms and effects. 
Thus, on the one hand DS ensures that gauge invariance 
is preserved in the $\Gamma_t \to 0$ limit, at variance with DR.
On the other hand, it introduces a possible 
dependence on how the on-shell reshuffling is implemented, which is not 
present in the DR approach and needs to be carefully assessed.
To our knowledge, this problem has not been discussed in depth in the 
literature; a more detailed study is underway and will be reported elsewhere.
In this work, we adopt the reshuffling employed by \textsc{MC@NLO}
and \textsc{POWHEG}~\cite{Frixione:2008yi,Re:2010bp}, where
the recoil is shared democratically among the initial-state particles,
also rescaling by the difference in parton luminosities due to the change 
of the partonic centre-of-mass energy. 

Finally, we comment on the introduction 
of a non-zero top-quark width in the DR2 and DS methods.
In order to regularise the singularity of $\mathcal{A}_{2t}$, we have to
modify the denominator of the resonant top-quark propagators as
\begin{align}
\label{eq:top_prop_reg}
 \frac{1}{p_{Wb}^2 - m_t^2} \, \to \, 
 \frac{1}{p_{Wb}^2 - m_t^2 + i m_t \Gamma_t} \,.
\end{align}
At variance with the case of a physical resonance, here
$\Gamma_t$ is just a mathematical regulator that does not necessarily need
to be equal to the physical top-quark width.%
\footnote{
A modified version of DS (DS$^*$), which requires to know the analytic structure of the poles over
each integration channel, was proposed in~\cite{Gavin:2013kga} to guarantee gauge invariance 
already with a finite width. In practice, there is no difference between DS and DS$^*$ if $\Gamma_t$ is small enough.
}
In fact, one can set it to any number that satisfies $\Gamma_t / m_t \ll 1$
without affecting the numerical result in a significant way~\cite{Binoth:2011xi,Gavin:2013kga}.
We have checked that the NLO DR2 and DS codes provide stable results
with $\Gamma_t$ in the interval between 1.48~GeV and 0.001~GeV.%
\footnote{However, the computational time does depend on this regulator,
because the smaller is $\Gamma_t$ the larger are the numerical instabilities, 
resulting in a slower convergence of the integration. 
For this reason, the results presented in the paper have been generated 
setting this regulator close to the physical value of the 
top width at LO, $\Gamma_t \simeq 1.48$~GeV.}

After all the technical details exposed in this section, we summarise the key points in order to 
clearly illustrate our rationale in assessing the results in the next sections:
\begin{itemize}
\item 
Our starting point is to assume the (common) case where results for $t \bar t (H)$ 
production are generated with on-shell top quarks.
Resonance profile and correlation among production and decay
are partially recovered from the off-shell LO amplitudes with decayed top quarks, 
following the procedure illustrated in~\cite{Frixione:2007zp}. 
In particular, after this procedure the on-shell production cross section is not changed. \\

\item 
The GS procedure is gauge invariant and ensures that all and just the on-shell $t \bar t (H)$ 
contribution is subtracted.
Thus, under the working assumptions in the previous point, 
GS provides a consistent definition of the missing $tW(H)$ cross section,
that can be combined with $t \bar t(H)$ without double countings and 
including all the remaining effects, such as interference. 
A local subtraction scheme should return a cross section close to the GS result
\emph{if off-shell and gauge-dependent effects are small}. \\

\item
DS is gauge invariant by construction.
The difference between the GS and DS cross sections can thus quantify off-shell effects
in the decayed $t \bar t(H)$ amplitudes.
From fig.~\ref{fig:mwb} and the related discussion, we already find DS2 to provide a better 
treatment than DS1 in the subtraction of the off-shell $t \bar t (H)$ contribution;
the difference between DS1 and DS2 quantifies the impact of different off-shell profiles. \\

\item
DR is in general gauge dependent.
The difference between GS and DR2 amounts to the impact of possible gauge-dependent
contributions and off-shell effects.
As it will be shown, for the $tW$ and $tWH$ processes this difference is tiny.
Finally, the difference between DR2 and DR1 amounts to the interference effects between $t \bar t(H)$ and $tW(H)$;
the single-top process is well defined \emph{per se} only if the impact of interference is small.
\end{itemize}
As a last comment, we argue that in practice gauge dependence in DR should not be an issue in our case.
When using a covariant gauge and only transverse external gluons,
any gauge-dependent term decouples from the $gg \to tWb$ amplitudes~\cite{Frixione:2008yi},
and this remains valid also after adding a Higgs.
An independent constraint on gauge-dependent effects comes also from the off-shell profiles in fig.~\ref{fig:mwb}.
In the $q \bar q$ channel, $|\mathcal{A}_{2t}|^2$ is free from gauge dependence and validates the 
$\mathcal{C}_{2t}$ DS2 off-shell profile for $tW(H)$; 
the gauge-invariant DS2 counterterm continues to agree with $|\mathcal{A}_{2t}|^2$ 
also in the $gg$ channel, which in turn limits the size of alleged gauge-dependent effects in DR2.
Moreover, even in the case of a significant gauge dependence, 
its effects should cancel out in a consistent combination of $t \bar t(H)$ and $tW(H)$ events,
if the off-shell amplitudes used to decay $t \bar t(H)$ have been computed in the same gauge as $tW(H)$.

\section{Setup for NLO+PS simulation}
\label{sec:nlosetup}

The code and events for $tW$ production at hadron colliders at NLO-QCD
accuracy can be generated in the {\sc MadGraph5\_aMC@NLO} framework
by issuing the following commands:
\vspace*{0.5em}
\begin{verbatim}
 > import model loop_sm-no_b_mass
 > generate p p > t w- [QCD]
 > add process p p > t~ w+ [QCD]
 > output
 > launch
\end{verbatim}
\vspace*{0.5em}
and similarly for $tWH$ production:
\vspace*{0.5em}
\begin{verbatim}
 > import model loop_sm-no_b_mass
 > generate p p > t w- h [QCD]
 > add process p p > t~ w+ h [QCD]
 > output
 > launch
\end{verbatim}
\vspace*{0.5em}
The output of these commands contains, among the NLO real emissions, 
the $tWb$ amplitudes that have to be treated with DR or DS.
The technical implementation of DR1 (no interference) in the NLO code 
simply amounts to edit the relevant \texttt{matrix\_*.f} files,
setting to zero the top-pair amplitudes.
To implement DR2, on the other hand, one subtracts the square of the top-pair 
amplitudes from the full matrix element.
A subtlety is that the top-pair amplitudes (and only those) need to be regularised by introducing 
a non-zero width in the top-quark propagator. 
Note that, as we have already remarked in sec.~\ref{sec:DRandDS}, 
this width is just a mathematical regulator.
The DS is more complicated, since it also requires the implementation of the momenta reshuffling
to put the top quark on-shell before computing the subtraction
term $\mathcal{C}_{2t}$. 
The automation of such on-shell subtraction in the {\sc MadGraph5\_aMC@NLO} framework is underway, 
and will be become publicly available in the near future. 

In our numerical simulations we set the mass of the Higgs boson to
$m_H=125.0$~GeV and the mass of the top quark to $m_t=172.5$~GeV,
which are the reference values used by the ATLAS and CMS collaborations
at the present time in Monte Carlo generations.
We renormalise the top Yukawa coupling on-shell by setting it to
$y_t/\sqrt{2}=m_t/v$, where $v \simeq 246$~GeV is the electroweak vacuum
expectation value, computed from the Fermi constant 
$G_F=1.16639 \cdot 10^{-5}$~GeV$^{-2}$; the electromagnetic coupling is also
fixed to $\alpha=1/132.507$.
The $W$ and $Z$ boson masses are set to $m_W=80.419$~GeV and $m_Z=91.188$~GeV.
In the 5FS the bottom-quark mass is set to zero in the matrix-element, 
while $m_b=4.75$~GeV determines the threshold of the bottom-quark parton distribution function (PDF),
which affects the parton luminosities.%
\footnote{In the 4FS simulations presented in \ref{app:twb_lo_4FS}
$m_b$ enters the calculation of the hard-scattering matrix elements 
and the phase space.
}
We have found the contributions proportional to the bottom Yukawa coupling to be negligible,
therefore we have set $y_b=0$ as well.

The proton PDFs and their uncertainties are evaluated employing reference sets and error replicas
from the NNPDF3.0 global fit~\cite{Ball:2014uwa}, at LO or NLO as well as in the 5FS or 4FS
(4FS numbers are shown in the appendix). 
The value of the strong coupling constant at LO and
NLO is set to $\alpha_s^{\mathrm{(5F,LO)}}(m_Z) = 0.130$ 
and, respectively, $\alpha_s^{\mathrm{(5F,NLO)}}(m_Z) = 0.118$. 

The factorisation and renormalisation scales ($\mu_F$ and $\mu_R$)
are computed dynamically on an event-by-event basis, 
by setting them equal to the reference scale $\mu_0^{d}=H_T/4 $, 
where $H_T$ is the sum of the transverse masses of all outgoing particles 
in the matrix element.
The scale uncertainty in the results is estimated varying $\mu_F$ and $\mu_R$ 
independently by a factor two around $\mu_0$.
Additionally, we also show total cross sections computed with a 
static scale, which we fix to $\mu_0^{s} = (m_t+m_W)/2 $ for $tW$ production
and to $\mu_0^{s} = (m_t+m_W+m_H)/2 $ for $tWH$.

We use a diagonal CKM matrix with $V_{tb}=1$, ignoring
any mixing between the third generation and the first two.
In particular, this means that the top quark always decays to a bottom quark
and a $W$ boson, $\mathrm{Br}(t \to b W) = 1$, with a width computed at LO in the 5FS
equal to $\Gamma_t = 1.4803$~GeV.%
\footnote{In the 4FS, due to a non-zero bottom mass, the LO width is slightly
reduced to $\Gamma_t = 1.4763$~GeV.
}
Spin correlations can be preserved by decaying the events 
with {\sc MadSpin}~\cite{Artoisenet:2012st}, following the procedure presented in~\cite{Frixione:2007zp}.
We choose to leave the $W$ bosons stable, because we focus on the 
behaviour of the $b$ jets stemming either from the top decay or from the 
initial-state gluon splitting.

Short-distance events are matched to the \textsc{Pythia8} 
parton shower~\cite{Sjostrand:2007gs} by
using the \textsc{MC@NLO} method~\cite{Frixione:2002ik}.
Jets are defined using the anti-$k_T$ algorithm~\cite{Cacciari:2008gp} 
implemented in \textsc{FastJet}~\cite{Cacciari:2011ma}, with radius $R=0.4$,
and required to have
\begin{align}
 p_T(j)>20\,{\rm GeV}\,, \quad |\eta(j)|<4.5\,.
\label{eq:jet_definition}
\end{align}
A jet is $b$-tagged if a $b$ hadron is found among its constituents
(we ideally assume 100\% $b$-tagging efficiency in our studies). 
The same kinematic cuts are applied for $b$ jets as for light flavour 
jets in the inclusive study. 
In the fiducial phase space, on the other hand, a requirement on the pseudorapidity of 
\begin{align}
|\eta(j_b)|<2.5
\end{align}
is imposed, resembling acceptances of $b$-tagging methods employed by 
the experiments.

\section{$\boldsymbol{tW}$ production}
\label{sec:tW_nlo}

In this section we (re-)compute NLO+PS calculations for
$tW$ production at the LHC,
running with a centre-of-mass energy $\sqrt{s}=13$~TeV.
With the shorthand $tW$ we mean the sum of the two processes
$pp \to tW^{-}$ and $pp \to \bar t W^{+}$, which have the same rates 
and distributions at the LHC.
We carefully quantify the impact of theoretical systematics in the 
event generation.
Our discussion is split in two parts, focusing first on the 
inclusive event generation and the related theoretical issues, and
then on what happens when fiducial cuts are applied.

\subsection{Inclusive results}
\label{sec:tW_nlo_inclusive}

\begin{figure*}
\center
 \includegraphics[width=0.48\textwidth]{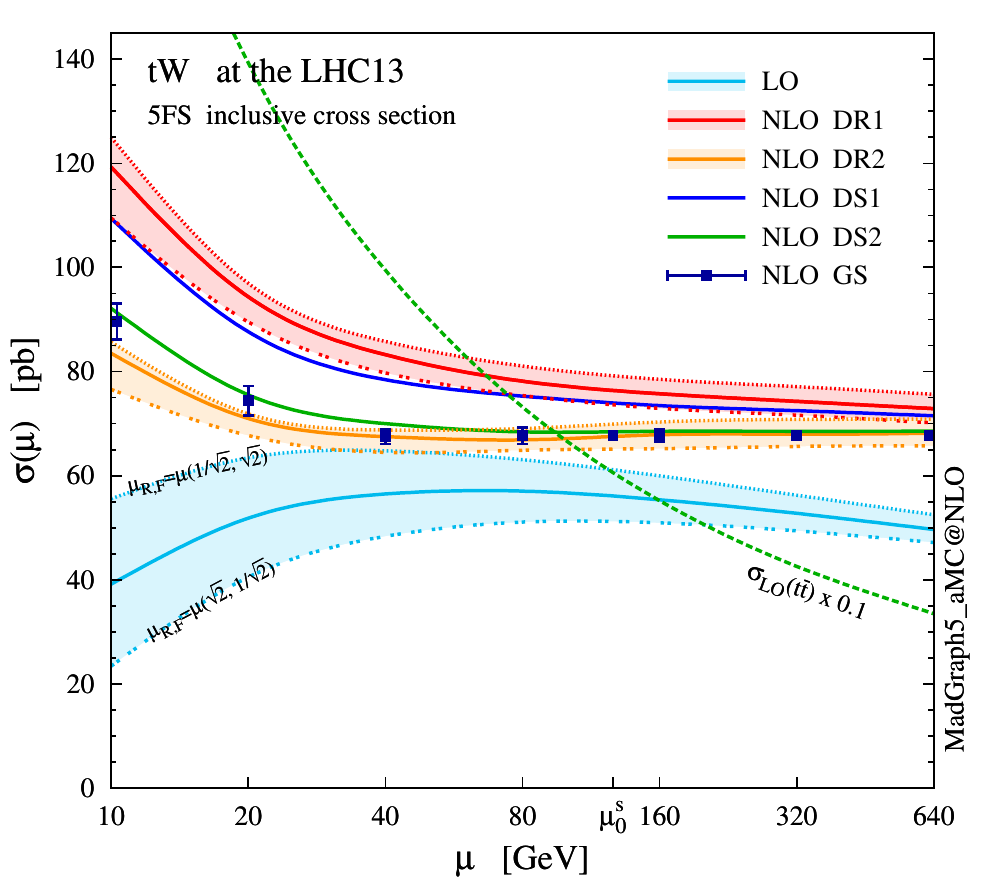}
 \hspace*{0.5em}
 \includegraphics[width=0.48\textwidth]{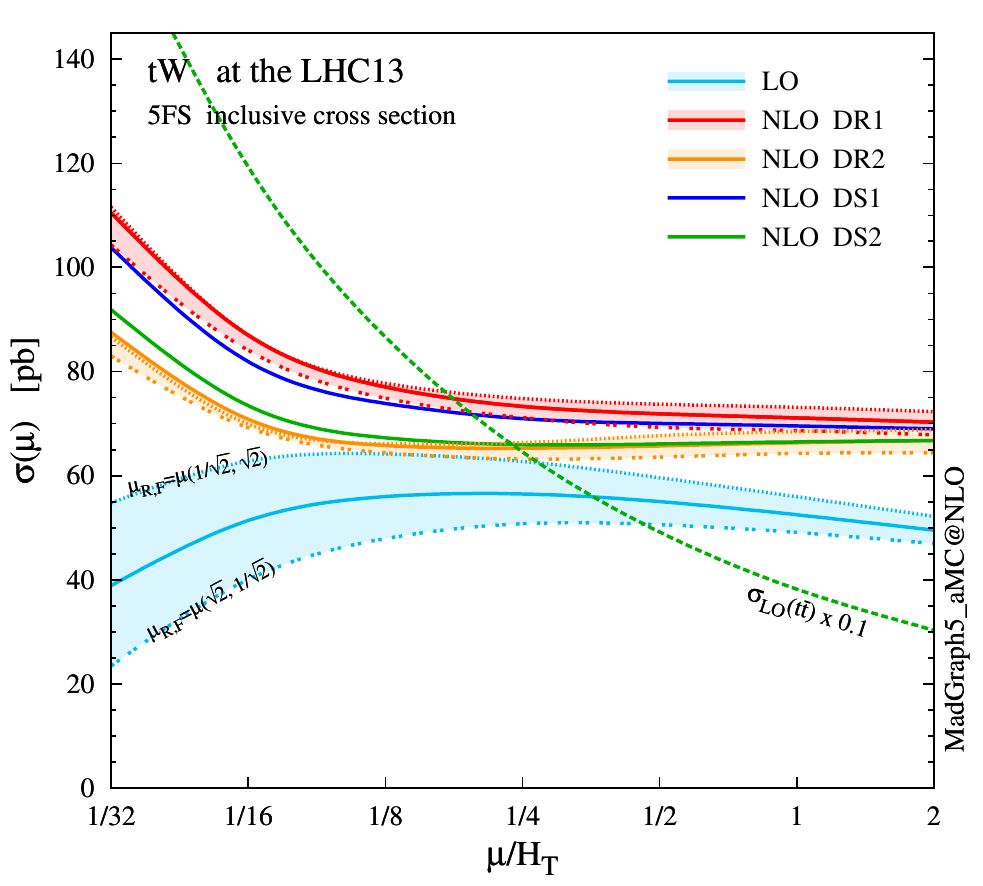}
 \caption{Scale dependence of the total cross section for
 $pp \to tW^{-}$ and $\bar tW^{+}$ at the 13-TeV LHC,
 computed in the 5FS at LO and NLO accuracy,
 presented for $\mu_F = \mu_R \equiv \mu$ using a static scale (left) 
 and a dynamic scale (right).
 The NLO $tWb$ channels are treated using DR and DS, see 
 sec.~\ref{sec:DRandDS} for more details.
 Furthermore, we show NLO results from GS (only for a static scale),
 and two off-diagonal profiles of the scale dependence,
 $( \mu_R=\sqrt{2}\mu\,,\,  \mu_F=\mu/\sqrt{2} )$ and
 $( \mu_R=\mu/\sqrt{2}\,,\,  \mu_F=\sqrt{2}\mu )$, 
 for LO and NLO DR.
 Finally, the scale dependence of $pp \to t \bar t$ at LO is also
 reported for comparison.}
\label{fig:tW_scalediag}
\end{figure*}

We start by showing in fig.~\ref{fig:tW_scalediag} the renormalisation and 
factorisation scale dependence of the $pp \to tW$ cross section,
computed at LO and NLO accuracy, keeping the $t$ stable. 
Results are obtained by employing the static and dynamic scales $\mu_0^s$ and $\mu_0^d$ 
(defined in sec.\ref{sec:nlosetup}) in the left and right plot respectively.
We show results where we simultaneously vary the renormalisation and factorisation scales on the diagonal
$\mu_R = \mu_F$; on top of this, for LO and NLO DR results, we also 
present two off-diagonal profiles where $\mu_R = \sqrt 2 \mu_F$ and  $\mu_R = \mu_F /\sqrt 2$.
In the two plots we present predictions  obtained employing both DR, neglecting (DR1, red) or taking into account (DR2, orange) the interference with $t \bar t$, 
and DS, with the two Breit--Wigner forms in eq.~\eqref{ds1} (DS1, blue)
or in eq.~\eqref{ds2} (DS2, green).
We also report results using global subtraction (GS, squares) for the static scale choice.
The details for the various NLO schemes can be found in sec.~\ref{sec:DRandDS}.
We remark that we have validated our NLO DR1 and DS1 codes against the \textsc{MC@NLO} code, 
finding very good agreement.
The values of the total rate computed at the central scale $\mu_0$ are also quoted
in table~\ref{tab:tW_NLO_xsect_1}.
Unlike in fig.~\ref{fig:tW_scalediag}, in this case scale variations are computed by varying 
$\mu_F$ and $\mu_R$ independently by a factor two around $\mu_0$.

\begin{table*}
\center \small 
\begin{tabular}{lllllllll}
 \hline
 \rule{0pt}{3ex}
 $tW$ (13~TeV) \hspace*{0.5em}
 & $\sigma(\mu_0^{s})$~[pb]
 & $\delta^\perc_{\mu}$
 & $\delta^\perc_{\mathrm{PDF}}$
 & $K$ 
 & $\sigma(\mu_0^{d})$~[pb]
 & $\delta^\perc_{\mu}$
 & $\delta^\perc_{\mathrm{PDF}}$
 & $K$
 \\[0.7ex]
 \hline
 \rule{0pt}{3ex}  LO
   &  56.07(3)  &  $^{+18.2}_{-17.4}$  &  \scriptsize{$\pm 8.4$ }   &  -     
   &  56.50(6)  &  $^{+21.9}_{-20.9}$  &  \scriptsize{$\pm 8.4$ }   &  -     \\[0.3ex]
 \rule{0pt}{3ex}  NLO DR1 
   &  76.46(9)  &  $^{+6.9}_{-8.1}$    &  \scriptsize{$\pm 2.0$ }   &  \footnotesize 1.36
   &  73.22(9)  &  $^{+5.1}_{-6.7}$    &  \scriptsize{$\pm 2.0$ }   &  \footnotesize 1.30  \\[0.3ex]
 \rule{0pt}{3ex}  NLO DR2 
   &  67.49(9)  &  $^{+6.3}_{-8.1}$    &  \scriptsize{$\pm 2.0$ }   &  \footnotesize 1.20
   &  65.12(9)  &  $^{+2.8}_{-6.8}$    &  \scriptsize{$\pm 2.0$ }   &  \footnotesize 1.15  \\[0.3ex]
 \rule{0pt}{3ex}  NLO DS1 
   &  73.80(9)  &  $^{+6.7}_{-8.1}$    &  \scriptsize{$\pm 1.9$ }   &  \footnotesize 1.32
   &  70.93(9)  &  $^{+4.0}_{-6.7}$    &  \scriptsize{$\pm 2.0$ }   &  \footnotesize 1.26  \\[0.3ex]
 \rule{0pt}{3ex}  NLO DS2 
   &  68.28(8)  &  $^{+6.6}_{-8.3}$    &  \scriptsize{$\pm 2.1$ }   &  \footnotesize 1.22
   &  66.09(9)  &  $^{+2.8}_{-6.8}$    &  \scriptsize{$\pm 1.9$ }   &  \footnotesize 1.17  \\[0.3ex]
 \rule{0pt}{3ex} NLO GS 
   &  67.8(7)    &  -  &  -  &  \footnotesize 1.21(1)  &  &  &  &    \\[0.7ex]
 \hline
\end{tabular}
\caption{Total cross sections for 
 $pp \to t W^-$ and $\bar t W^+$ at the 13-TeV LHC,  in the 5FS at LO and NLO accuracy with different  schemes, 
 computed with a static scale $\mu_0^{s} = (m_t+m_W)/2$
 and a dynamic scale $\mu_0^{d} = H_T/4$.
 We also report the scale and PDF uncertainties and the NLO-QCD $K$ factors;
 the numerical uncertainty affecting the last digit is quoted in parentheses.}
\label{tab:tW_NLO_xsect_1}
\end{table*}

As expected, NLO corrections visibly reduce the scale dependence with respect to LO predictions. 
Comparing DR1 and DR2, we see that interference effects are negative at this centre-of-mass energy, 
and reduce significantly the NLO cross section, by about 13\%.
Also, the cross section scale dependence is different, in particular for very small scales.
This effect is driven by the LO scale dependence in $t \bar t$ amplitudes, which is larger at low scales.
Moving to DS, we find that DS1 and DS2 predictions show a 8\% difference.
Therefore, the dependence on the subtraction scheme is large, being comparable to the scale uncertainty 
or even larger.

We note that the total rate predictions obtained with DR2 and DS2 agree rather well within uncertainties, 
especially at the reference scale choice, and also agree with the predictions from the GS scheme. 
This result is quite satisfactory because it supports some important observations.
First, that the off-shell effects of the top-quark resonant diagrams are small, 
and indeed well described by the (gauge invariant) parametrisation of eq.~\eqref{eq:ds2}.
Second, that possible gauge dependence in DR2 is in practice not an issue if one uses a covariant gauge,
where the subtraction of $|\mathcal{A}_{2t}|^2$ turns out to be very close to an on-shell 
gauge-invariant subtraction.
On the other hand, DR1, which does not include the interference in the definition of the signal, 
and DS1, which has a different profile over the virtuality of the intermediate top quark, 
do not describe well the NLO effects and extrapolate to a biased total cross section, 
even in the $\Gamma_t \to 0$ limit.
Thus, a third observation is that interference terms are not negligible, and it is mandatory 
to keep them in the definition of the $tW$ process in order to have a complete simulation.
Finally, a fourth point is that to include interference effects is not enough, but one also
needs to subtract the top-pair process with an adequate profile over the phase space.
This picture is confirmed at the level of differential distributions in the following discussion, 
and also at the total cross section level in the 4FS, see \ref{app:twb_lo_4FS}.
 
\begin{figure*}
\center
 \includegraphics[width=0.325\textwidth]{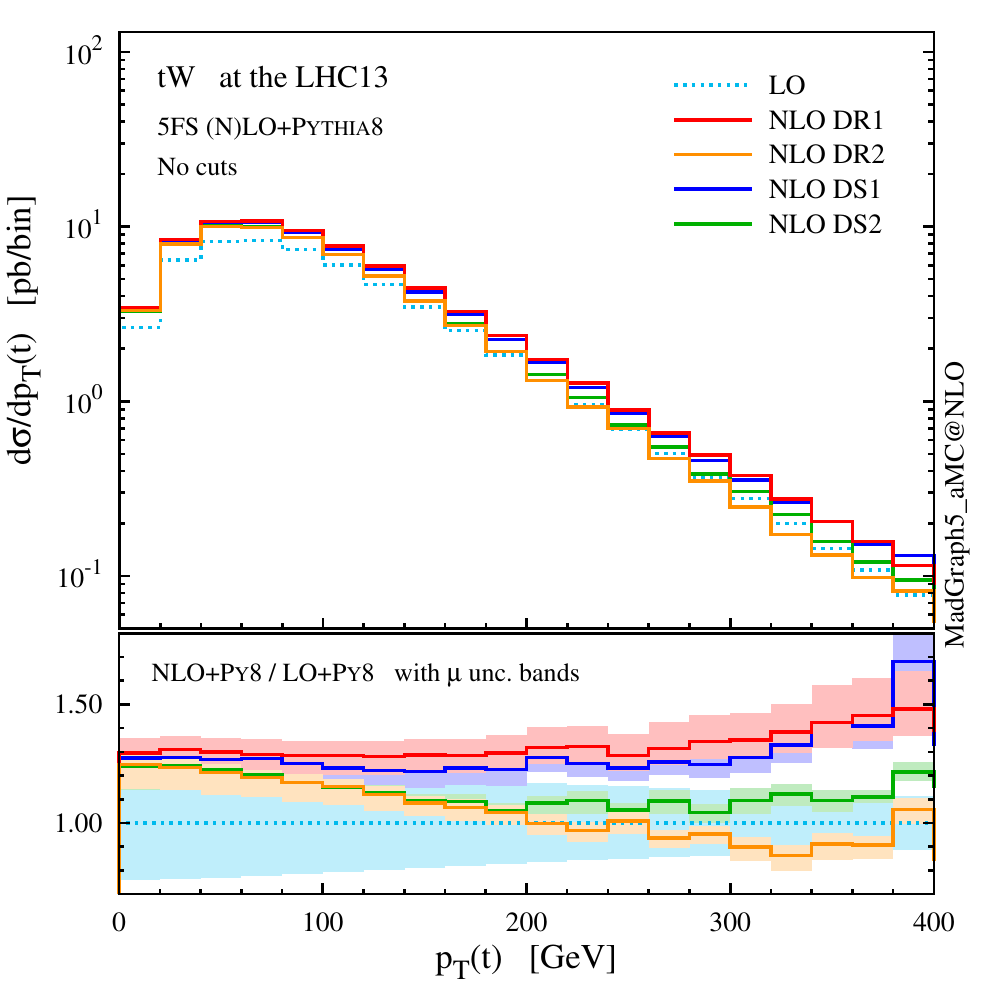}\qquad
 \includegraphics[width=0.325\textwidth]{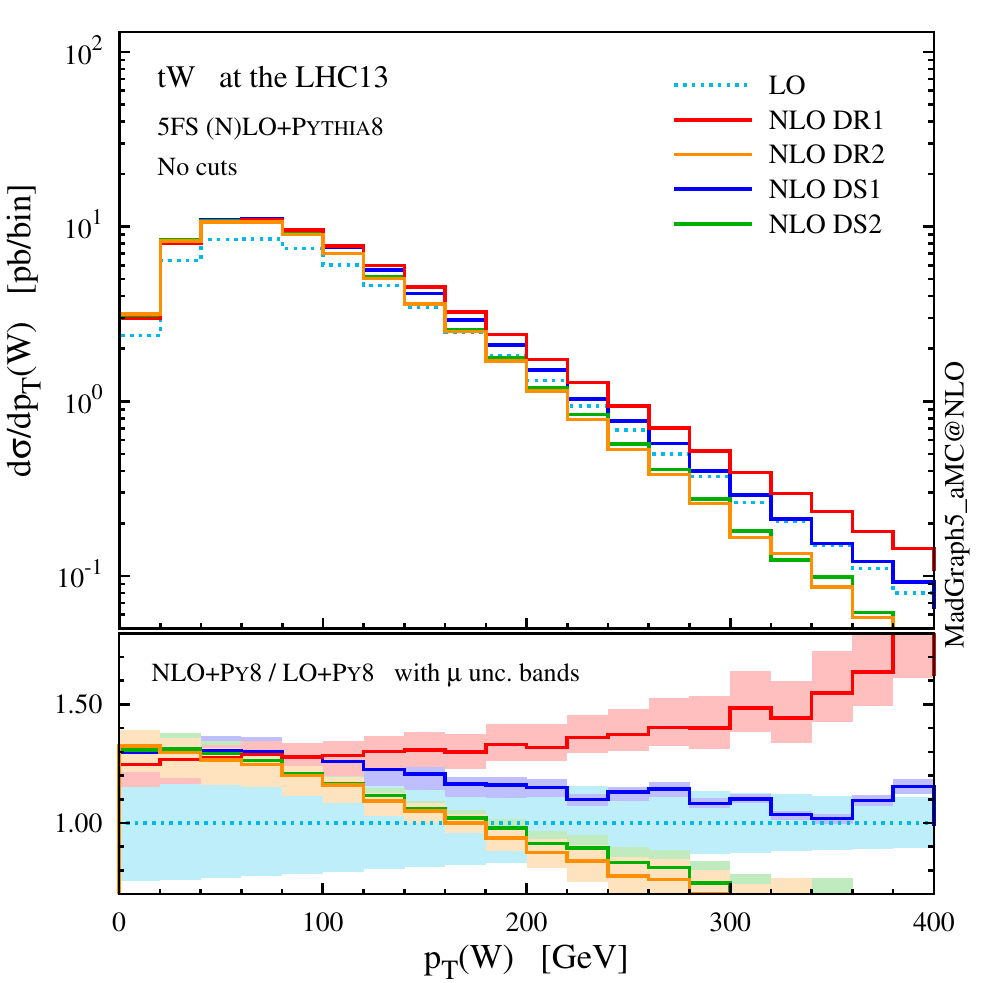}\\
 \includegraphics[width=0.325\textwidth]{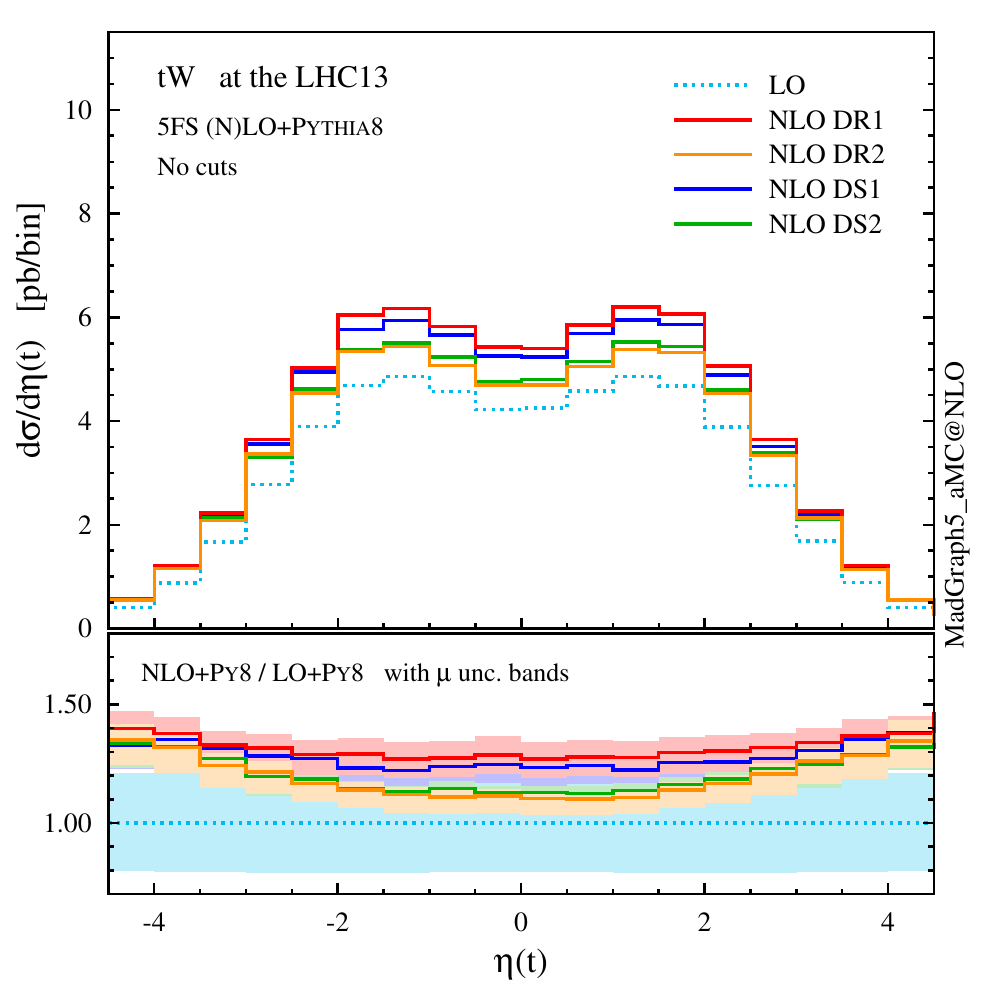}\qquad
 \includegraphics[width=0.325\textwidth]{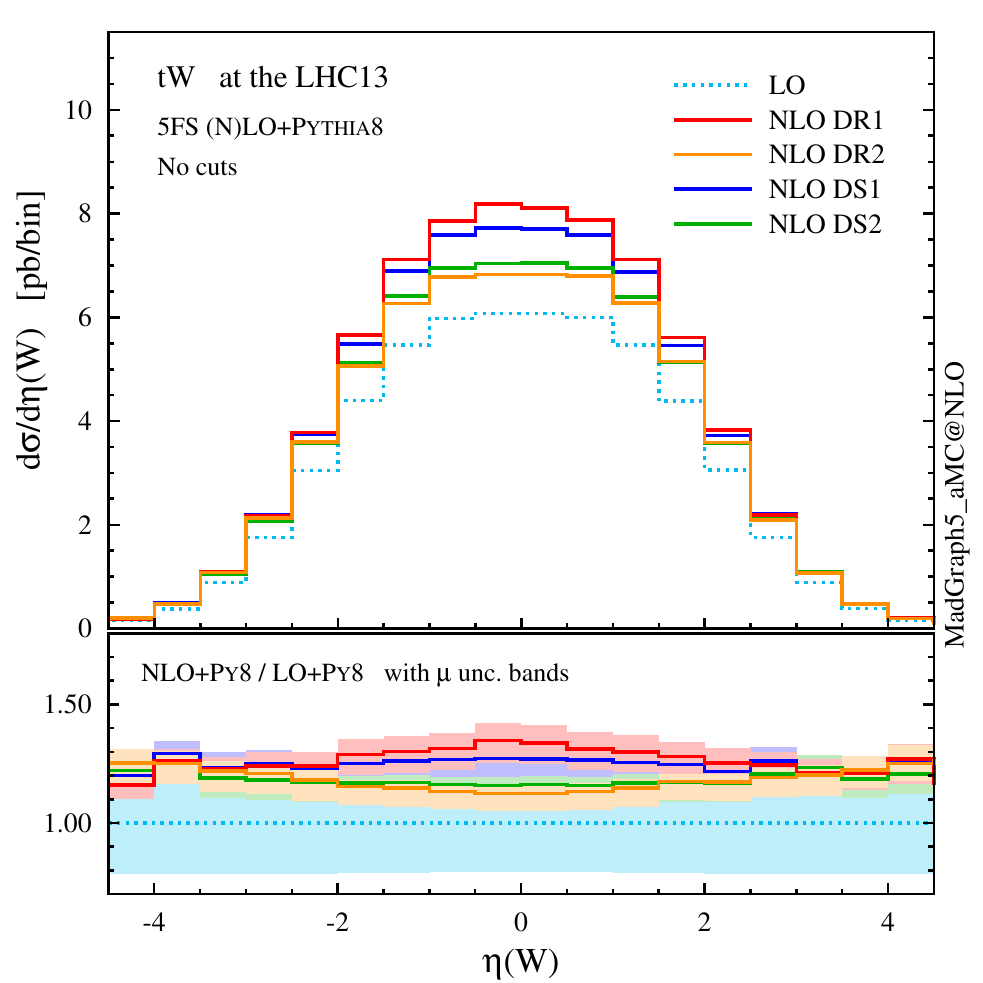} 
\caption{\label{fig:tW_5FNLO_dist_1}
$p_T$ and $\eta$ distributions for the top quark and the $W$ boson at NLO+PS accuracy in $tW$ production at the 13-TeV LHC. The lower panels provide information on the differential $K$ factors with the scale uncertainties.  
}

\center
 \includegraphics[width=0.325\textwidth]{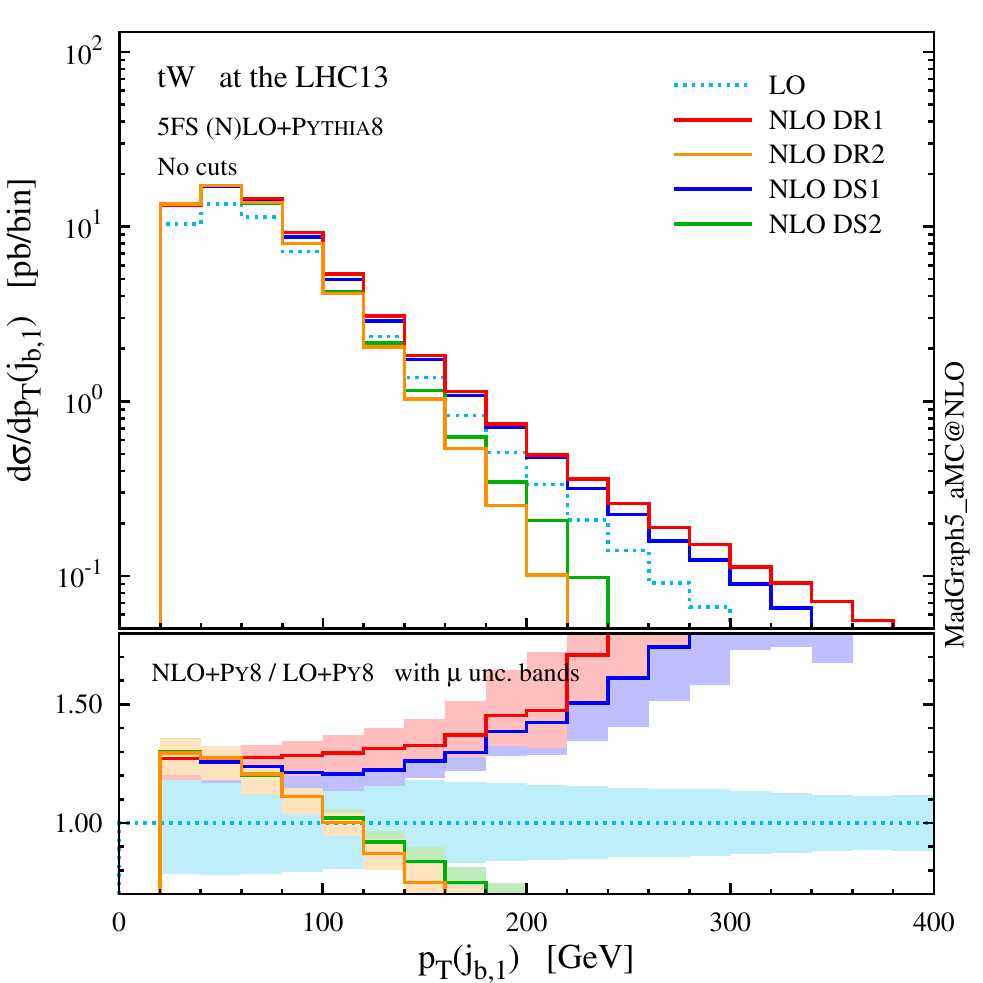}\qquad
 \includegraphics[width=0.325\textwidth]{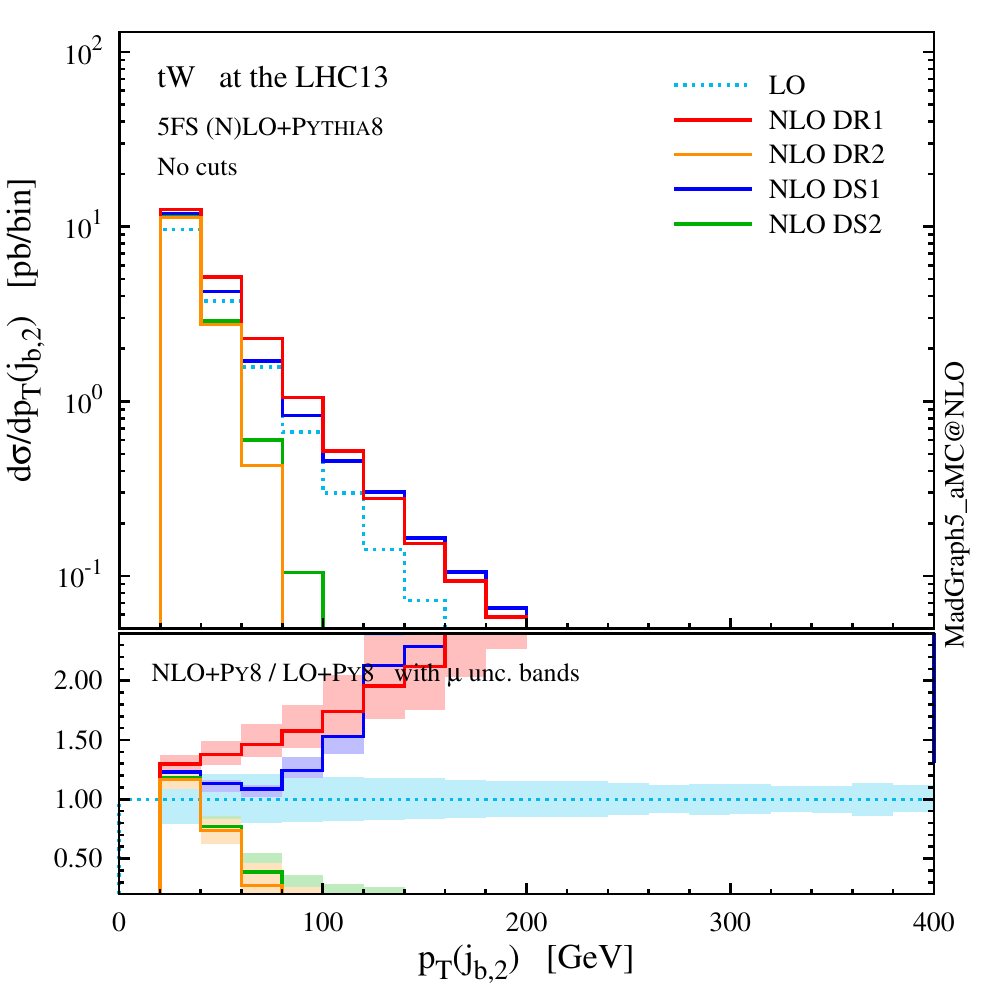}\\
 \includegraphics[width=0.325\textwidth]{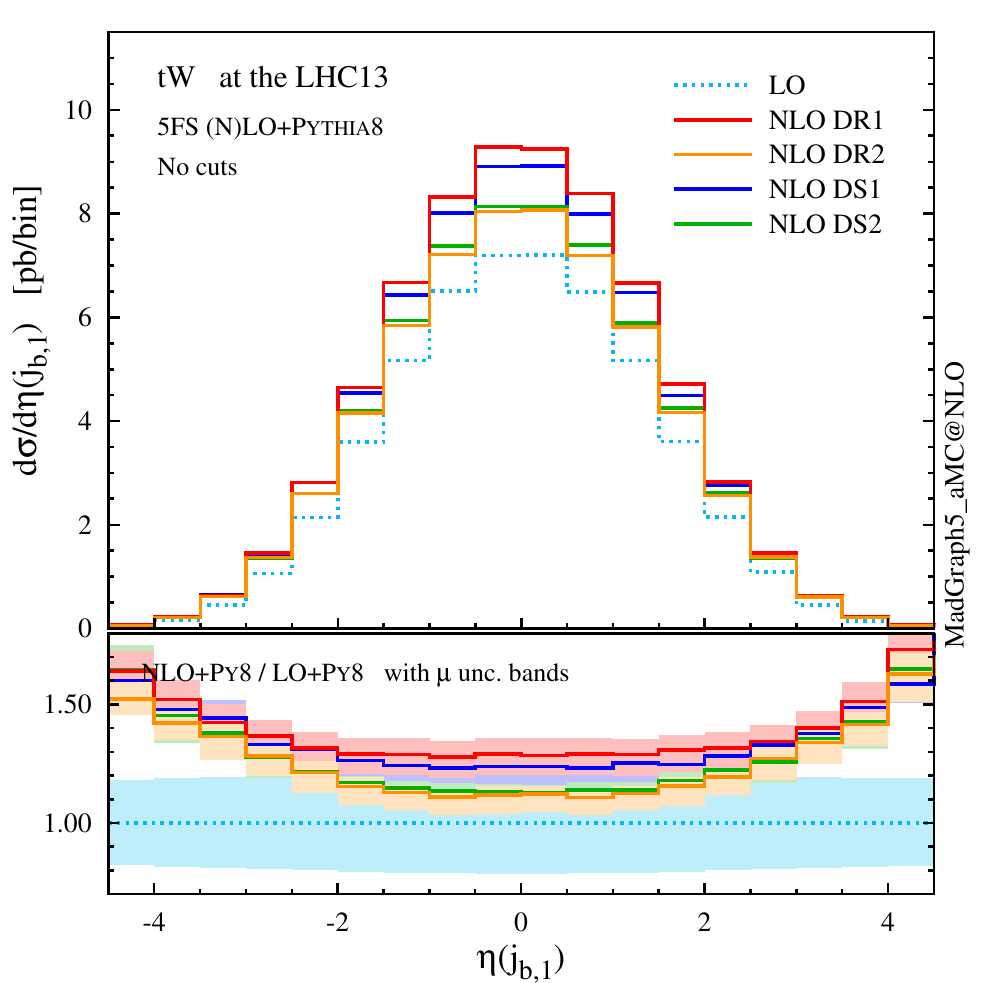}\qquad
 \includegraphics[width=0.325\textwidth]{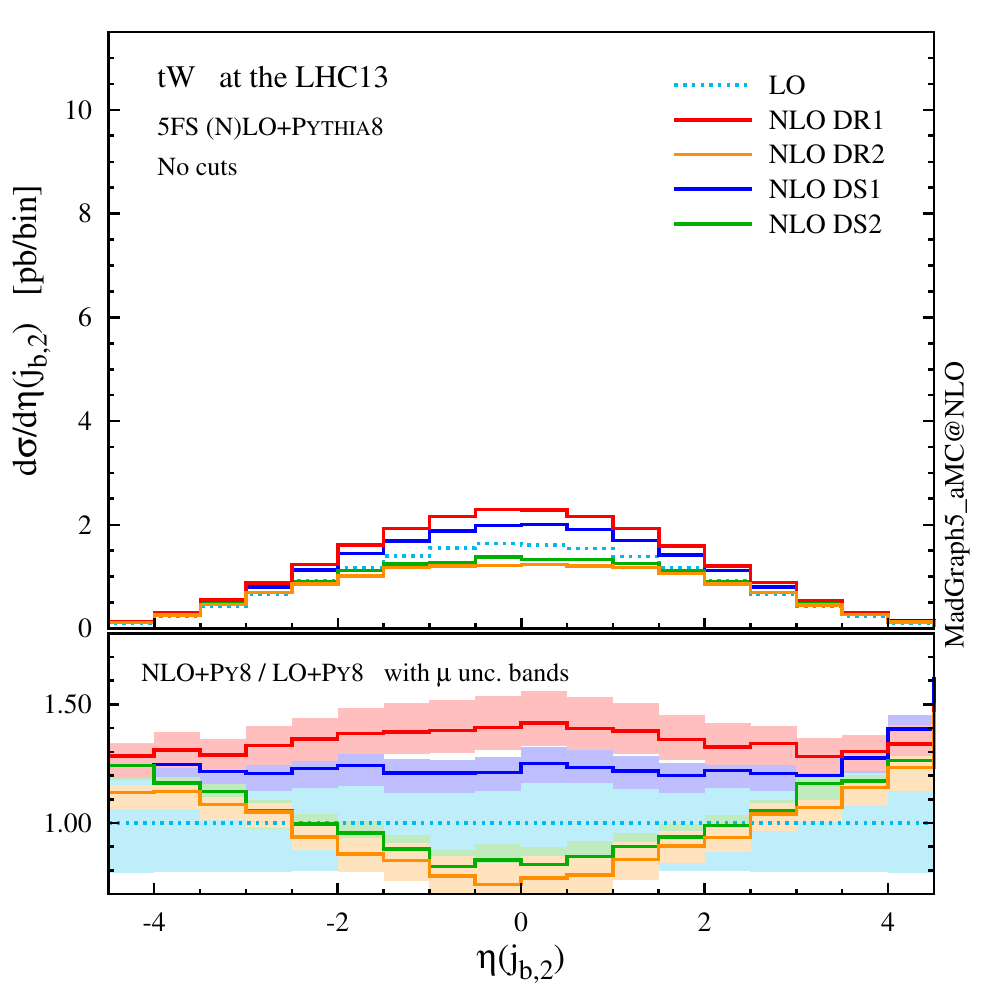}
\caption{\label{fig:tW_5FNLO_dist_1b}
Same as fig.~\ref{fig:tW_5FNLO_dist_1}, but for the $b$-tagged jets.
Note that the second-hardest $b$ jet is described by the parton shower at LO, while by the matrix element at NLO.
}
\end{figure*}

We now turn to differential distributions, and we show some relevant observables in figs.~\ref{fig:tW_5FNLO_dist_1} and \ref{fig:tW_5FNLO_dist_1b}.
Here, we employ a dynamical scale choice, $\mu_0 = H_T/4$ and we do not impose any cut on the final-state particles.
Note that, for simplicity and after the shorthand $tW$, we label as $t$ 
both the undecayed top quark in $tW^{-}$ production 
and the antitop in $\bar t W^+$; similarly, $W$ indicates the $W^{-}$ in
the first process and $W^+$ in the second one, i.e. the boson produced 
in association with $t$, and not the one coming from the $t$ decay. 
Particles (not) coming from the top decay are identified
by using the event-record information. 
We see that the DR1 and DS1 simulations tend to produce harder and more central distributions,
while the DR2 and DS2 results, very similar one another, tend to be softer and more forward. 
In any case, NLO corrections cannot be taken into account by the LO scale uncertainty, nor
be described by a $K$ factor, especially for the physics of $b$ jets.
The hardest $b$ jet ($j_{b,1}$) dominantly comes from the top decay, while
the second-hardest $b$ jet is significantly softer due to the initial-state $g\to b \bar b$ splitting.
As seen for DR2, the high-$p_T$ $W$ boson and $b$ jets are highly suppressed 
due to the negative interference with the $t \bar t$ process.
In fact, due to this interference the cross section can become negative in some corners
of the phase space, for example in the high-$p_T$ tail of the second $b$ jet.
We interpret this fact as a sign that $tW$ cannot be separated from $t \bar t$ in this region,
and the two contributions must be combined in order to obtain a physically observable (positive) cross section.

In summary, the $tW$--$t\bar t$ interference significantly 
affects the inclusive total rate as well as the shapes of various distributions at NLO. 
In particular, different schemes give rise to different NLO results, with ambiguities 
which in principle can be larger than the scale uncertainty. 
Such differences arise from two sources: the interference between resonant (top-pair) and non-resonant 
(single-top) diagrams, which is relevant and ought to be taken into account,
and (in the case of DS) the treatment of the off-shell tails of the top-pair contribution. 
These ambiguities are intrinsically connected to the attempt of separating two processes 
that cannot be physically separated in the whole phase space. 
On the other hand, we have also found that two of such schemes, DR2 and DS2, 
give compatible results among themselves and integrate up to the total cross section 
defined in a gauge invariant way in the GS scheme. 
We are now ready to explore whether a region of phase space (possibly accessible from the experiments) 
exists where the two processes can be separated in a meaningful way.

\subsection{Results with fiducial cuts}
\label{sec:tW_nlo_fiducial}

\begin{table*}
\center \small 
\begin{tabular}{lllllllllllll}
 \hline
 \rule{0pt}{3ex}  & \multicolumn{1}{c}{No cuts}  & \multicolumn{2}{c}{Fiducial cuts}  & \multicolumn{2}{c}{Fiducial cuts + top reco.}   \\[0.3ex]
 \rule{0pt}{3ex}    
 & $\sigma_{\mathrm{NLO}}\pm\delta^\perc_{\mu}\pm\delta^\perc_{\mathrm{PDF}}$
 & $\sigma_{\mathrm{NLO}}\pm\delta^\perc_{\mu}\pm\delta^\perc_{\mathrm{PDF}}$
 & eff.
 & $\sigma_{\mathrm{NLO}}\pm\delta^\perc_{\mu}\pm\delta^\perc_{\mathrm{PDF}}$
 & eff.
 \\[0.7ex] 
 \hline
 \rule{0pt}{3ex}  $t \bar t$ 
   &  744.1(9)\,$^{+4.8}_{-8.7}$\,\scriptsize{$\pm 1.7$ }  &   44.9(3)\enskip\,$^{+6.0}_{-9.5}$\,\scriptsize{$\pm 1.9$ }   &  0.06     &   44.9(3)\enskip\,$^{+6.0}_{-9.5}$\,\scriptsize{$\pm 1.9$ }   &  0.06  \\[0.3ex]
 \rule{0pt}{3ex}  $tW$ DR1                                                                                             
   &  73.22(9)\,$^{+5.1}_{-6.7}$\,\scriptsize{$\pm 2.0$ }  &  44.70(7)\,$^{+4.0}_{-6.7}$\,\scriptsize{$\pm 1.9$ }   &  0.61     &  41.70(7)\,$^{+3.8}_{-6.8}$\,\scriptsize{$\pm 1.9$ }   &  0.57  \\[0.3ex]
 \rule{0pt}{3ex}  $tW$ DR2                                                                                             
   &  65.12(9)\,$^{+2.8}_{-6.8}$\,\scriptsize{$\pm 2.0$ }  &  43.88(8)\,$^{+3.2}_{-7.0}$\,\scriptsize{$\pm 1.9$ }   &  0.67     &  41.85(8)\,$^{+3.7}_{-7.0}$\,\scriptsize{$\pm 1.9$ }   &  0.64  \\[0.3ex]
 \rule{0pt}{3ex}  $tW$ DS1                                                                                             
   &  70.93(9)\,$^{+4.0}_{-6.7}$\,\scriptsize{$\pm 2.0$ }  &  44.65(8)\,$^{+3.8}_{-6.8}$\,\scriptsize{$\pm 1.9$ }   &  0.63     &  41.90(8)\,$^{+3.8}_{-6.8}$\,\scriptsize{$\pm 1.9$ }   &  0.59  \\[0.3ex]
 \rule{0pt}{3ex}  $tW$ DS2                                                                                             
   &  66.09(9)\,$^{+2.8}_{-6.8}$\,\scriptsize{$\pm 1.9$ }  &  44.05(8)\,$^{+3.3}_{-6.9}$\,\scriptsize{$\pm 1.9$ }   &  0.67     &  41.91(8)\,$^{+3.8}_{-6.9}$\,\scriptsize{$\pm 1.9$ }   &  0.63  \\[0.7ex]
 \hline
\end{tabular}
\caption{Total cross sections in pb at the LHC 13~TeV for the processes $pp \to t \bar t$
 and $pp \to t W$, in the 5FS at NLO+PS accuracy.
 Results are presented before any cut (left), after fiducial cuts (center), and also adding top reconstruction 
 on the event sample (right). 
 We also report the scale and PDF uncertainties, as well as the cut efficiency with respect to the case with no cuts.
 All numbers are computed with the reference dynamic scale $\mu_0 = H_T/4$,
 and the numerical uncertainty affecting the last digit is reported in parentheses.}
\label{tab:tW_NLO_xsect_2}
\end{table*}

In this section we would like to investigate whether $tW$ can be defined separately from 
$t \bar t$ at least in some fiducial region of the phase space, in the sense that in such a region interference terms
between the two processes and thus theoretical ambiguities are suppressed.
In practice, this goal can be achieved by comparing results among different NLO schemes,
 since the difference among them provides a measure of interference effects
and related theoretical systematics (gauge dependence in DR, subtraction term in DS).
We remark that the following toy analysis is mainly for illustrative purposes,
since the same procedure can be applied to any set of fiducial cuts 
defined in a real experimental analysis, also imposing a selection on 
specific decay products of the $W$ bosons.

Motivated by the $b$-jet spectra in fig.~\ref{fig:tW_5FNLO_dist_1}
and by experimental $tW$ searches, a popular strategy 
to suppress the $t \bar t$ background
as well as $tW$--$t\bar t$ interference is to select events with exactly 
one central $b$ jet~\cite{Tait:1999cf,Campbell:2005bb,White:2009yt,Chatrchyan:2012zca,Chatrchyan:2014tua,Aad:2015eto}.
We define our set of ``fiducial cuts'' for $tW$ by selecting only
events with
\begin{enumerate}
 \item exactly one $b$ jet with \enskip $p_T(j_b)>20$~GeV \\ and \enskip $|\eta(j_b)|<2.5$\,, \\[-0.5em]
 \item exactly two central $W$ bosons with rapidity \\ $|y(W)|<2.5$\,. 
\end{enumerate}
In this regard we stress that the first selection is the key to suppress the contributions
from $t\bar{t}$ amplitudes, hence both the pure $t \bar t$ ``background'' as well as
the $tW$--$t\bar t$ interference (i.e. theoretical ambiguities).
Note that we would like to draw general conclusions about the generation of $tW$ 
events, therefore we have chosen to define a pseudo event category that does 
not depend on the particular decay channel of the $W$ bosons.
The second selection is added to mimic a good reconstructability of these bosons 
inside the detector regardless of their final-state daughters; it affects 
less than $7\perc$ of the events surviving selection 1.

Looking at table~\ref{tab:tW_NLO_xsect_2} we can see that, 
before any cut is applied, the event category is largely dominated 
by the $t \bar t$ contribution. 
Once the above fiducial cuts are applied, the $t \bar t$ contribution
is reduced by more than a factor 16, while the $tW$ rate shrinks by 
about just one third (for DR2 and DS2), bringing the 
signal-to-background ratio $\sigma(tW)/\sigma(t \bar t)$ close to unity, 
which is exactly the aim of $tW$ searches.
The impact of interference has been clearly reduced by the cuts; 
The fiducial cross sections computed with the different NLO schemes agree
much better with each other, than before selections are applied.
Still, there is a minor residual difference in the rates, 
which amounts to about $2\perc$.

\begin{figure*}
\center
 \includegraphics[width=0.325\textwidth]{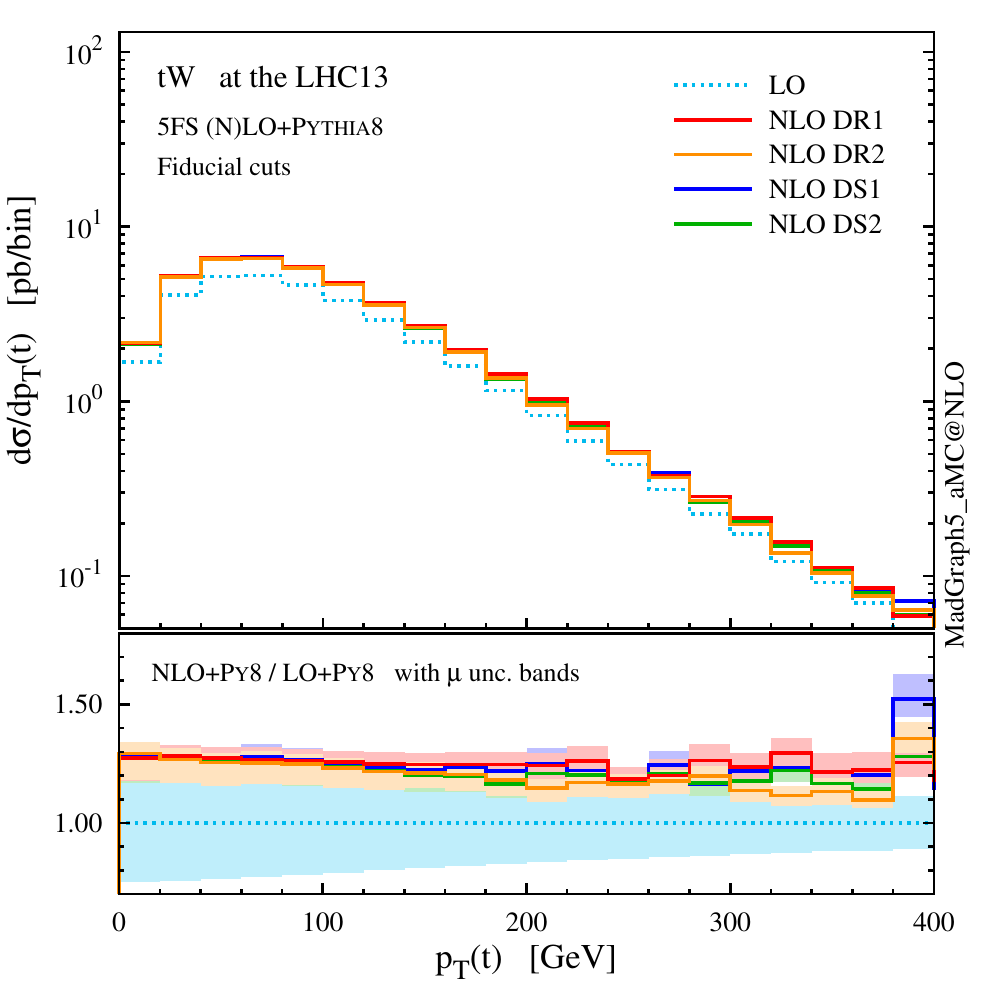}\qquad
 \includegraphics[width=0.325\textwidth]{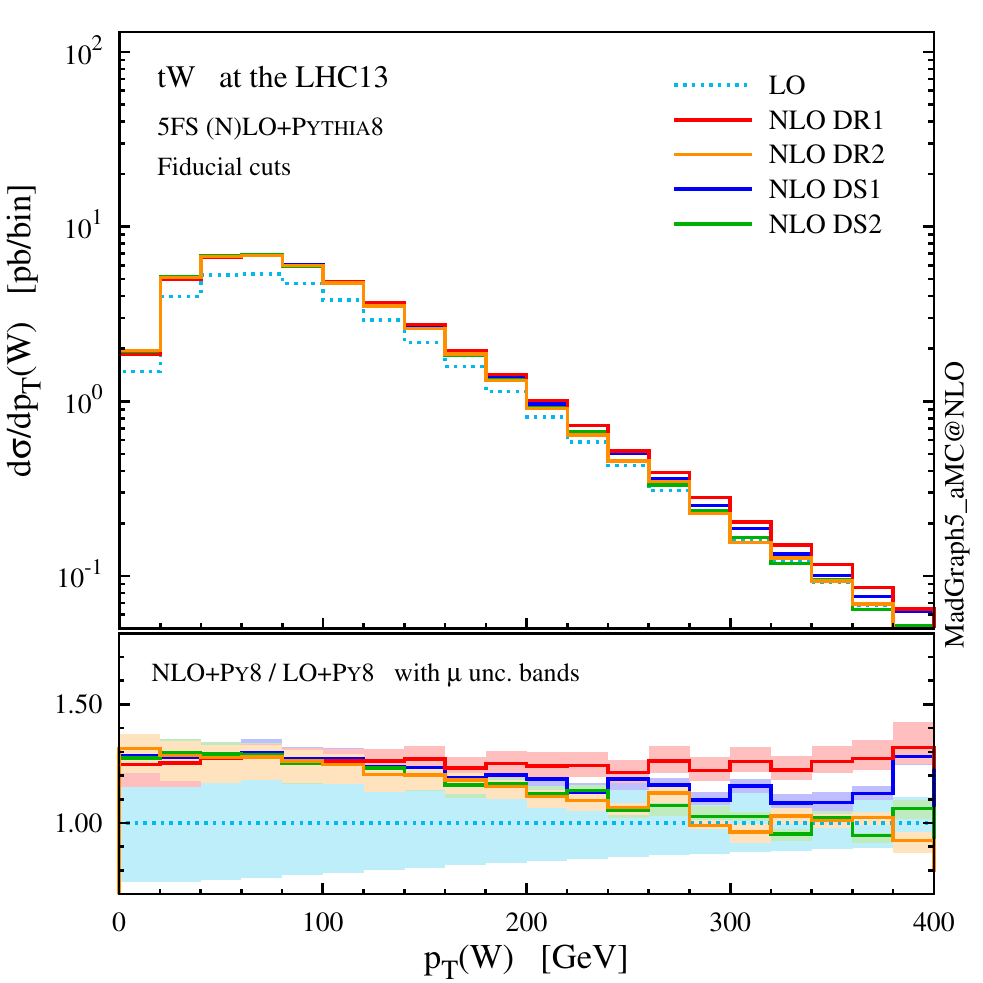}\\
 \includegraphics[width=0.325\textwidth]{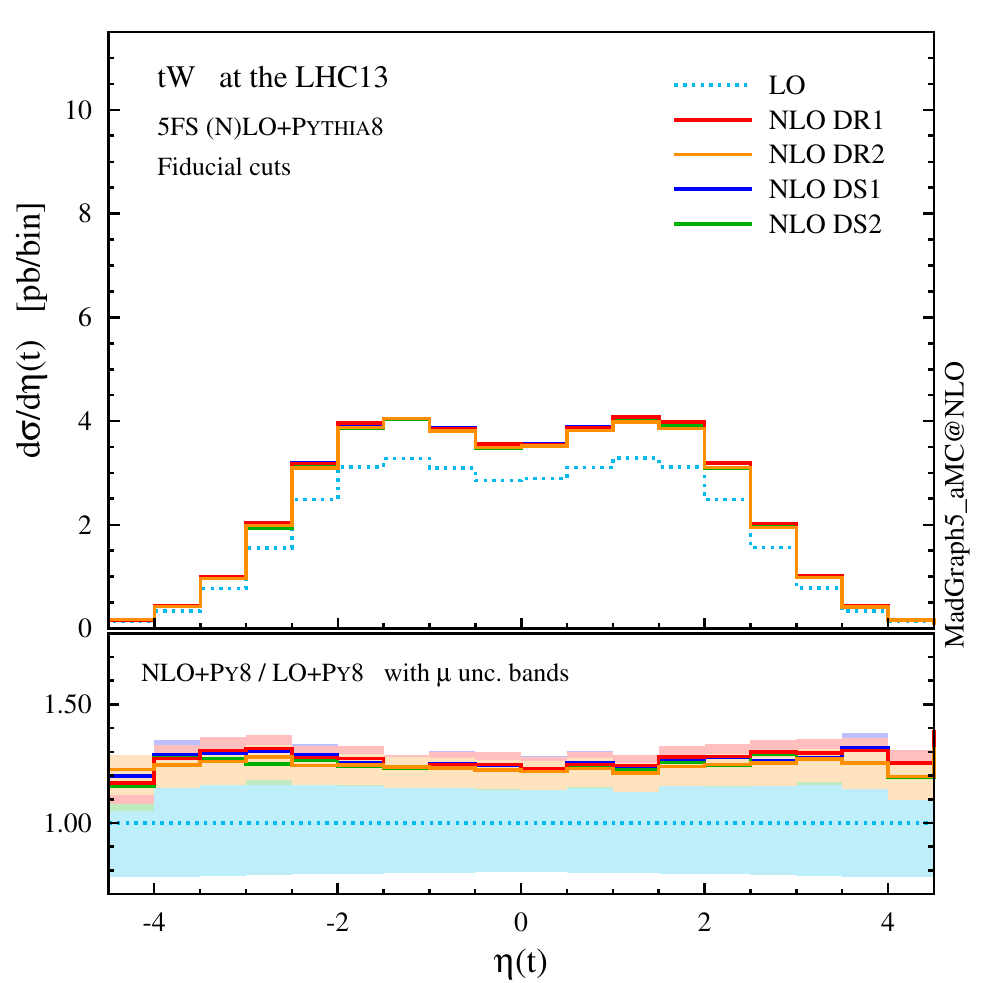}\qquad
 \includegraphics[width=0.325\textwidth]{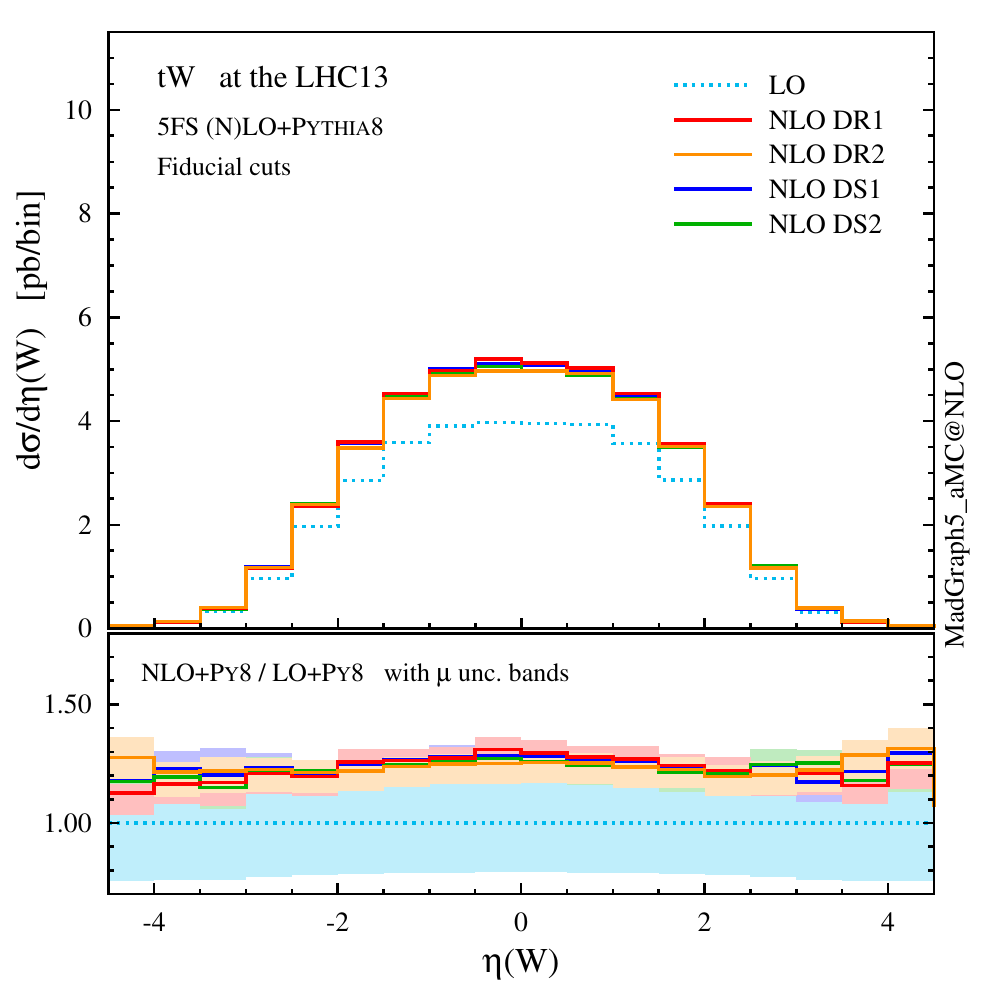}
\caption{\label{fig:tW_5FNLO_dist_2}
 $p_T$ and $\eta$ distributions the top quark and 
 the $W$ boson as in fig.~\ref{fig:tW_5FNLO_dist_1},
 but after applying the fiducial cuts to suppress interference between $tWb$
 and $t \bar t$.}

\vspace*{2em}

\center
 \includegraphics[width=0.325\textwidth]{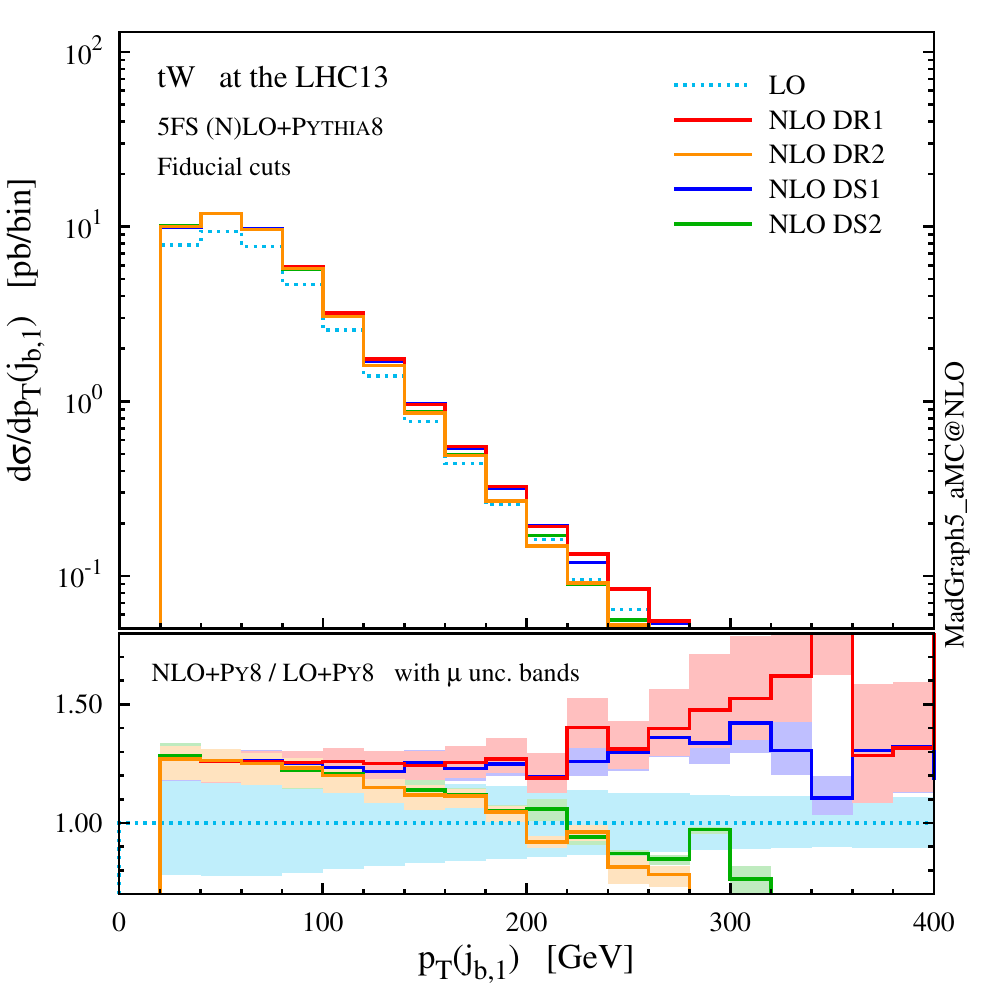}\qquad
 \includegraphics[width=0.325\textwidth]{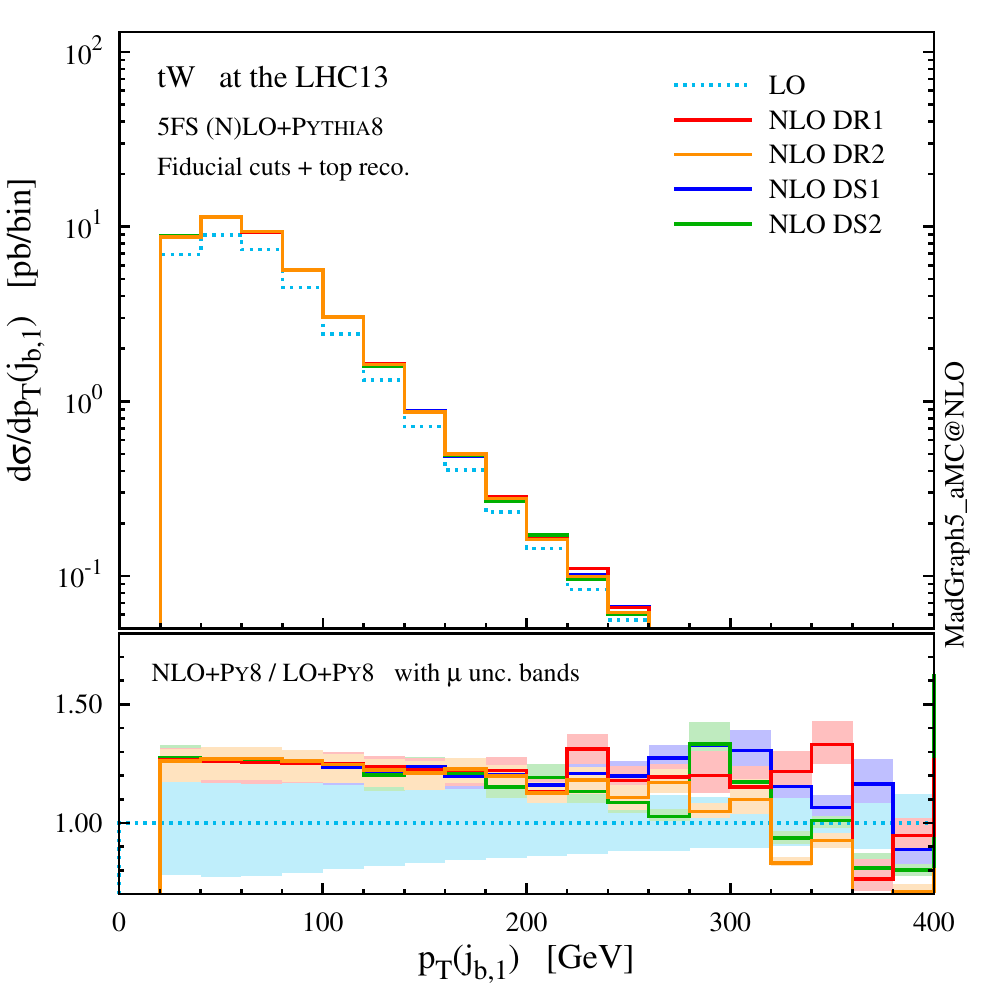}\\
 \includegraphics[width=0.325\textwidth]{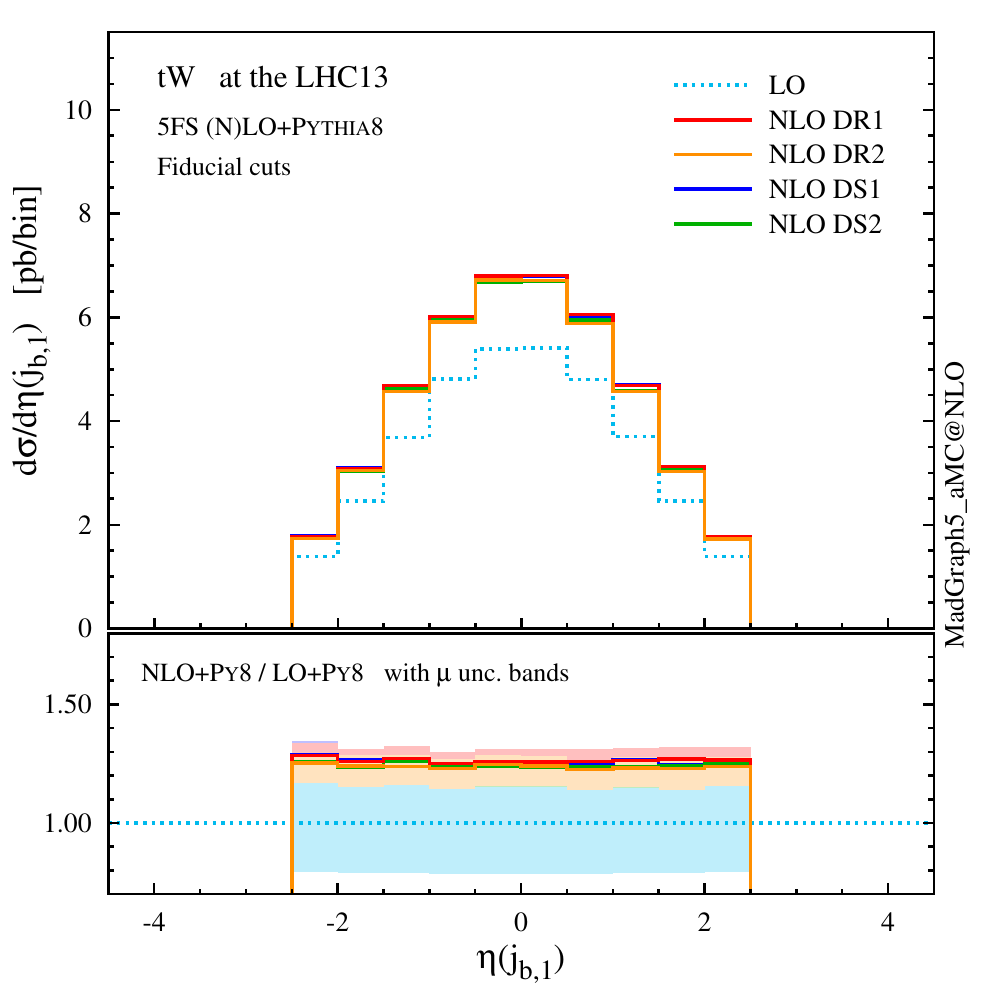}\qquad
 \includegraphics[width=0.325\textwidth]{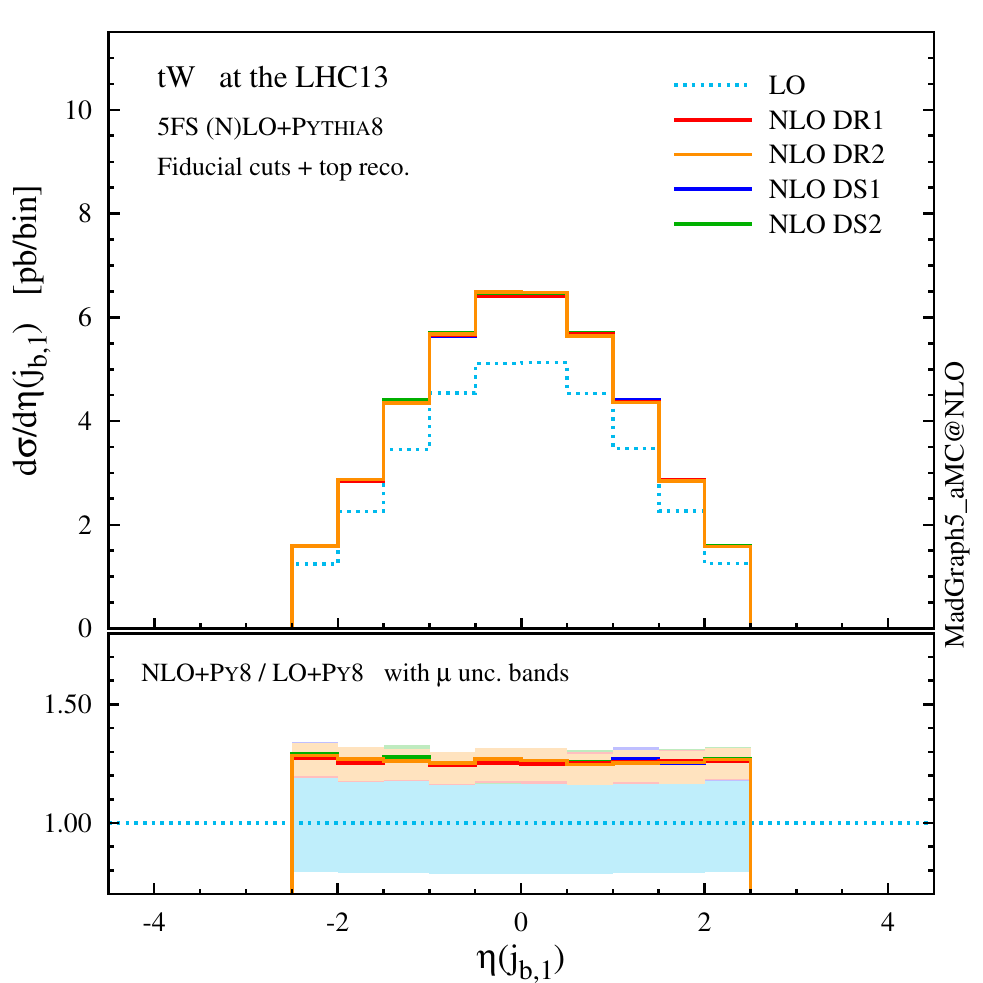}
\caption{\label{fig:tW_5FNLO_dist_2b}Same as fig.~\ref{fig:tW_5FNLO_dist_2}, but for the central $b$-tagged jet. 
For the right plot, in addition to the fiducial cuts, the top reconstruction is required.} 
\end{figure*}

\begin{figure*}
\center
\includegraphics[width=1.625\columnwidth]{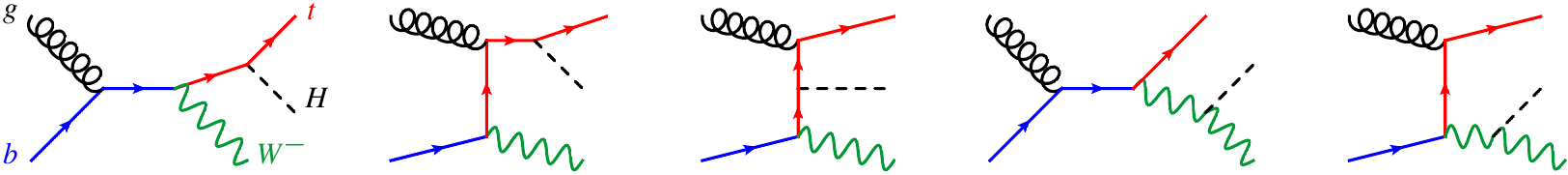}
\caption{LO Feynman diagrams for $tW^{-}H$ production in the 5FS.}
\label{fig:twh_5FS_diagrams}
\end{figure*} 

\begin{figure*}
\center 
\includegraphics[width=1.625\columnwidth]{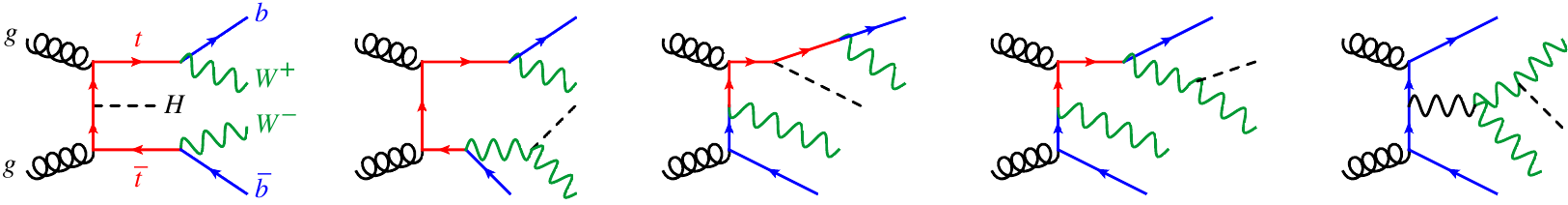}\\
($\mathcal{A}_{2t}$) 
\hspace*{5.8em} ($\mathcal{A}_{1t}$) \hspace*{5.8em} ($\mathcal{A}_{1t}$) 
\hspace*{5.8em} ($\mathcal{A}_{0t}$) \hspace*{5.8em} ($\mathcal{A}_{0t}$) 
\caption{Examples of doubly resonant (first on the left), 
singly resonant (second two) and 
non-resonant (last two) diagrams contributing to 
$WbWbH$ production. 
The first three diagrams (with the $t$ line cut) 
describe the NLO real-emission contribution to the $tW^{-}H$ process.
}
\label{fig:twbh_4FS_diagrams}
\end{figure*} 

From the distributions in figs.~\ref{fig:tW_5FNLO_dist_2} and \ref{fig:tW_5FNLO_dist_2b} we can see 
once more an improved agreement among the different NLO schemes in the fiducial region.
The lower panels show flatter and positive $K$ factors and a lower scale dependence in the high-$p_T$ tail 
than before the cuts, since we have suppressed the interference with LO $t \bar t$ amplitudes.
Although considerably mitigated, some differences are still visible among the four schemes in 
the high-$p_T$ region of the $b$-tagged jet ($j_{b,1}$).
Monte Carlo information shows that the central $b$ jet coincides 
with the one stemming from the top decay ($j_{b,t}$) for the vast majority 
of events. 
In the high-$p_T$ region, however, the $b$ jet can also be originated by a hard initial-state $g \to b \bar b$ splitting, similar to the case of $t$-channel $tH$ 
production~\cite{Demartin:2015uha}.

This suggests that, if on top of the fiducial cuts we also demand the central 
$b$ jet to unambiguously originate from the top quark, then we may be able 
to suppress even further the $tW$--$t \bar t$ interference and the related 
theoretical systematics.
In fact, we can see from table~\ref{tab:tW_NLO_xsect_2} and from the right plot in fig.~\ref{fig:tW_5FNLO_dist_2b}
that, after such a requirement is included in the event selection, 
the total rates as well as the distributions end up in almost perfect agreement, 
and one can effectively talk about $tW$ and $t \bar t$ as separate 
processes in this region: interference effects have been suppressed 
at or below the level of numerical uncertainty in the predictions.
A possible remark is that the top-reconstruction requirement shaves off
another $\sim 2$~pb of the cross section, i.e. more than the residual
discrepancy between the different NLO schemes before this last selection is applied.

To summarise, a naturally identified region of phase space exists where $tW$ is well defined, i.e. gauge invariant and  basically independent of the scheme used (either DR1, DR2, DS1, DS2) to  subtract the $t\bar t$ contribution. 
Given the fact that DS2 and DR2 also give consistent results outside the fiducial region 
and integrate to the same total cross section, equal to  the GS one, they can both be used in MC simulations. 
In practice, given the fact that the gauge-dependent effects are practically small when employing a covariant gauge,
and that the implementation in the code is rather easy, DR2 is certainly a very convenient scheme to use 
in simulations of $tW$ production in the 5FS, including the effects of interference with the $t \bar t$ contribution.
In addition, one can use the difference between DR1 and DR2 (i.e. the amount of $tW$--$t \bar t$ interference) 
to assess whether the fiducial region where the measurements are performed is such that the process-definition 
uncertainties are under control (smaller than the missing higher order uncertainties), 
and to estimate the residual process-definition systematics. 
We have seen that requiring the presence of exactly one central $b$ jet is a rather effective way 
to identify such a fiducial region. 
We have also found that, especially in DR2 and DS2 schemes, the perturbative series for the $tW$ process 
is well-behaved, NLO-QCD corrections mildly affect the shape of distributions but reduce the scale dependence 
considerably with respect to LO. 
A further handle to suppress process-definition systematics can be given by a reconstruction of the top quark, 
identifying the central $b$ jet as coming from its decay. Top-tagging techniques are being developed 
(theoretical and experimental reviews can be found at \cite{Plehn:2011tg} 
and \cite{Caudron:2110201,CMS-PAS-JME-13-007}), and may help to define 
a sharper fiducial region, although this may depend on the trade-off between the top-tagging efficiency 
and the amount of residual process-definition ambiguities to be suppressed.

\section{$\boldsymbol{tWH}$ production}
\label{sec:tWH_nlo}

In this section we present novel NLO+PS results for $tWH$ production in the 5FS
at the 13-TeV LHC.
Similar to what we have done for $tW$ in the previous section, 
we address the theoretical systematics both at the inclusive level
and with fiducial cuts.
We anticipate that our findings for $tWH$ are qualitatively similar 
to the ones for $tW$, but the larger numerical ratio between the 
top-pair and single-top contributions enhances the impact of interference 
effects and exacerbates theoretical systematics in the simulation,
which are clearly visible in the $t$, $W$, $H$ and $b$-jet observables.
We will see that this can be alleviated after applying suitable cuts.
Finally, we investigate the impact of non-SM couplings of the Higgs
boson on this process.

\subsection{Inclusive results}
\label{sec:tWH_nlo_inclusive}

\begin{figure*}
\center 
 \includegraphics[width=0.48\textwidth]{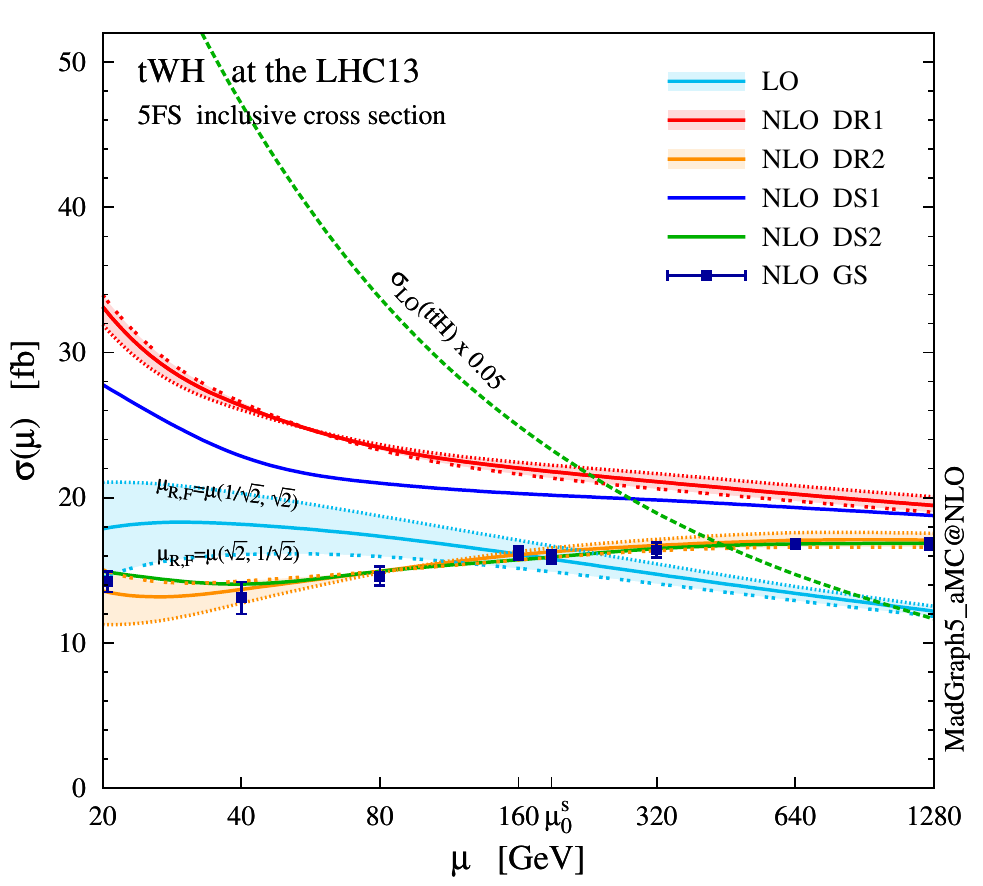} 
 \hspace*{0.5em}
 \includegraphics[width=0.48\textwidth]{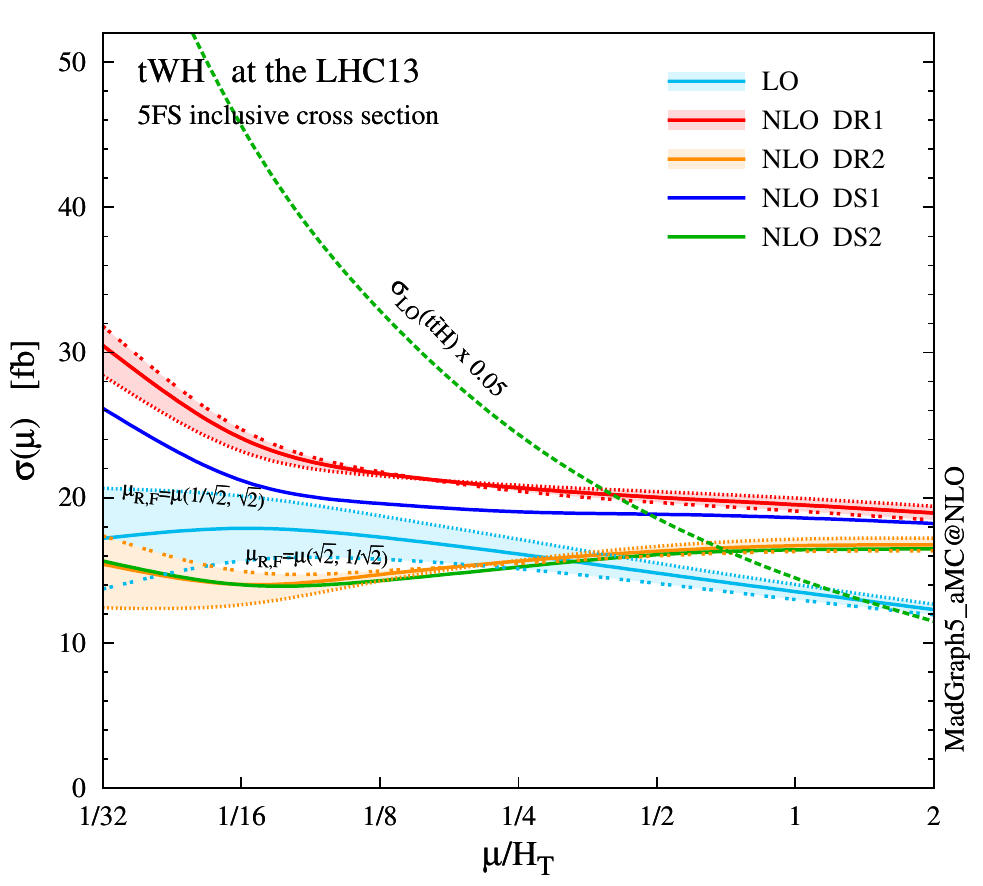} 
\caption{Scale dependence of the total cross section for
 $pp \to tW^{-}H$ and $\bar tW^{+}H$ at the 13-TeV LHC,
 computed in the 5FS at LO and NLO accuracy,
 presented for $\mu_F = \mu_R \equiv \mu$ using a static scale (left) 
 and a dynamic scale (right).
 The NLO $tWbH$ channels are treated using DR and DS, see 
 sec.~\ref{sec:DRandDS} for more details.
 Furthermore, we show NLO results from GS (only for a static scale),
 and two off-diagonal profiles of the scale dependence,
 $( \mu_R=\sqrt{2}\mu\,,\,  \mu_F=\mu/\sqrt{2} )$ and
 $( \mu_R=\mu/\sqrt{2}\,,\,  \mu_F=\sqrt{2}\mu )$, 
 for LO and NLO DR.
 Finally, the scale dependence of $pp \to t \bar tH$ at LO is also
 reported as a reference.}
 \label{fig:tWH_scalediag}
\end{figure*} 

\begin{table*}
\center \small 
\begin{tabular}{lllllllll}
 \hline
 \rule{0pt}{3ex}   
 $tWH$ (13~TeV) \hspace*{0.5em}
 & $\sigma(\mu_0^{s})$~[fb]
 & $\delta^\perc_{\mu}$
 & $\delta^\perc_{\mathrm{PDF}}$
 & $K$ 
 & $\sigma(\mu_0^{d})$~[fb]
 & $\delta^\perc_{\mu}$
 & $\delta^\perc_{\mathrm{PDF}}$
 & $K$
 \\[0.7ex] 
 \hline
 \rule{0pt}{3ex}  LO
   &  15.77(1)  &  $^{+11.3}_{-11.1}$  &  \scriptsize{$\pm 11.2$}   &  -     
   &  16.14(2)  &  $^{+12.9}_{-12.8}$  &  \scriptsize{$\pm 11.1$}   &  -     \\[0.3ex]
 \rule{0pt}{3ex}  NLO DR1 
   &  21.72(2)  &  $^{+5.8}_{-4.3}$    &  \scriptsize{$\pm 3.0$ }   &  \footnotesize 1.38     
   &  20.72(2)  &  $^{+5.0}_{-3.1}$    &  \scriptsize{$\pm 3.0$ }   &  \footnotesize 1.28  \\[0.3ex]
 \rule{0pt}{3ex}  NLO DR2 
   &  16.28(4)  &  $^{+4.6}_{-6.2}$    &  \scriptsize{$\pm 2.7$ }   &  \footnotesize 1.03  
   &  15.68(3)  &  $^{+4.5}_{-5.9}$    &  \scriptsize{$\pm 2.7$ }   &  \footnotesize 0.97  \\[0.3ex]
 \rule{0pt}{3ex}  NLO DS1 
   &  20.17(3)  &  $^{+4.0}_{-3.9}$    &  \scriptsize{$\pm 3.2$ }   &  \footnotesize 1.28     
   &  19.11(3)  &  $^{+2.3}_{-2.3}$    &  \scriptsize{$\pm 2.9$ }   &  \footnotesize 1.18  \\[0.3ex]
 \rule{0pt}{3ex}  NLO DS2 
   &  16.00(3)  &  $^{+4.8}_{-6.9}$    &  \scriptsize{$\pm 2.5$ }   &  \footnotesize 1.01     
   &  15.31(3)  &  $^{+5.1}_{-6.7}$    &  \scriptsize{$\pm 2.5$ }   &  \footnotesize 0.95  \\[0.3ex]
 \rule{0pt}{3ex}  NLO GS  
   &  15.9(5)   &  -  &  -  &  \footnotesize 1.01(3)  &  &  &  &  \\[0.7ex]
 \hline
\end{tabular}
\caption{Total cross sections for 
 $pp \to t W^-H$ and $\bar t W^+H$ at the 13-TeV LHC,  in the 5FS at LO and NLO accuracy with different  schemes, 
 computed with a static scale $\mu_0^{s} = (m_t+m_W+m_H)/2$
 and a dynamic scale $\mu_0^{d} = H_T/4$.
 We also report the scale and PDF uncertainties and the NLO-QCD $K$ factors;
 the numerical uncertainty affecting the last digit is quoted in parentheses.}
\label{tab:tWH_NLO_xsect_1}
\end{table*}

As for $tW$, we start by showing the renormalisation and factorisation scale 
dependence of the $tWH$ cross section in fig.~\ref{fig:tWH_scalediag}, 
both at LO and NLO accuracy, using different schemes to treat the $tWbH$ real-emission channels 
(the details for the various NLO schemes can be found in sec.~\ref{sec:DRandDS}).
The values of the total rate computed at the central scale $\mu_0$ are also quoted
in table~\ref{tab:tWH_NLO_xsect_1}.
Unlike in fig.~\ref{fig:tWH_scalediag}, in this case scale variations are computed by varying 
$\mu_F$ and $\mu_R$ independently by a factor two around $\mu_0$.

The same pattern we have found for $tW$ is repeated.
Comparing DR results obtained by neglecting (DR1, red) or taking into account 
(DR2, orange) interference with $t \bar t H$, we observe again that 
these interference effects are negative,
but their relative impact on the cross section is even more sizeable.
The interference reduces the NLO rate
by about 5~fb, which amounts to a hefty $-25\perc$, leading to a $K$ factor close to 1.
Since interference effects are driven by the LO $t \bar t H$ contribution,
they grow larger for lower scale choices.
The cross sections obtained employing the two DS techniques, DS1 (blue) and DS2 (green), 
show large differences which go beyond the missing higher orders estimated by scale variations,
and can be traced back to the different Breit-Wigner prefactor in the subtraction term $\mathcal{C}_{2t}$.
As it has been the case for $tW$ production, we find that DR2 and DS2 are in good agreement with GS.

In complete analogy with the case of the $tWb$ channel in $tW$ production
at NLO, we perform a study of the theoretical systematics in the 
modelling of the $tWbH$ channel 
(employing the 4FS to isolate this contribution),
which can be found in \ref{app:twb_lo_4FS}. 

\begin{figure*}
\center
 \includegraphics[width=0.325\textwidth]{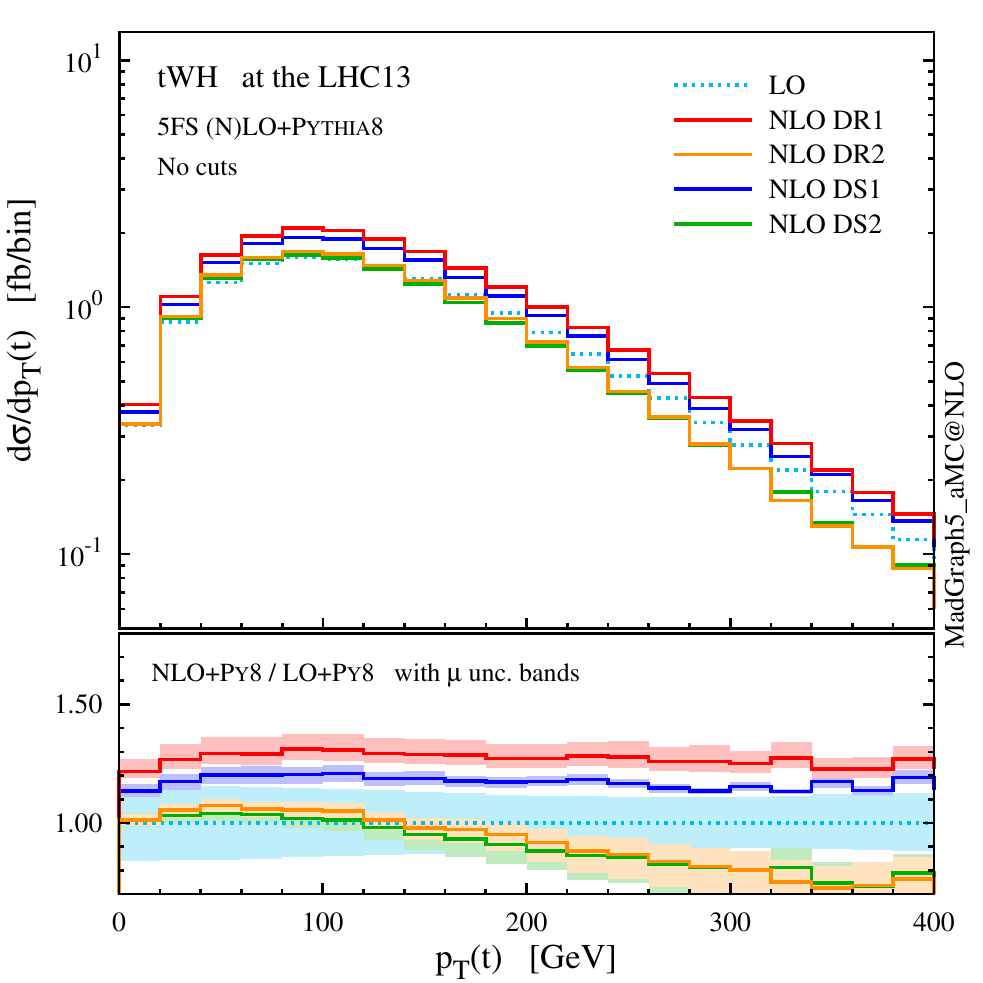}
 \includegraphics[width=0.325\textwidth]{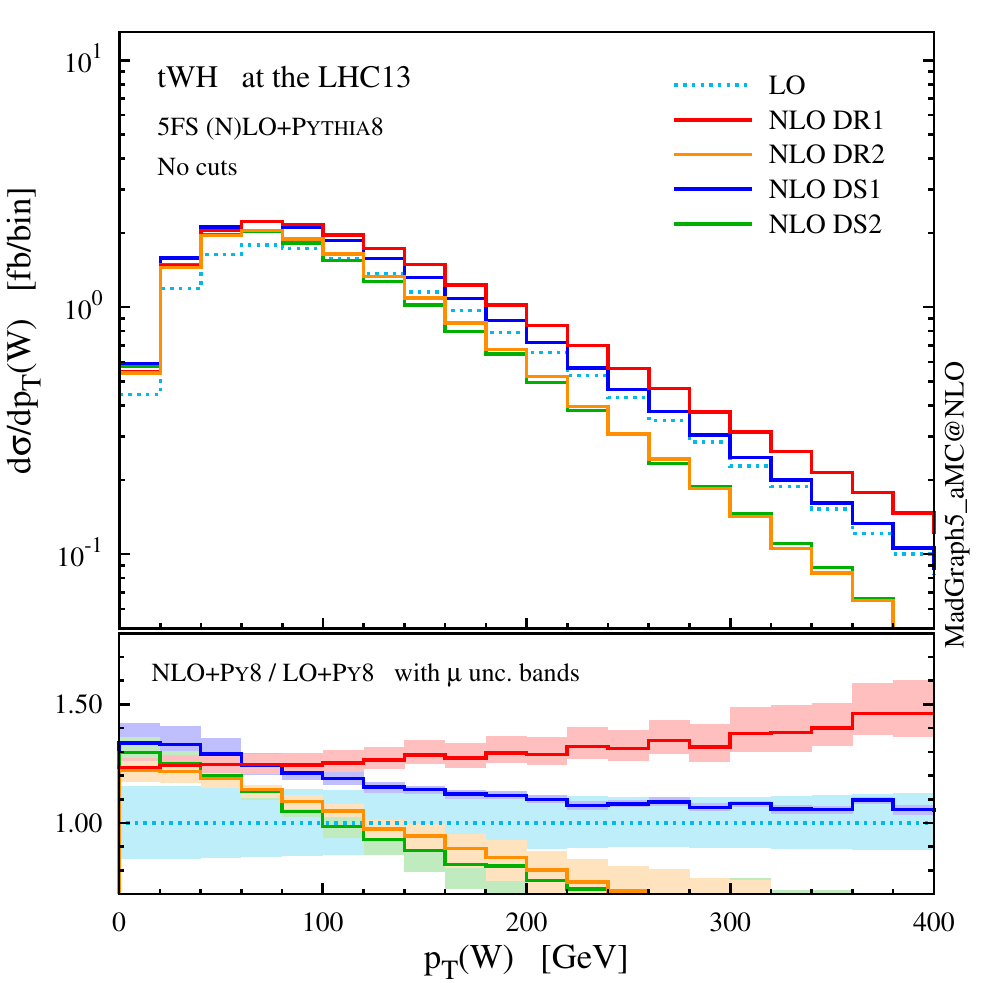}
 \includegraphics[width=0.325\textwidth]{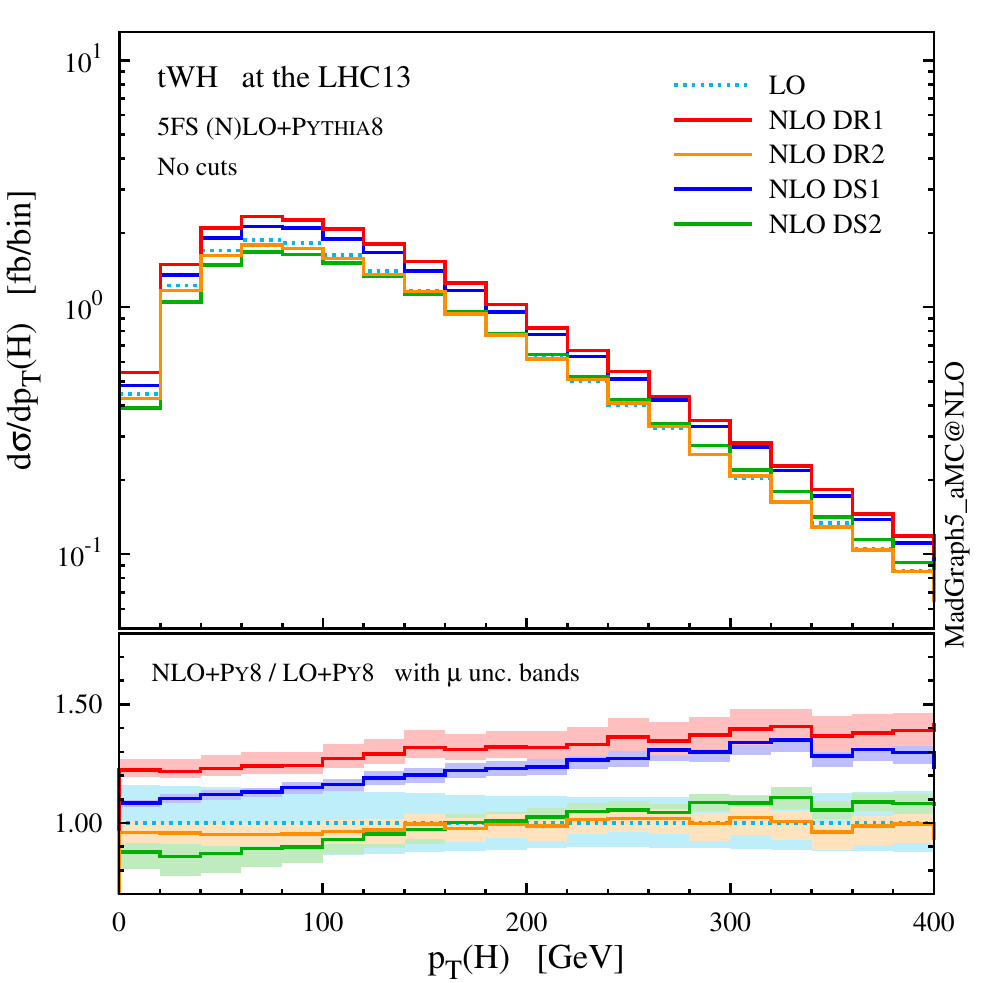}\\
 \includegraphics[width=0.325\textwidth]{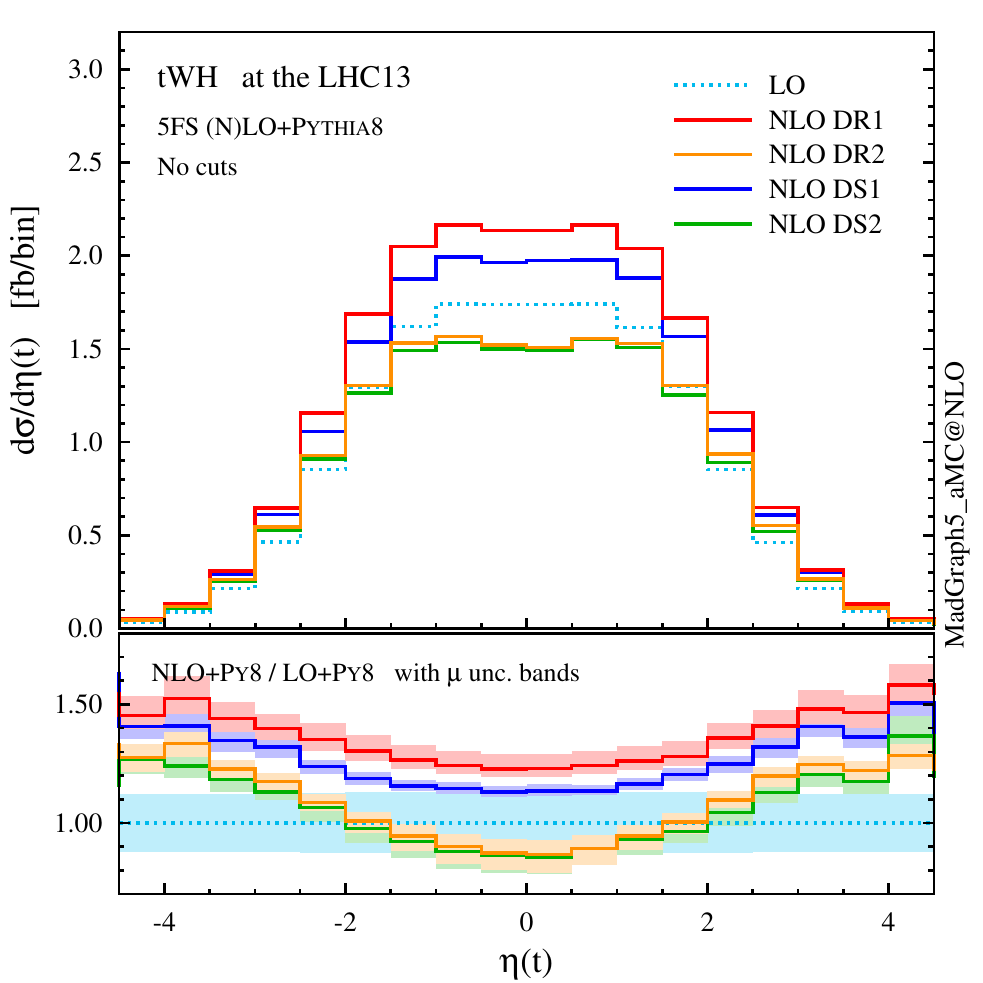} 
 \includegraphics[width=0.325\textwidth]{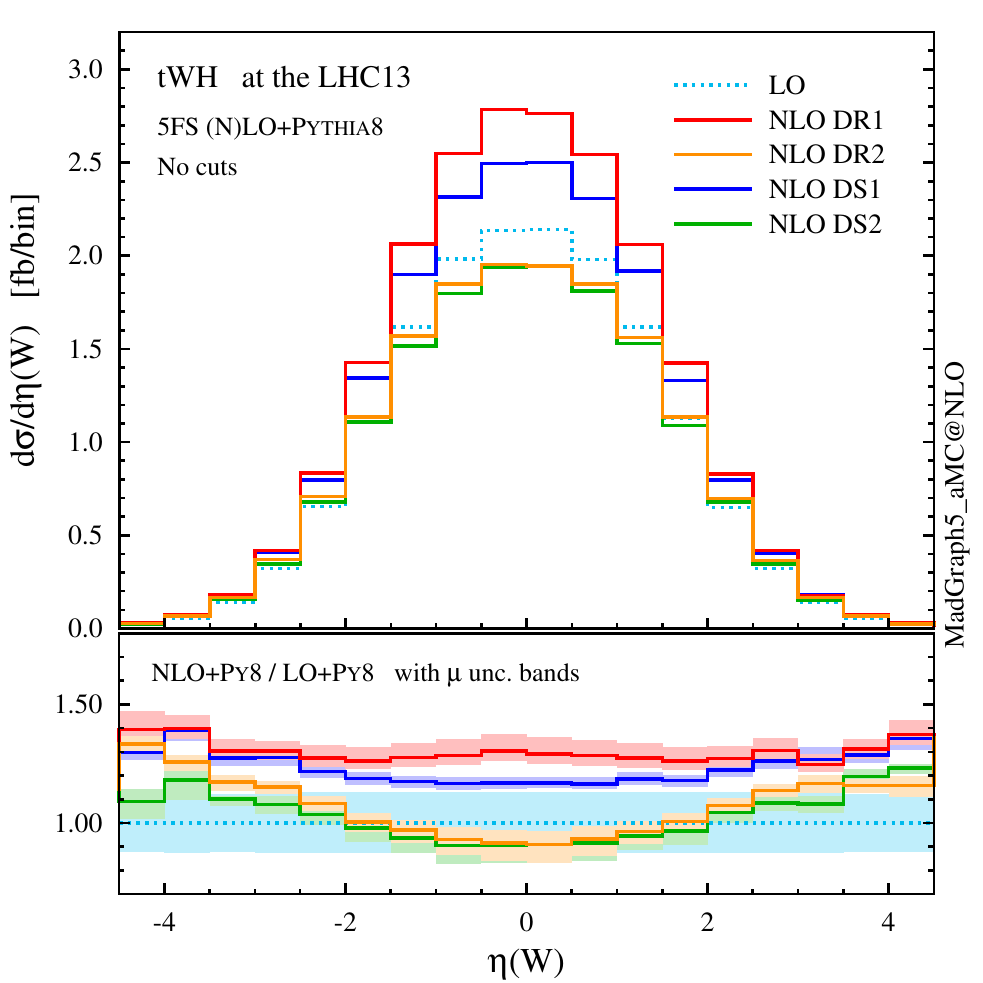} 
 \includegraphics[width=0.325\textwidth]{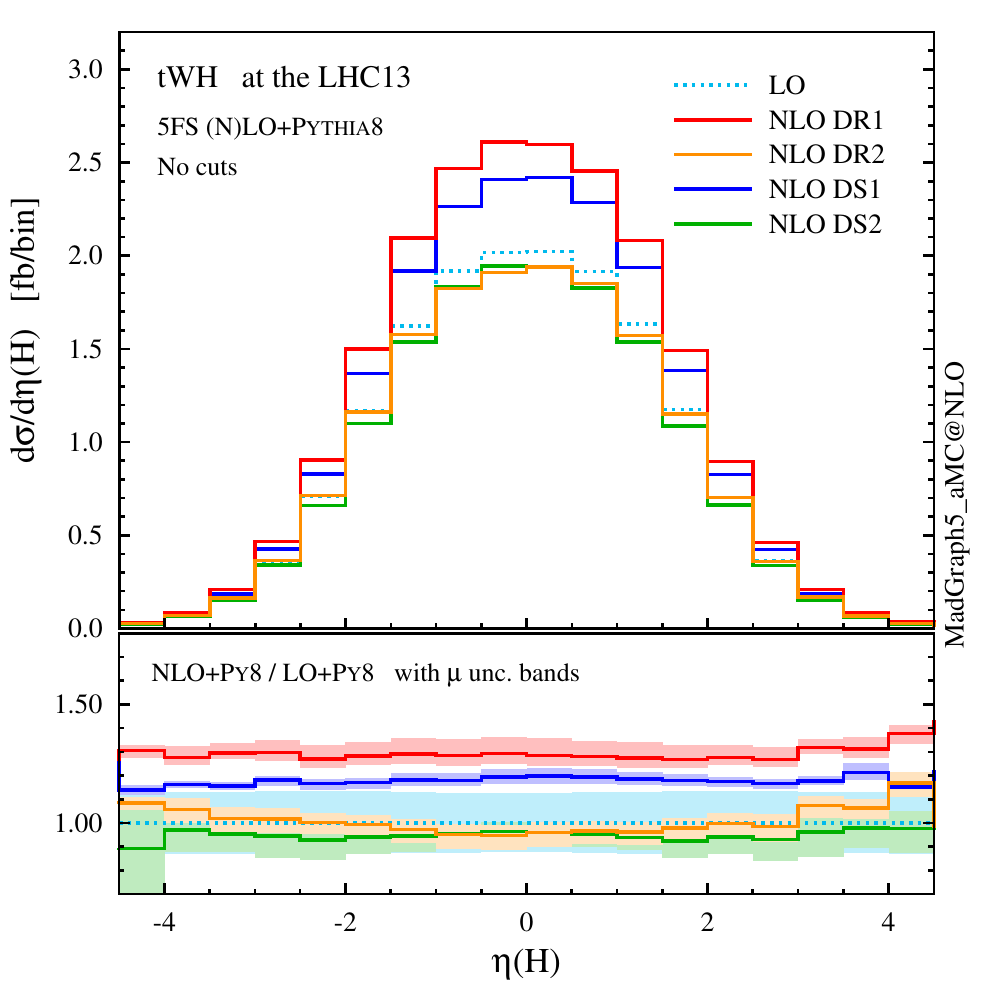}
\caption{$p_T$ and $\eta$ distributions for the top quark, the $W$ boson and the Higgs boson at NLO+PS accuracy in $tWH$ production at the 13-TeV LHC. The lower panels provide information on the differential $K$ factors with the scale uncertainties.} 
\label{fig:tWH_5FNLO_dist_1}

\center
 \includegraphics[width=0.325\textwidth]{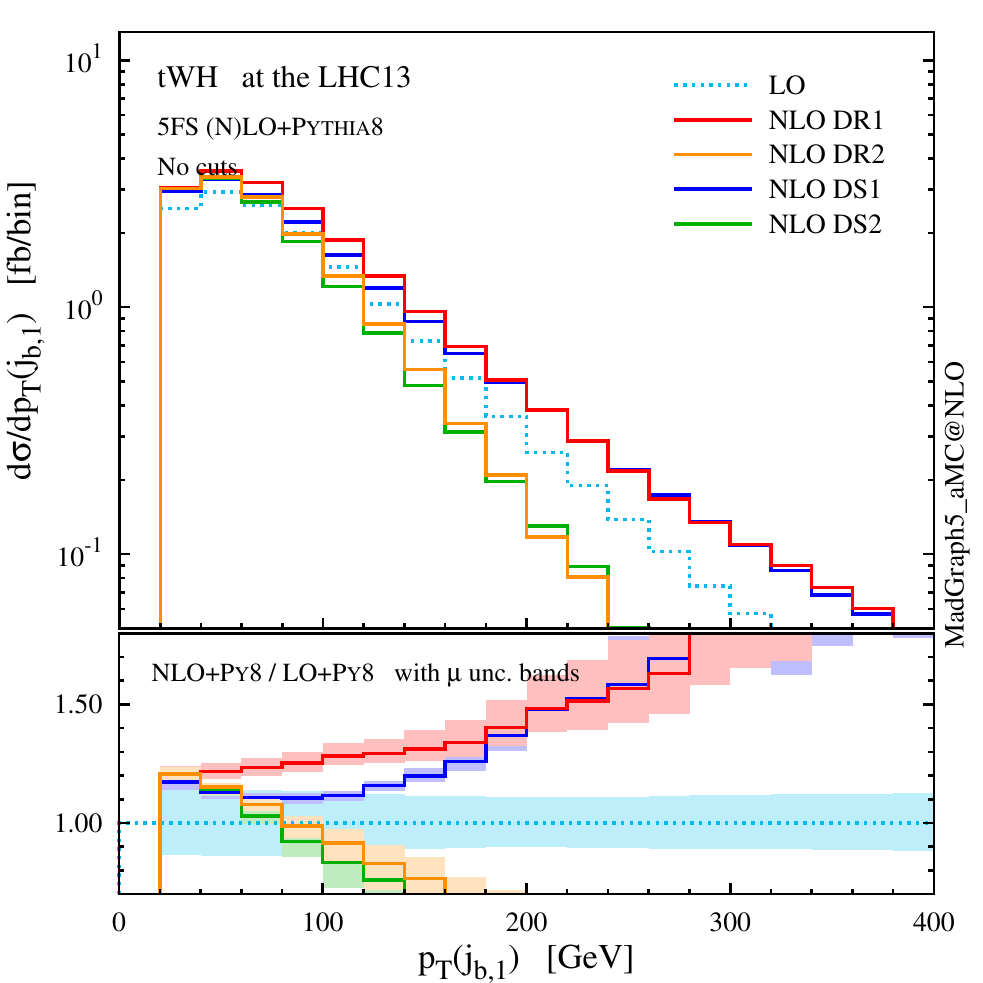}\qquad
 \includegraphics[width=0.325\textwidth]{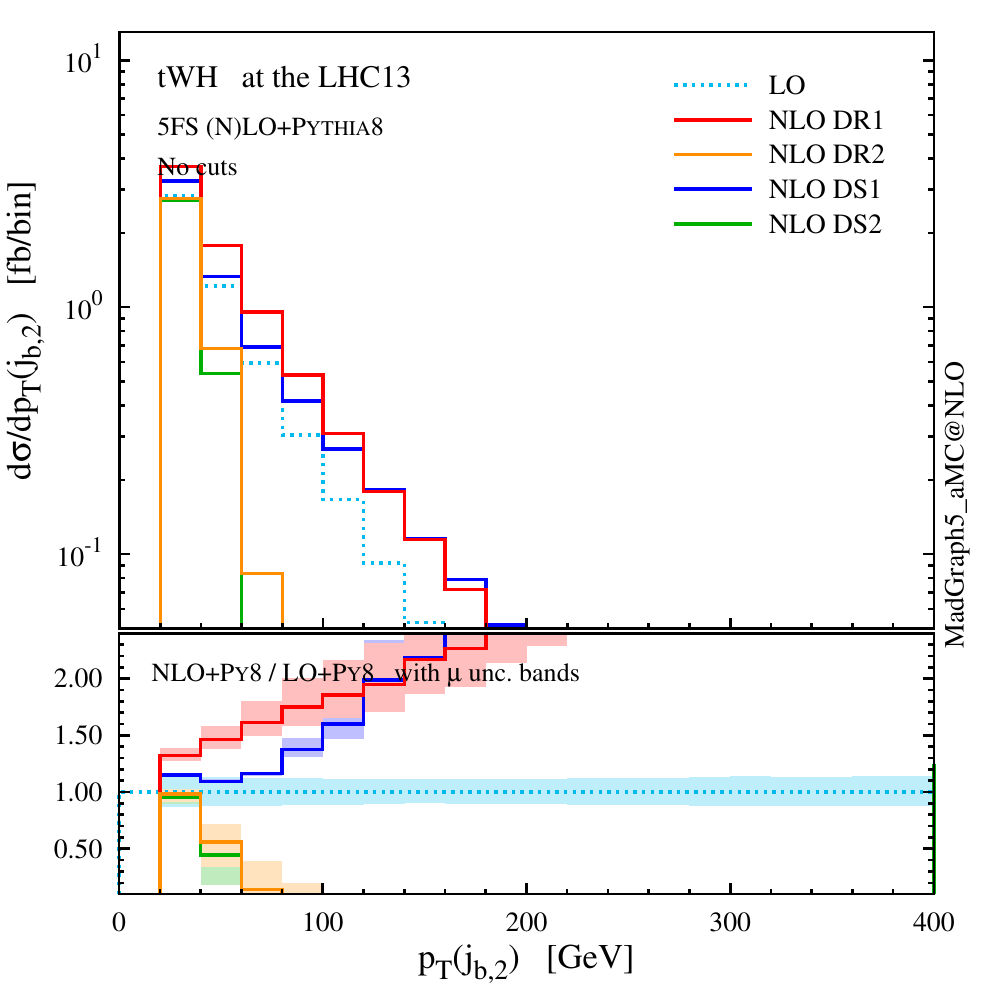}\\ 
 \includegraphics[width=0.325\textwidth]{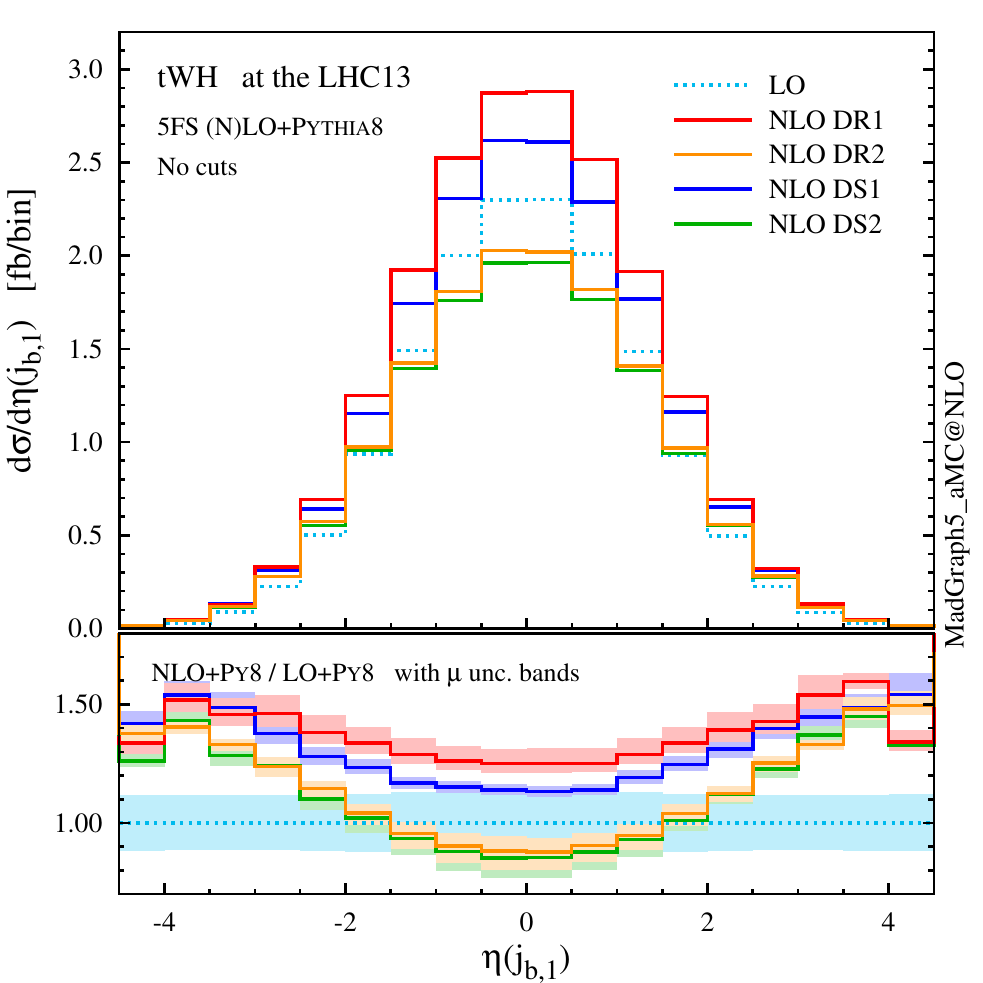}\qquad
 \includegraphics[width=0.325\textwidth]{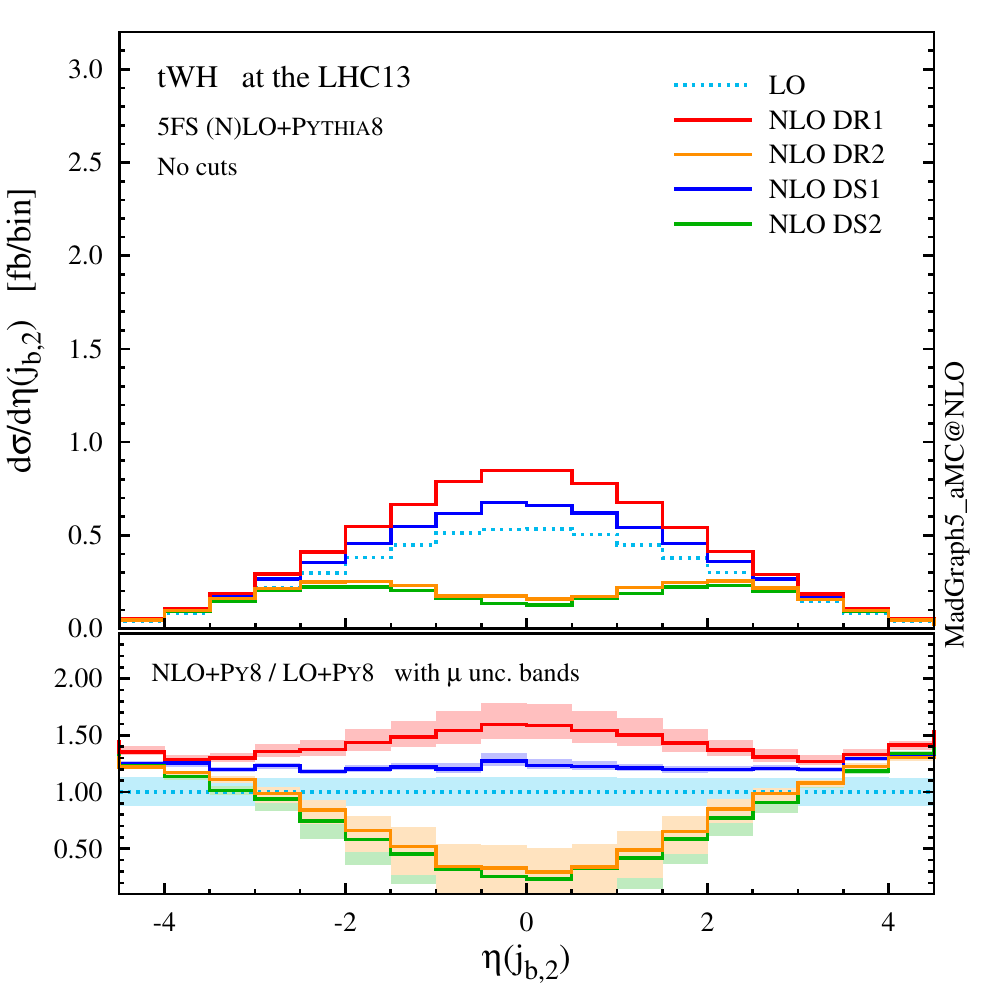}
\caption{Same as fig.~\ref{fig:tWH_5FNLO_dist_1}, but for the $b$-tagged jets. 
Note that the second-hardest $b$ jet is described by the parton shower at LO, while by the matrix element at NLO.
}
\label{fig:tWH_5FNLO_dist_1b}
\end{figure*} 

In figs.~\ref{fig:tWH_5FNLO_dist_1} and \ref{fig:tWH_5FNLO_dist_1b} we collect some differential distributions.
Observables related to the Higgs boson can essentially be described by a constant 
$K$ factor for each subtraction scheme. 
On the other hand, similar to the $tW$ case, 
the NLO distributions for the top quark and the $W$ boson are quite different among the four NLO techniques.
As we know, these differences are driven essentially by whether the interference with $t \bar t H$ is included or not
(in DR), and by the profile of the subtraction term (in DS). 
These NLO effects are quite remarkable for the $b$~jets, since the negative interference 
with $t \bar t H$ drastically suppresses central hard $b$ jets.

Summarising, in analogy with the $tW$ process, effects due to the interference between 
$t \bar t H$ and $tWH$ which appear in NLO corrections of the latter process are significant,
and hence the details on how the $t\bar tH$ contribution is subtracted enormously affect 
the predictions for both the total rate and the shape of distributions.
On the one hand, a LO description of $tWH$ in the 5FS is apparently not sufficient.
On the other hand, the NLO prediction strongly depends on the subtraction scheme employed.
This last point is only a relative issue, if we take into account the fact that DR2 and DS2 results 
are quite consistent with each other and integrate to the same total cross section as GS, 
which suggests that they provide a better description of the physics not included in $t \bar t H$ 
than DR1 and DS1.
Nevertheless, as in the case of $tW$ production, it is clear that fiducial cuts are crucial 
to obtain a meaningful separation of $tWH$ from $t \bar t H$, 
and their effects will be discussed in the next subsection.

\subsection{Results with fiducial cuts}
\label{sec:tWH_nlo_fiducial}

\begin{table*}
\center \small 
\begin{tabular}{lllllllllllll}
 \hline
 \rule{0pt}{3ex}  & \multicolumn{1}{c}{No cuts}  & \multicolumn{2}{c}{Fiducial cuts}  & \multicolumn{2}{c}{Fiducial cuts + top reco.}   \\[0.3ex]
 \rule{0pt}{3ex}    
 & $\sigma_{\mathrm{NLO}}\pm\delta^\perc_{\mu}\pm\delta^\perc_{\mathrm{PDF}}$ 
 & $\sigma_{\mathrm{NLO}}\pm\delta^\perc_{\mu}\pm\delta^\perc_{\mathrm{PDF}}$ 
 & eff.
 & $\sigma_{\mathrm{NLO}}\pm\delta^\perc_{\mu}\pm\delta^\perc_{\mathrm{PDF}}$
 & eff.
 \\[0.7ex] 
 \hline
 \rule{0pt}{3ex}  $t \bar t H$ 
   &  485.0(9)\,$^{+1.3}_{-5.3}$\,\scriptsize{$\pm 1.8$ } &   21.5(2)\enskip\,$^{+2.0}_{-6.8}$\,\scriptsize{$\pm 2.7$ }   &  0.04  &  21.5(2)\enskip\,$^{+2.0}_{-6.8}$\,\scriptsize{$\pm 2.7$ }  &  0.04 \\[0.3ex]
 \rule{0pt}{3ex}  $tWH$ DR1                        
   &  20.72(2)\,$^{+5.0}_{-3.1}$\,\scriptsize{$\pm 3.0$ } &  12.12(2)\,$^{+2.7}_{-2.3}$\,\scriptsize{$\pm 2.5$ }   &  0.58  &  11.18(2)\,$^{+2.2}_{-2.3}$\,\scriptsize{$\pm 2.5$ }   &  0.54  \\[0.3ex]
 \rule{0pt}{3ex}  $tWH$ DR2                        
   &  15.68(3)\,$^{+4.5}_{-5.9}$\,\scriptsize{$\pm 2.7$ } &  11.43(2)\,$^{+1.6}_{-2.4}$\,\scriptsize{$\pm 2.4$ }   &  0.73 &  11.04(2)\,$^{+1.8}_{-2.4}$\,\scriptsize{$\pm 2.4$ }   &  0.70  \\[0.3ex]
 \rule{0pt}{3ex}  $tWH$ DS1                        
   &  19.11(3)\,$^{+2.3}_{-2.3}$\,\scriptsize{$\pm 2.9$ } &  11.79(2)\,$^{+1.8}_{-2.3}$\,\scriptsize{$\pm 2.5$ }   &  0.62 &  11.02(2)\,$^{+1.7}_{-2.3}$\,\scriptsize{$\pm 2.5$ }   &  0.58  \\[0.3ex]
 \rule{0pt}{3ex}  $tWH$ DS2                        
   &  15.31(3)\,$^{+5.1}_{-6.7}$\,\scriptsize{$\pm 2.5$ } &  11.37(2)\,$^{+1.6}_{-2.3}$\,\scriptsize{$\pm 2.4$ }   &  0.74 &  11.05(2)\,$^{+1.8}_{-2.4}$\,\scriptsize{$\pm 2.4$ }   &  0.72  \\[0.7ex]
 \hline
\end{tabular}
\caption{Total cross sections in fb at the LHC 13~TeV for the processes $pp \to t \bar tH$
 and $pp \to t WH$, in the 5FS at NLO+PS accuracy.
 Results are presented before any cut (left), after fiducial cuts (center), and also adding top reconstruction 
 on the event sample (right). 
 We also report the scale and PDF uncertainties, as well as the cut efficiency with respect to the case with no cuts.
 All numbers are computed with the reference dynamic scale $\mu_0 = H_T/4$,
 and the numerical uncertainty affecting the last digit is reported in parentheses.}
\label{tab:tWH_NLO_xsect_2}
\end{table*}

We now move to investigate whether the separation between $tWH$ and 
$t \bar t H$ can become meaningful in a fiducial region,
where interference between the two processes and theoretical systematics 
are suppressed.
The problem is exactly analogous to the $tW$--$t \bar t$ separation.
In practice, for any selection defined by suitable cuts, 
one needs to quantify the residual difference among different subtraction schemes 
and see if it is small enough.

Motivated by the same rationale behind our $tW$ discussion,
we define our set of ``fiducial cuts'' for $tWH$ selecting only events with
\begin{enumerate}
 \item exactly one $b$ jet with \enskip $p_T(j_b)>20$~GeV \\ and \enskip $|\eta(j_b)|<2.5$\,, \\[-0.5em]
 \item exactly two central $W$ bosons with \enskip $|y(W)|<2.5$\,, \\[-0.5em]
 \item exactly one central Higgs boson with~$|y(H)|<2.5$\,.  
\end{enumerate}
We recall that the first selection is the key to suppress the double-top amplitudes and hence
$tWH$--$t \bar t H$ interference and theoretical ambiguities.
We do not assume any particular decay channel for the heavy bosons and
hence the second and third selections are added to mimic a good reconstructability of 
the $W$ and $H$ bosons in the detector. However, they are not crucial since they 
affect just $5\perc$ of the events after surviving selection 1.
Our pseudo event category is defined mainly for illustrating the issues
behind the simulation of the $tWH$ signal,
but the same procedure can be applied to any realistic set of fiducial cuts in 
experimental analyses, including a selection on specific decay products 
of the $W$ and $H$ bosons.

\begin{figure*}
\center
 \includegraphics[width=0.325\textwidth]{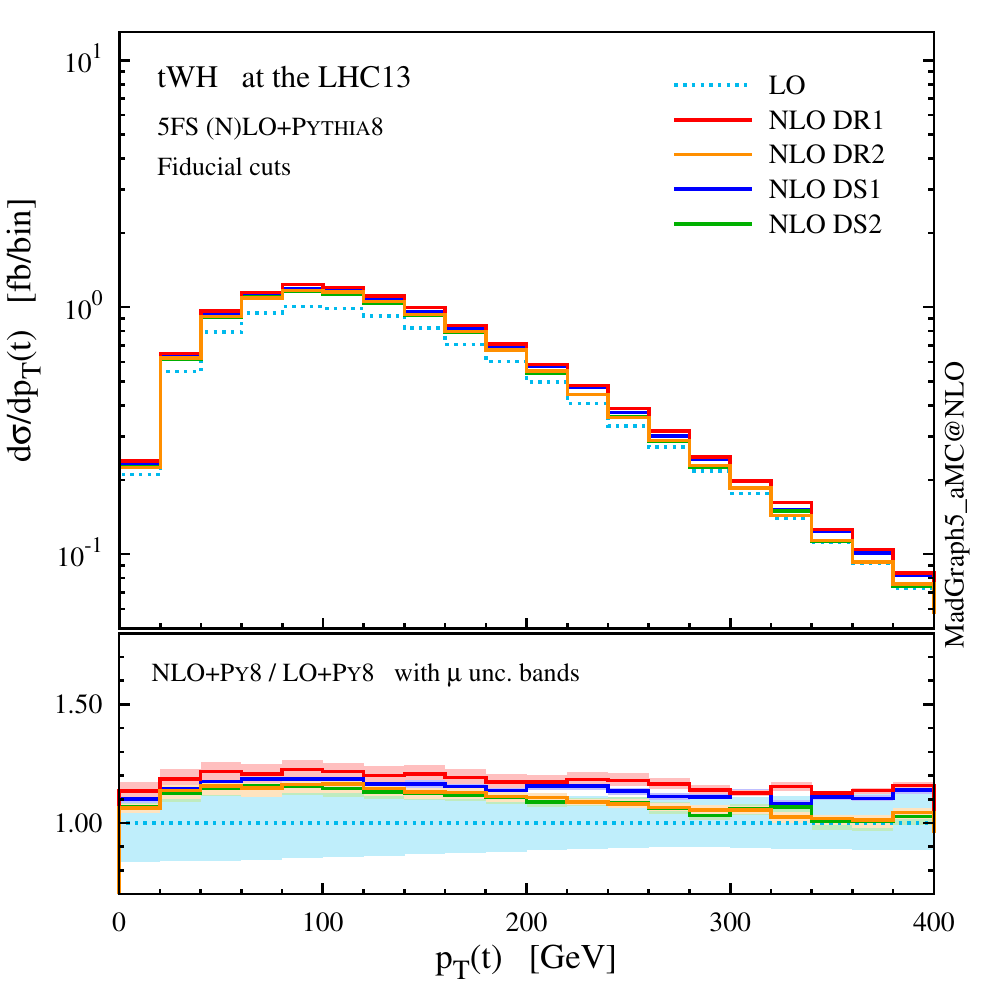}
 \includegraphics[width=0.325\textwidth]{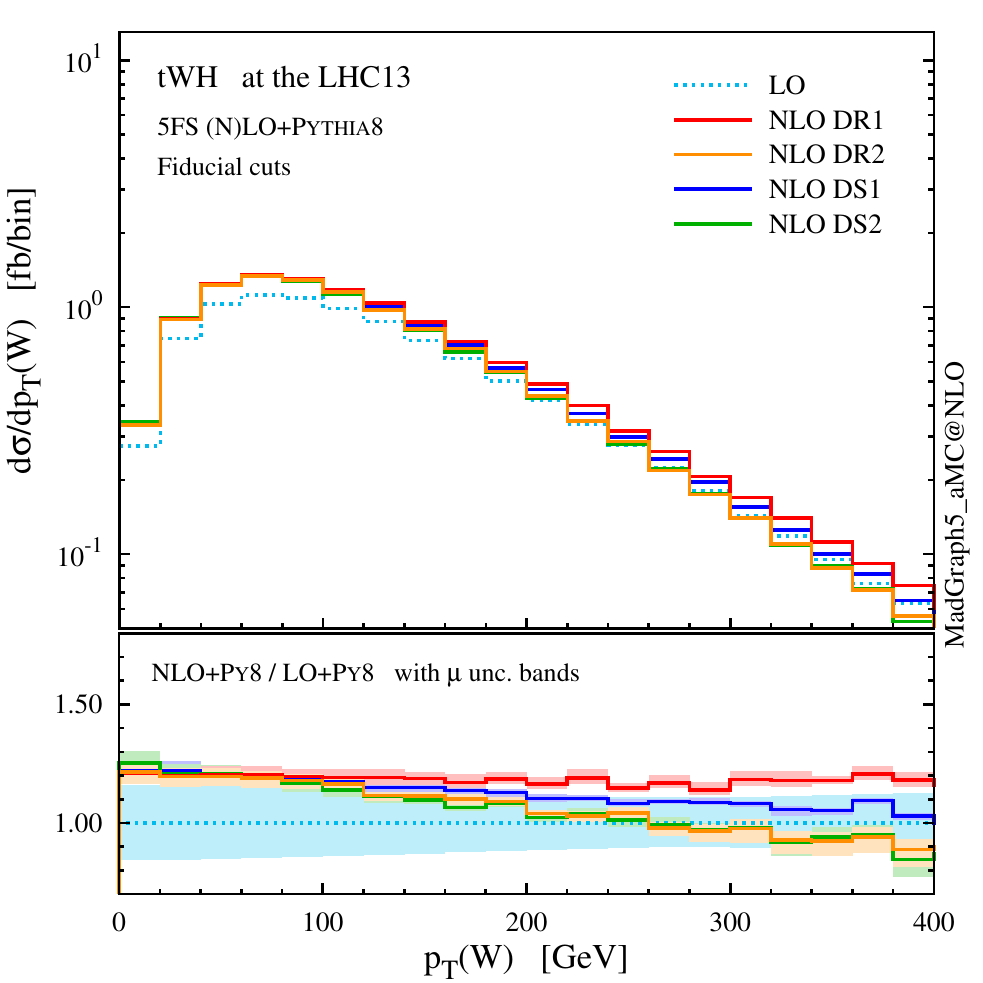}
 \includegraphics[width=0.325\textwidth]{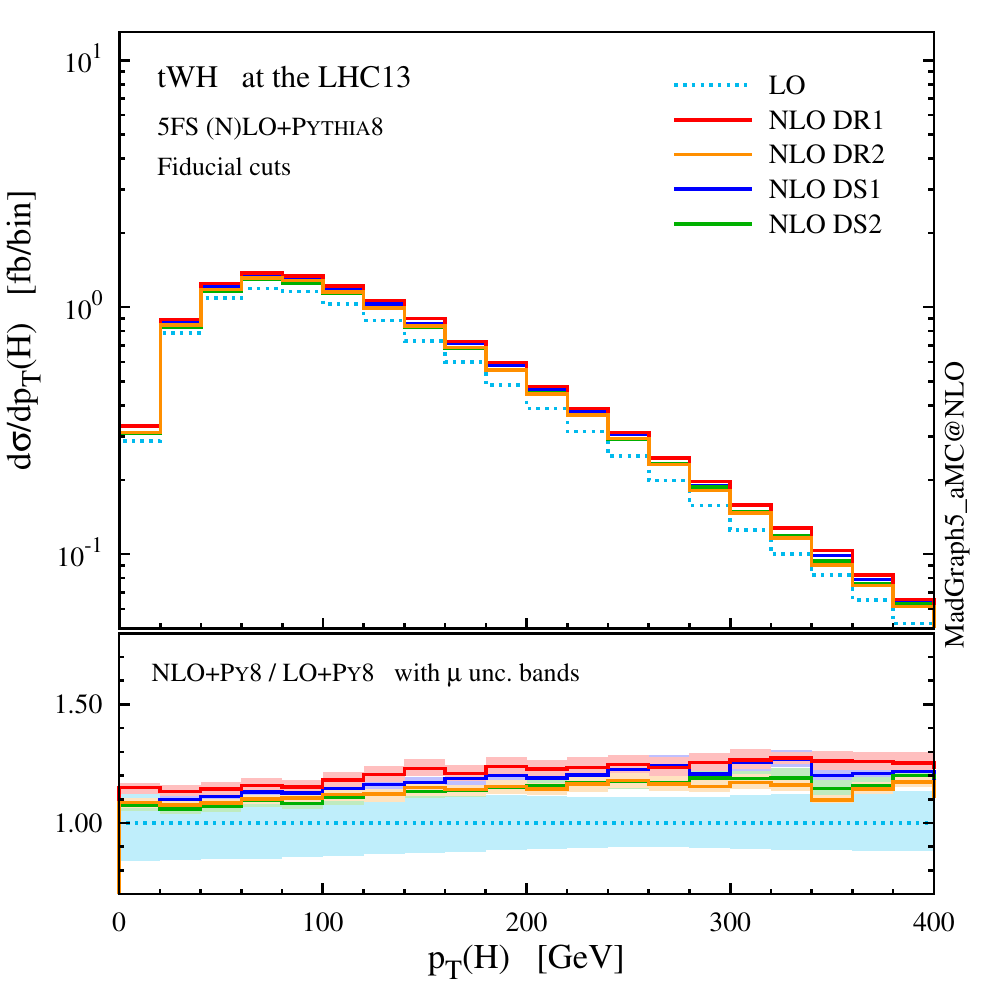}\\
 \includegraphics[width=0.325\textwidth]{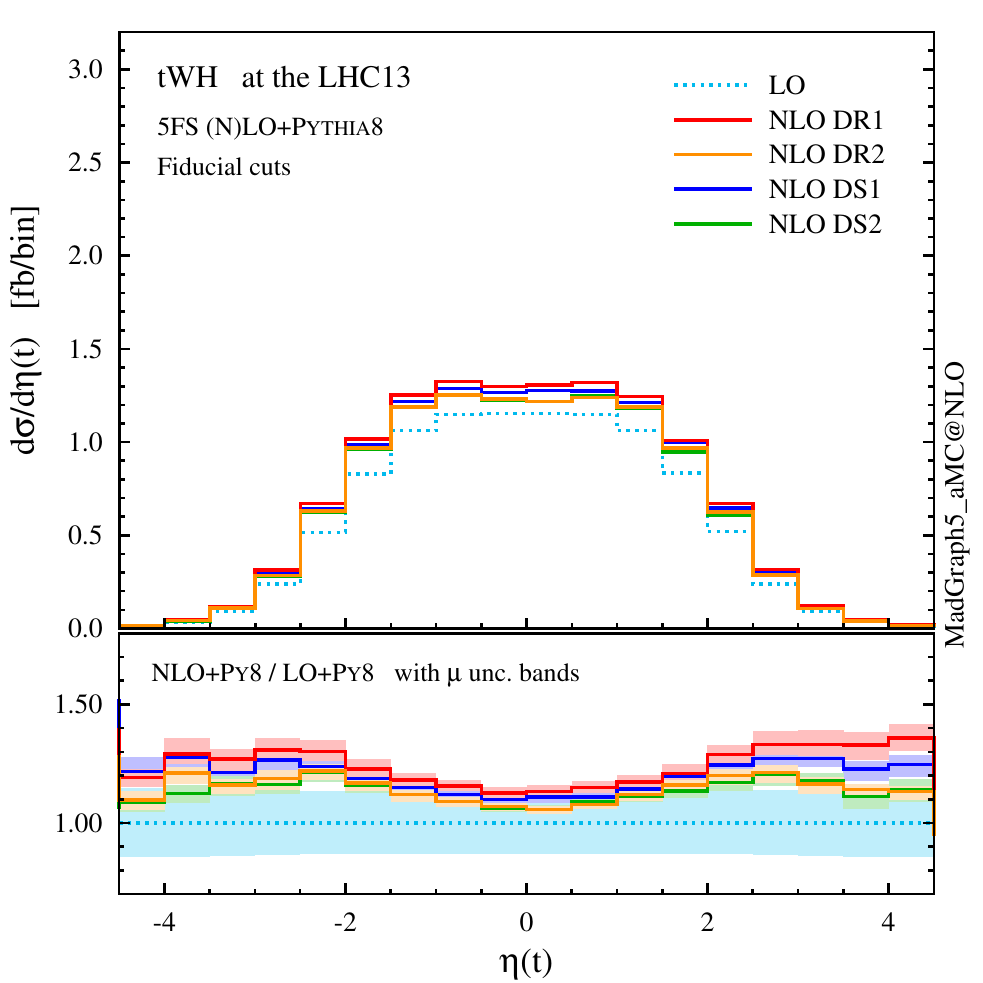}
 \includegraphics[width=0.325\textwidth]{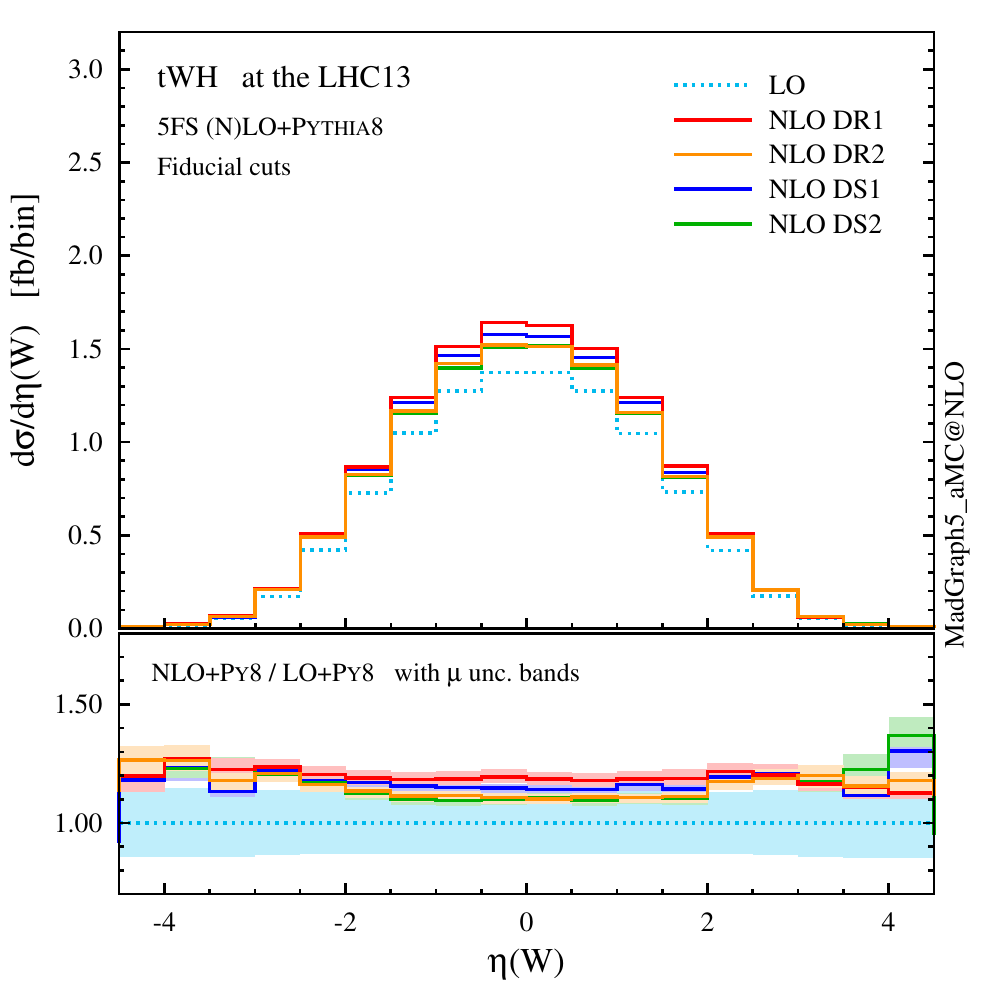}
 \includegraphics[width=0.325\textwidth]{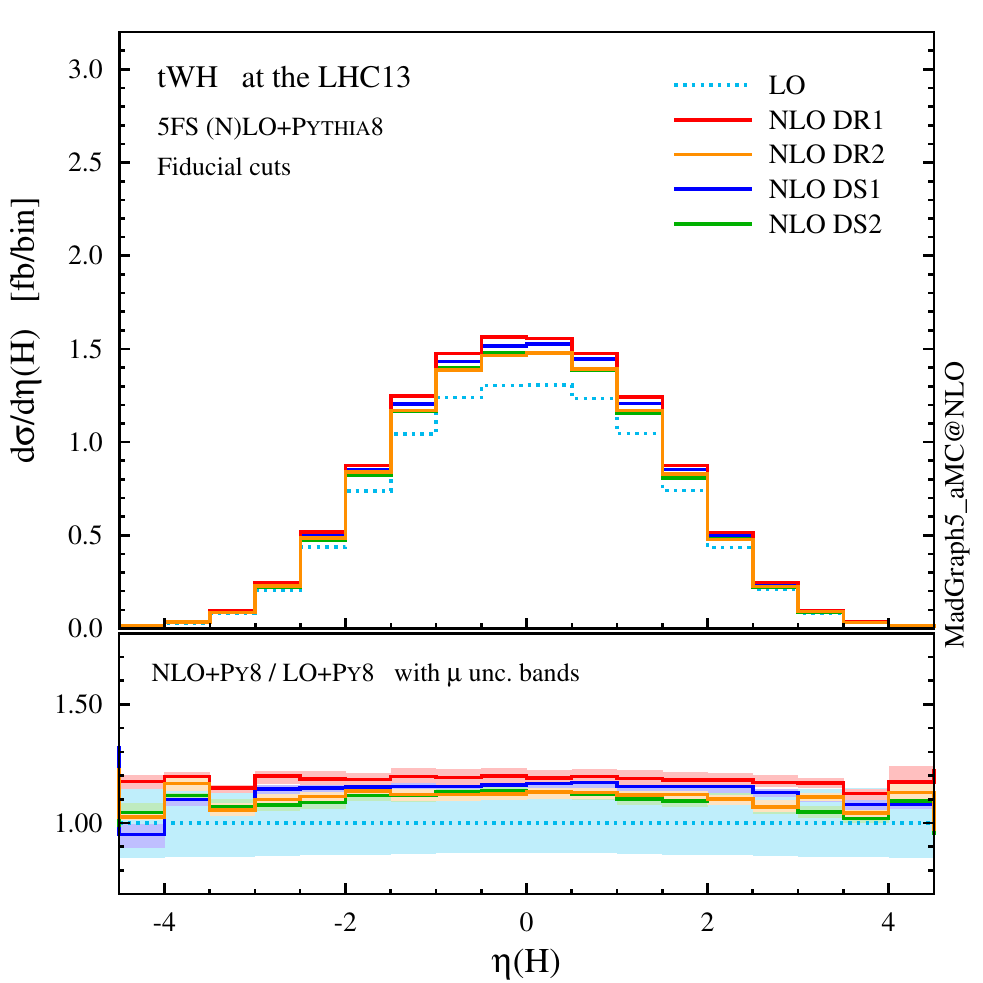}
\caption{$p_T$ and $\eta$ distributions the top quark, 
 the $W$ boson and the Higgs boson as in fig.~\ref{fig:tWH_5FNLO_dist_1},
 but after applying the fiducial cuts to suppress interference between $tWHb$
 and $t \bar tH$.} 
\label{fig:tWH_5FNLO_dist_2}

\vspace*{2em}

\center
 \includegraphics[width=0.325\textwidth]{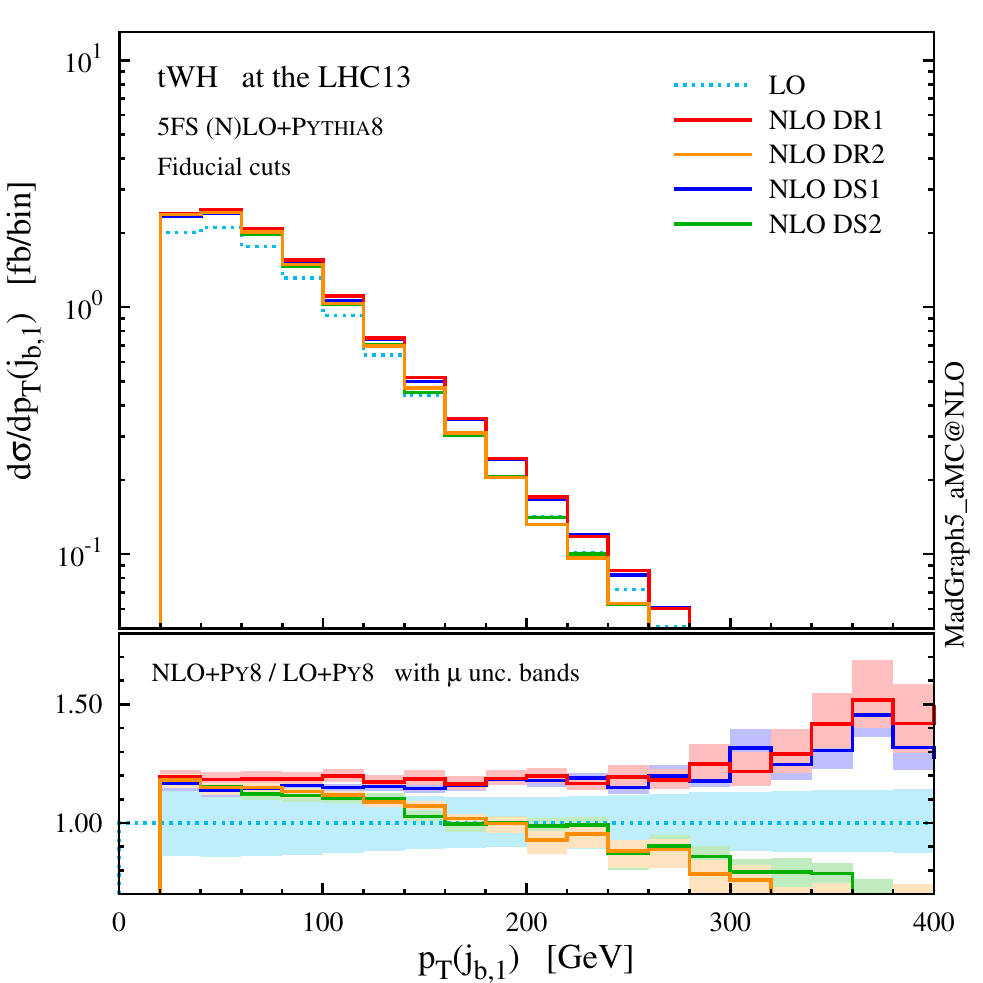}\qquad
 \includegraphics[width=0.325\textwidth]{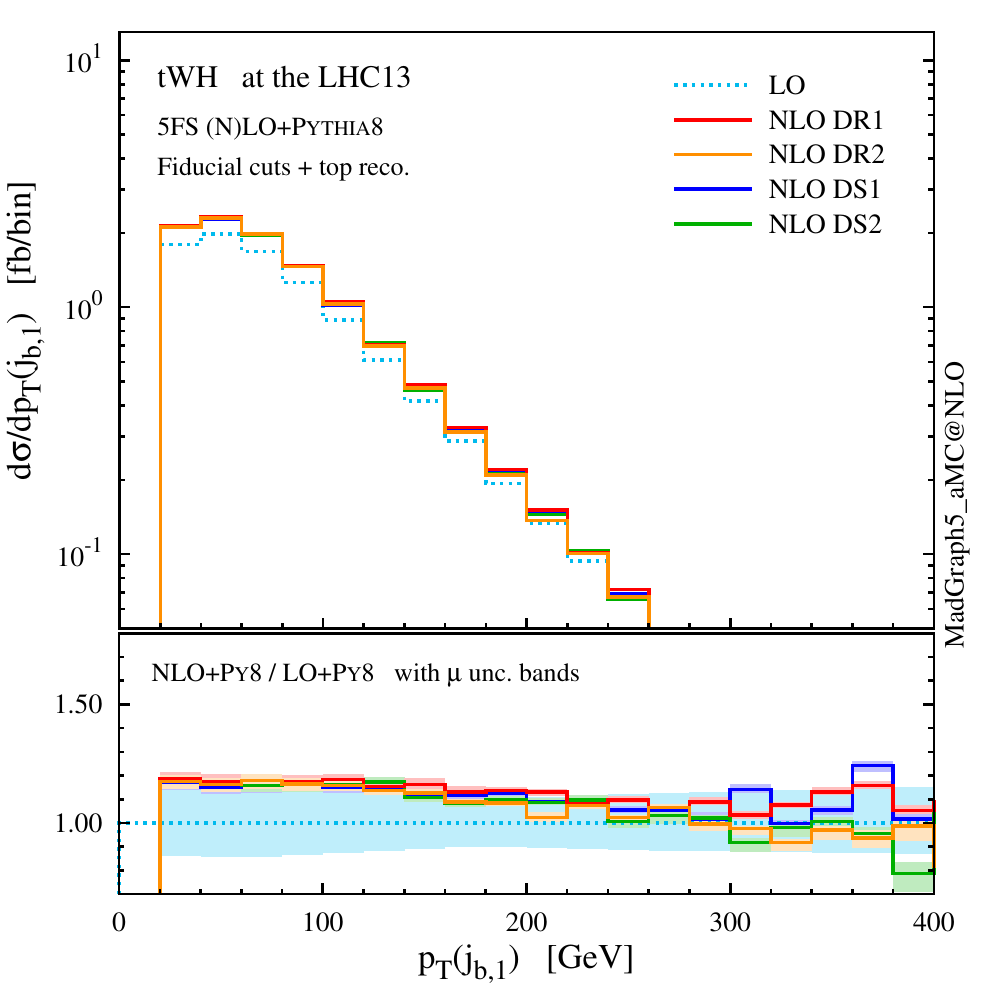}\\
 \includegraphics[width=0.325\textwidth]{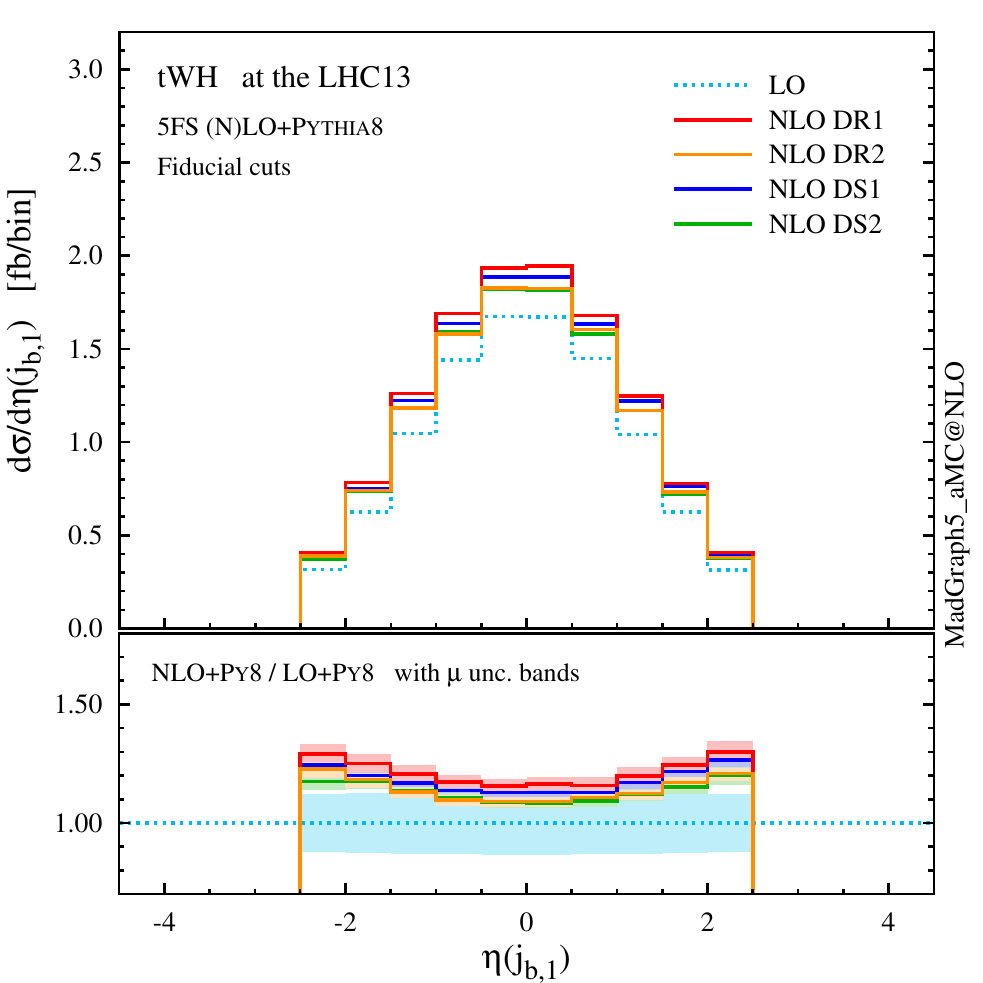}\qquad
 \includegraphics[width=0.325\textwidth]{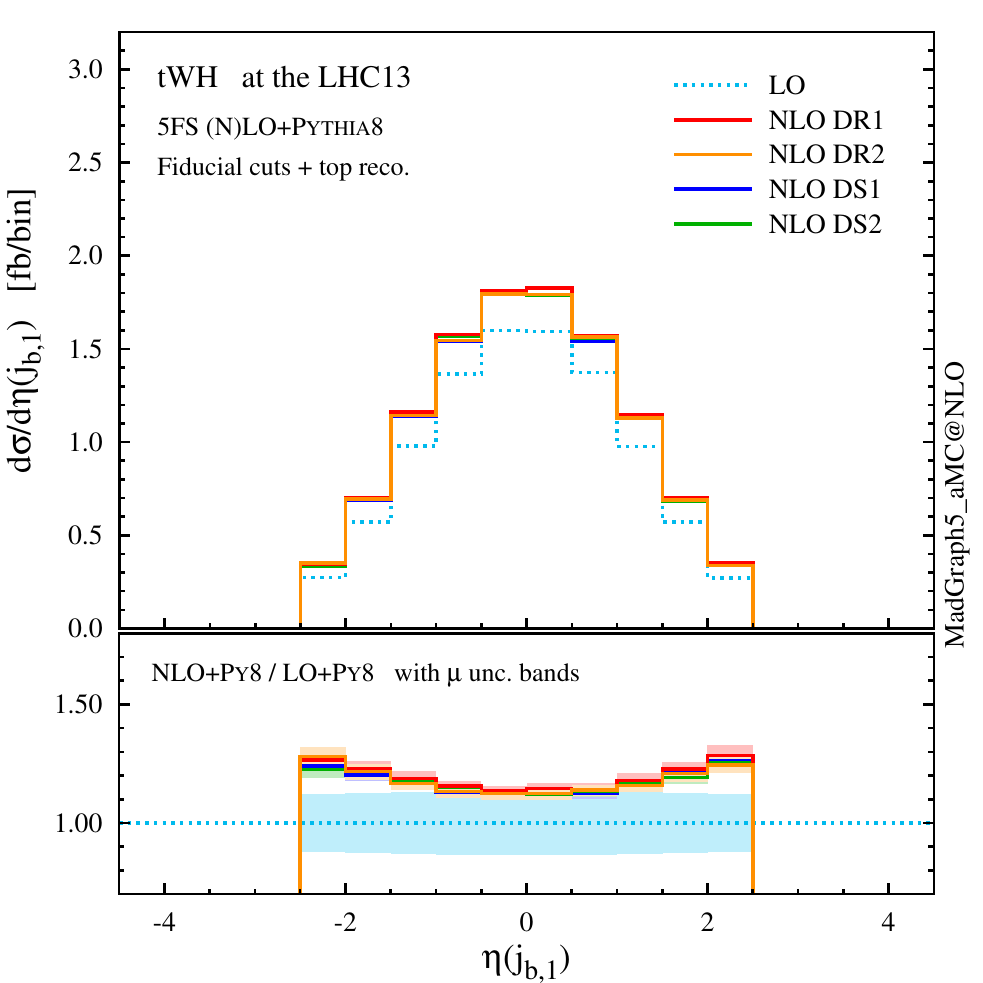}
\caption{Same as fig.~\ref{fig:tWH_5FNLO_dist_2}, but for the central $b$-tagged jet. 
 For the right plot, in addition to the fiducial cuts, the top reconstruction is required.}
\label{fig:tWH_5FNLO_dist_2b}
\end{figure*}

Looking at table~\ref{tab:tWH_NLO_xsect_2}, we can see that the situation for 
$tWH$ is very similar to the one we have already seen for $tW$.
Before the fiducial cuts, the category is largely dominated by $t \bar t H$ events.
Once the fiducial cuts are applied, the contribution from $t \bar t H$ 
is reduced by more than a factor 20, while the one from $tWH$ just by
about 1/4 (for DR2), enhancing the signal-to-background ratio ($tWH/t \bar t H$) 
to about 0.5, which is encouraging from the search point of view.
The interference with LO $t \bar t H$ amplitudes has been visibly reduced, 
with fiducial cross sections among the four techniques agreeing much better than in 
the inclusive case; this is also apparent in the differential distributions 
of figs.~\ref{fig:tWH_5FNLO_dist_2} and \ref{fig:tWH_5FNLO_dist_2b}, and in particular in the much smaller
scale dependence in the tails of $tWH$ distributions at NLO.

Nevertheless, a residual difference of about $6\perc$ (0.7 fb) is present
between the DR1 and DR2 fiducial cross sections, and this discrepancy is
also visible in the shape of some $p_T$ distributions. 
Once again, if we use MC information to additionally require 
the central $b$ jet to come unambiguously from the top quark,
the residual interference effects are further reduced to less than $1\perc$
at a tiny cost on the signal efficiency.
This brings the differential predictions in excellent agreement among the four schemes  
and with this selection one can effectively consider $tWH$ and $t \bar t H$ 
as separate processes. 

Finally, we briefly comment on the possibility to observe the $tWH$ signal
at the LHC.
Naturally, one may wonder whether it will be possible to observe it 
over the (already quite rare) $t \bar t H$ process, in an experimental 
analysis that applies a selection similar to our fiducial cuts. 
For example, the LHC Run II is expected to deliver an integrated luminosity 
in the 100~fb$^{-1}$ ballpark. 
In our pseudo event category (with top reconstruction), the difference between 
including or excluding the $tWH$ contribution amounts to\\[1em] 
\begin{tabular}{rl}
 $t \bar t H$ only : 
 & $2147 \, \pm 46$ (stat.)   $^{+101}_{-204}$ (theo.) events, \\[1em]
 $t \bar t H + tWH$ : 
 & $3251 \, \pm 57$ (stat.)   $^{+147}_{-257}$ (theo.) events. \\[1em]
\end{tabular}
Unfortunately, once branching ratios of the Higgs and $W$ bosons and 
realistic efficiencies are taken into account, these numbers disfavour 
the possibility to observe $tWH$ over $t \bar t H$ at the Run II.
On top of that, there are many more background processes contributing to our 
event category than just $t \bar t H$. This makes the searches for the SM $tWH$ 
signal extremely challenging, and the high-luminosity upgrade of the
LHC is definitely needed in order to have a sufficient number of events.

On the other side, simulated $tWH$ events should be taken 
into account in other searches for Higgs boson and top quark associated 
production, which are not necessarily going to apply $tWH$-specific 
fiducial cuts, in order to complete the MC modelling. 
In particular, this will be relevant in searches for the $t\bar{t}H$ signal,
and also for the $t$-channel $tH$ process (also called $tHq$ by experiments) 
with Higgs decay into a pair of bottom quarks ($H\to b\bar{b}$), where 
semileptonic $tWH$ events can lurk in the signal region 
defined by a large ($b$-)jet multiplicity. 
In fact, including the $tWH$ simulation in the signal definition 
(as opposed to considering it a background) in the case of either 
$t\bar{t}H$ or $t$-channel $tH$ searches will lead to a more 
comprehensive view on Higgs boson and top-quark associated production,
e.g. being relevant when setting limits or measuring the signal strength.%

\subsection{Higgs characterisation}
\label{sec:HC}

In this section we explore the sensitivity of $tWH$ production 
to beyond the Standard Model (BSM) physics in the Higgs sector.
In particular, we start by studying the total production rate in the 
so-called ``$\kappa$-framework''
\cite{LHCHiggsCrossSectionWorkingGroup:2012nn,Heinemeyer:2013tqa} 
where the SM Higgs interactions are simply 
rescaled by a dimensionless constant $\kappa$. Then, we move to
characterising the Yukawa interaction between the Higgs boson 
and the top quark, which in general can be a mixture of CP-even 
and CP-odd terms, similar to what has been done for $t$-channel 
$tH$ production in sec.~5 of~\cite{Demartin:2015uha}.
To describe the Yukawa interaction, we consider the following Lagrangian
for a generic spin-0 mass eigenstate $X_0$ that couples to both scalar and
pseudoscalar fermionic currents
\begin{align}
 {\cal L}_0^t 
   = -\bar\psi_t\big(
         c_{\alpha}\kappa_{\sss Htt}g_{\sss Htt} 
       +i s_{\alpha}\kappa_{\sss Att}g_{\sss Att}\, \gamma_5 \big)
      \psi_t\, X_0 \,,
\label{eq:0ff}
\end{align}
where $c_{\alpha}\equiv\cos\alpha$ and $s_{\alpha}\equiv\sin\alpha$ 
are the cosine and sine of the CP-mixing phase $\alpha$; 
$\kappa_{\sss Htt,Att}$ are real dimensionless parameters that rescale the
magnitude of the CP-even and CP-odd couplings, and
$g_{\sss Htt}=g_{\sss Att}=m_t/v\,(=y_t/\sqrt{2})$, with $v\simeq 246$~GeV.  
While redundant (only two independent real quantities are needed to parametrise 
the most general CP-violating interaction between a spin-0 particle and
the top quark at dimension four), this parametrisation has 
the practical advantage of easily interpolating between 
the purely CP-even ($c_{\alpha}=1,s_{\alpha}=0$) and
purely CP-odd  ($c_{\alpha}=0,s_{\alpha}=1$) cases, 
as well as to easily recover the SM when $c_{\alpha}=1 \,,\, \kappa_{\sss Htt}=1 \,$.
In the $\kappa$-framework $c_{\alpha}=1$, and only the part proportional
to $\kappa_{\sss Htt}$ is considered.
On the other hand, the SM-like interactions between the Higgs and the 
EW vector bosons is described by
\begin{align}
 {\cal L}_0^V =
 \kappa_{\rm SM}\big( \tfrac{1}{2}g_{\sss HZZ}\, Z_\mu Z^\mu 
                    +g_{\sss HWW}\, W^+_\mu W^{-\mu} \big) \, X_0 \,,
\label{eq:0vvSM}
\end{align}
where $g_{\sss HVV}=2m^2_V/v$ ($V=W,Z$).
For the full Higgs characterisation (HC) Lagrangian, including 
CP-even and CP-odd higher-dimensional $X_0VV$ operators,
we refer to~\cite{Artoisenet:2013puc,Maltoni:2013sma}.
The Feynman rules from these Lagrangians are coded in the publicly 
available~\texttt{HC\_NLO\_X0} model~\cite{FR-HC:Online}. 
The code and events for $tWX_0$ production at NLO can be generated
in a way completely analogous to SM $tWH$:
\vspace*{0.5em}
\begin{verbatim}
 > import model HC_NLO_X0-no_b_mass
 > generate p p > t w- x0 [QCD]
 > add process p p > t~ w+ x0 [QCD]
\end{verbatim}
\vspace*{0.5em}

\begin{figure*}
\center 
\includegraphics[height=0.4\textwidth]{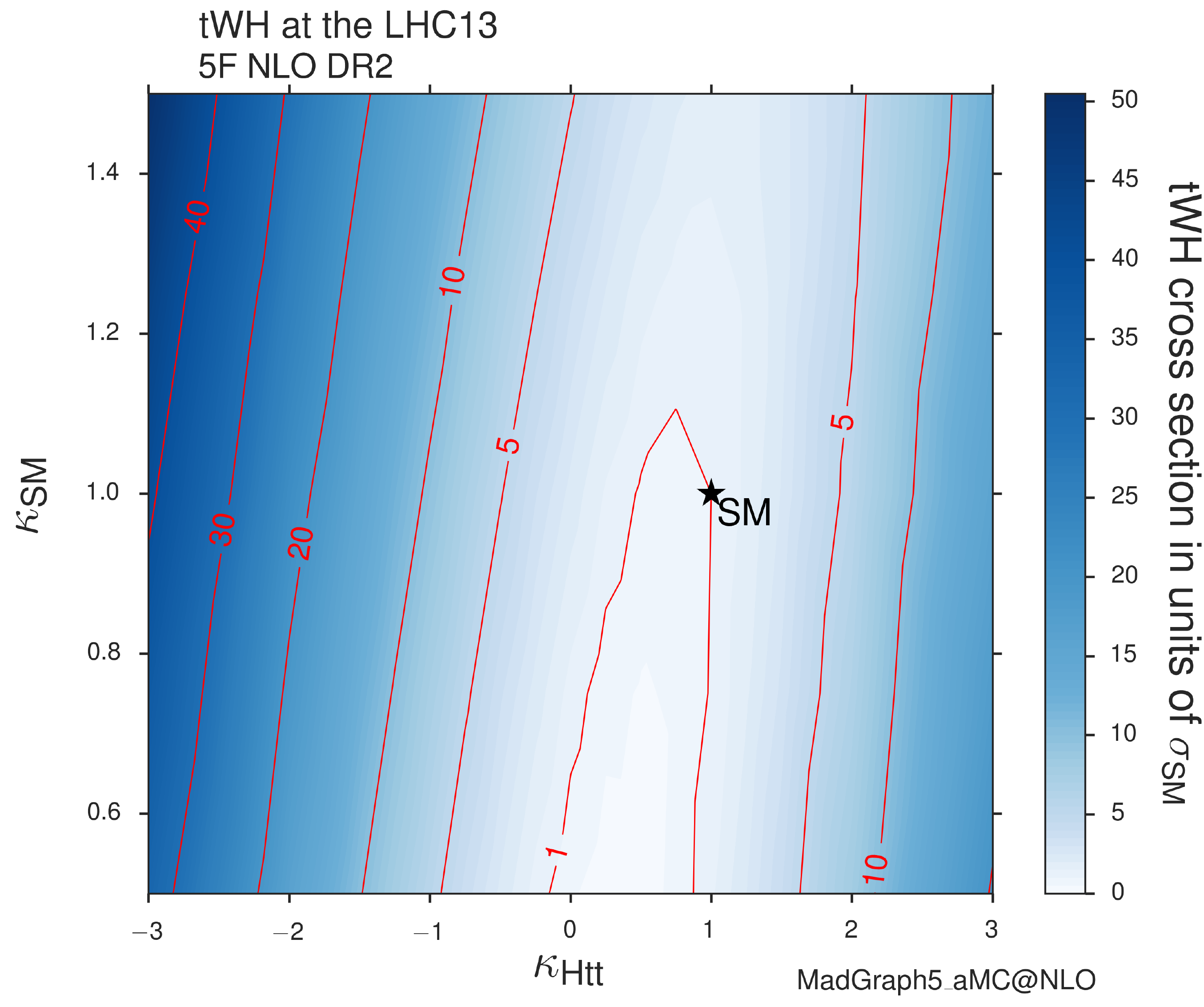}\qquad
\includegraphics[height=0.38\textwidth]{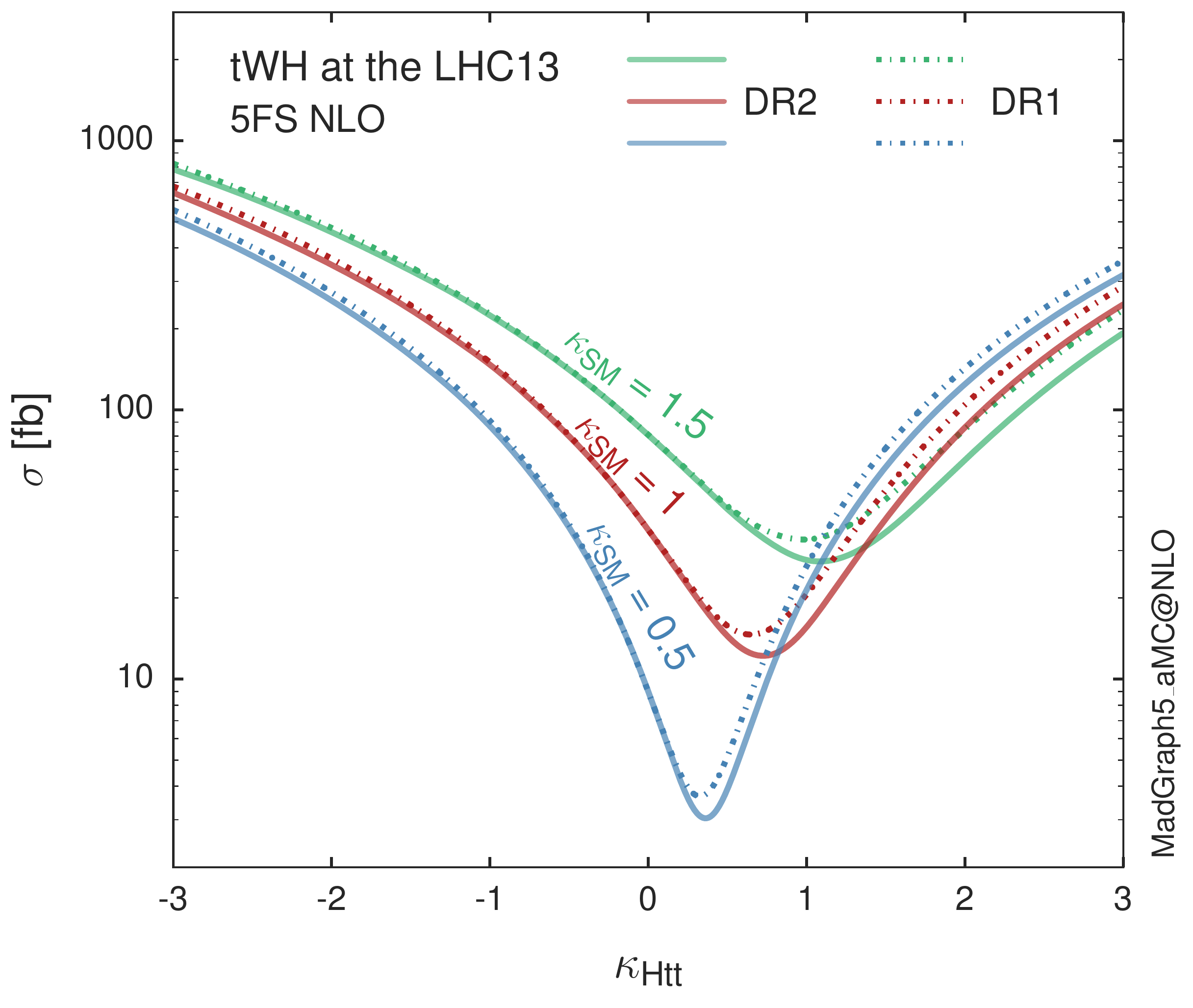}\\
\caption{Left: Inclusive $tWH$ cross sections with DR2 
 scanned over different values for $\kappa_{Htt}$ and $\kappa_{\rm SM}$.
 Note that the Standard Model configuration ($+1,+1$) almost lies in a minimum, 
 which means the process is suited for constraining this place due to 
 enhanced rates for deviations from the SM.
 Right: The $tWH$ cross section is shown for three different
 intensities of the $X_0WW$ coupling $\kappa_{\rm SM}$, as a function
 of $\kappa_{Htt}$, where DR1 results are also reported, to gauge the impact
 of interference with $t \bar t H$.} 
\label{fig:tWH_xsec_cfcv}
\end{figure*} 

\begin{figure}
\center 
 \includegraphics[width=0.95\columnwidth]{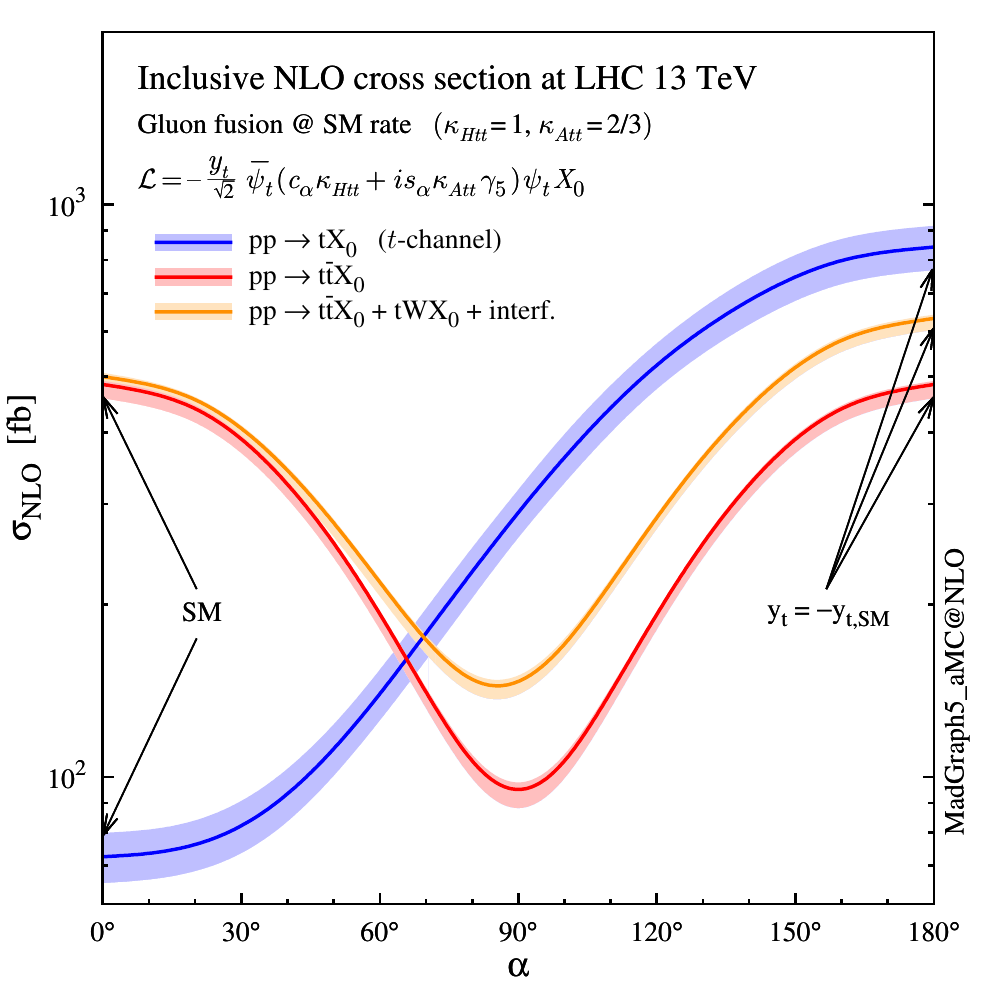}
\caption{
NLO cross sections (with scale uncertainties) for $pp \to t \bar t X_0$, $pp \to tWX_0$ (with DR2) and $pp \to tX_0$ ($t$-channel) at the 13-TeV LHC as a function of the CP-mixing angle $\alpha$, where $\kappa_{\sss Htt}$ and $\kappa_{\sss Att}$ are set to reproduce the SM gluon-fusion cross section for every value of $\alpha$.
 The $t \bar t X_0$ and $tWX_0$ processes have been computed using
 the dynamic scale $\mu_0=H_T/4$, while $tX_0$ results are 
 taken from~\cite{Demartin:2015uha}.
} 
\label{fig:HC_xsect}
\end{figure}

In this section we show results obtained only with the DR techniques.  
We start by showing results in the $\kappa$-framework in
fig.~\ref{fig:tWH_xsec_cfcv}.
We can see that a CP-even Higgs boson 
is highly sensitive to the relative sign of Higgs couplings to fermions ($t$)
and EW bosons ($W$). Depending on the ($\kappa_{Htt},\,\kappa_{SM}$) 
configuration, the inclusive $tWH$ rate 
(DR2, including interference with $t \bar t H$)
can be enhanced from $15$~fb to almost $800$~fb. 
The $tWH$ process can thus be exploited to further constrain the 
allowed regions in the 2-dimensional plane spanned by $\kappa_{Htt}$ 
and $\kappa_{SM}$ together with the already sensitive $tH$ production.

Given the experimental constraints after the LHC Run~I~\cite{Khachatryan:2016vau},
we can reasonably fix the Higgs interaction with the EW bosons to be
the SM one, and turn to study CP-mixing effects in the Higgs--fermion sector.
It is also reasonable to assume that gluon fusion is 
dominated by the top-quark loop, and consequently the $X_0$--top interaction 
must reproduce the SM gluon-fusion rate at NLO accuracy to comply with 
experimental results.
This fixes the values of the rescaling factors in eq.~\eqref{eq:0ff} to
\begin{align}
 \kappa_{\sss Htt} = 1 \,, \qquad \kappa_{\sss Att} 
 = |\, g_{\sss Hgg}/ g_{\sss Agg} \,| = 2/3 \,,
\label{eq:GFkappas}
\end{align}
leaving the value of the CP-mixing angle $\alpha$ free.

\begin{figure*}
\center
 \includegraphics[width=0.2425\textwidth]{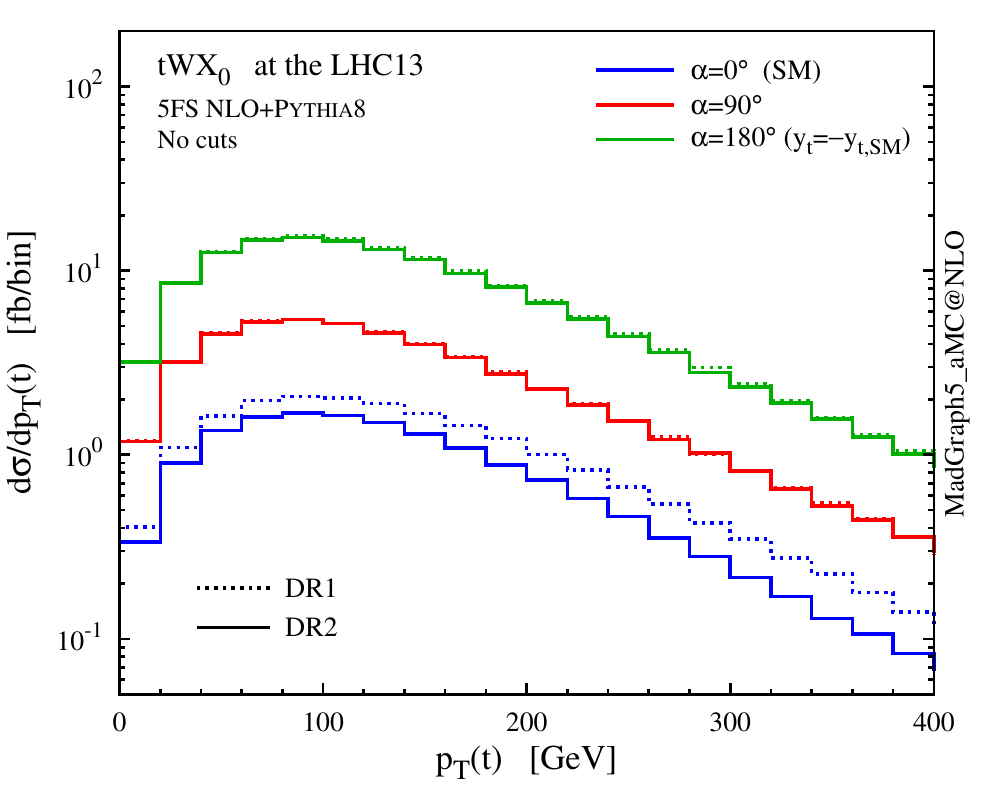}
 \includegraphics[width=0.2425\textwidth]{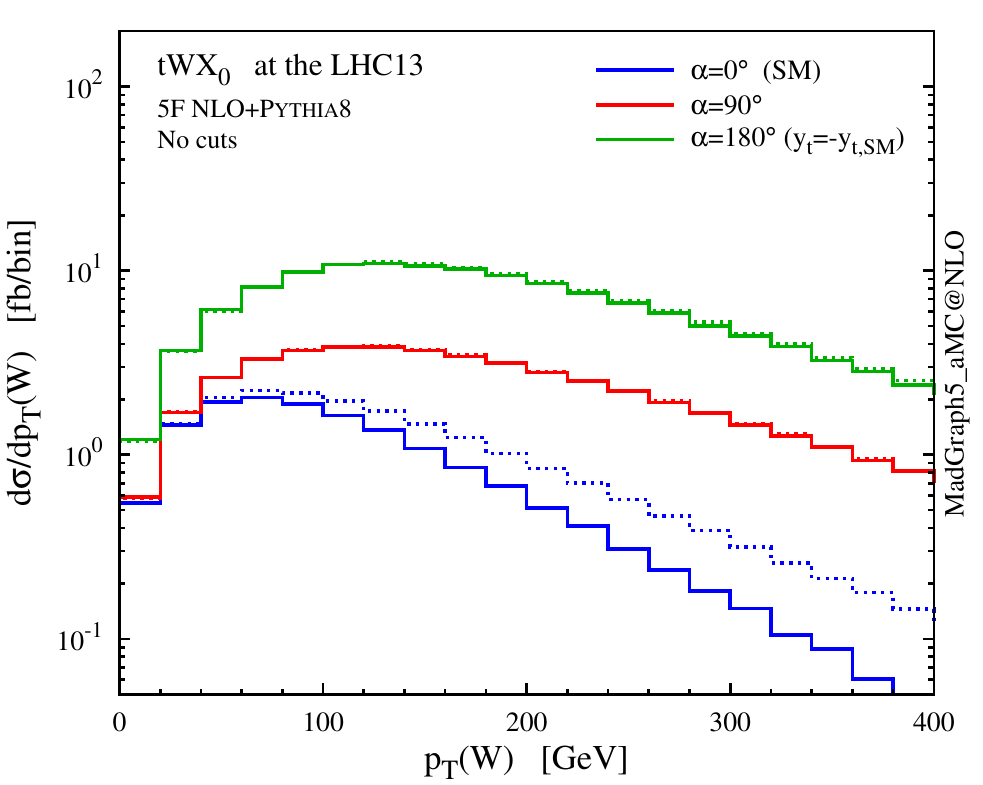} 
 \includegraphics[width=0.2425\textwidth]{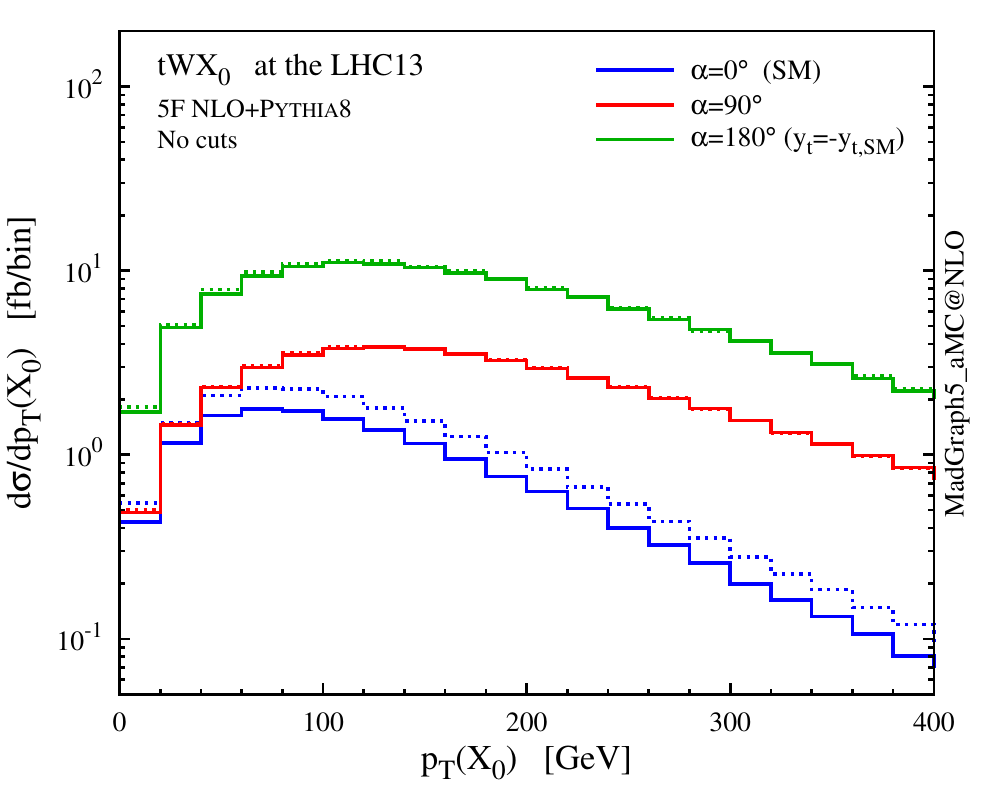}
 \includegraphics[width=0.2425\textwidth]{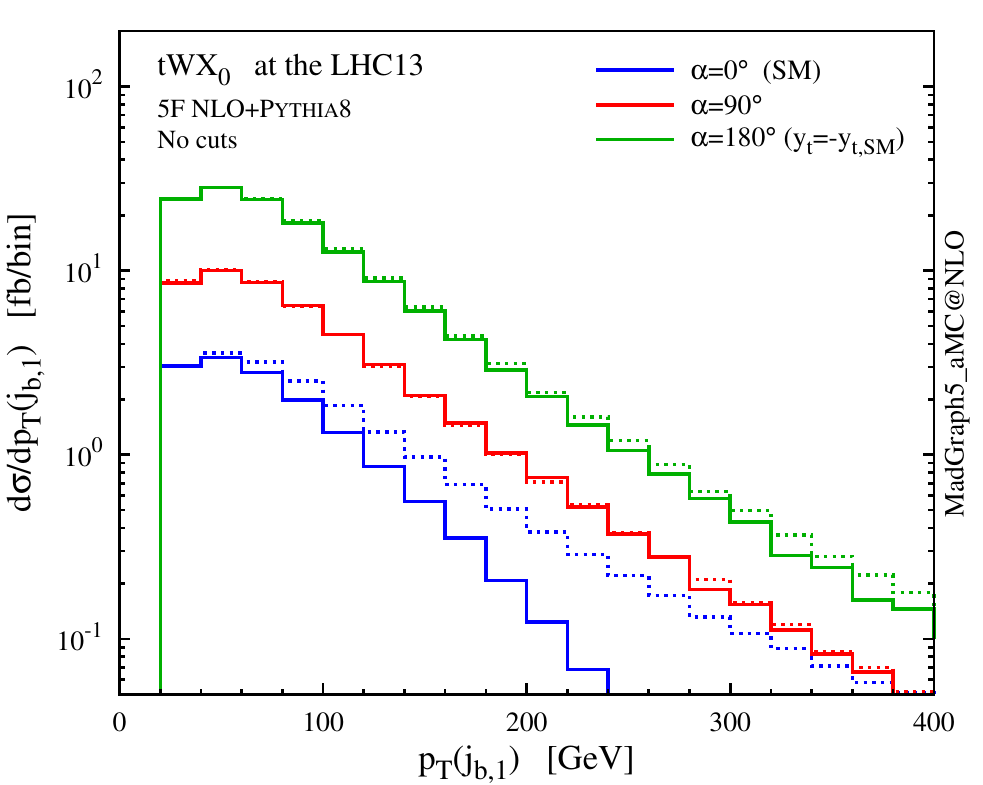}\\
 \includegraphics[width=0.2425\textwidth]{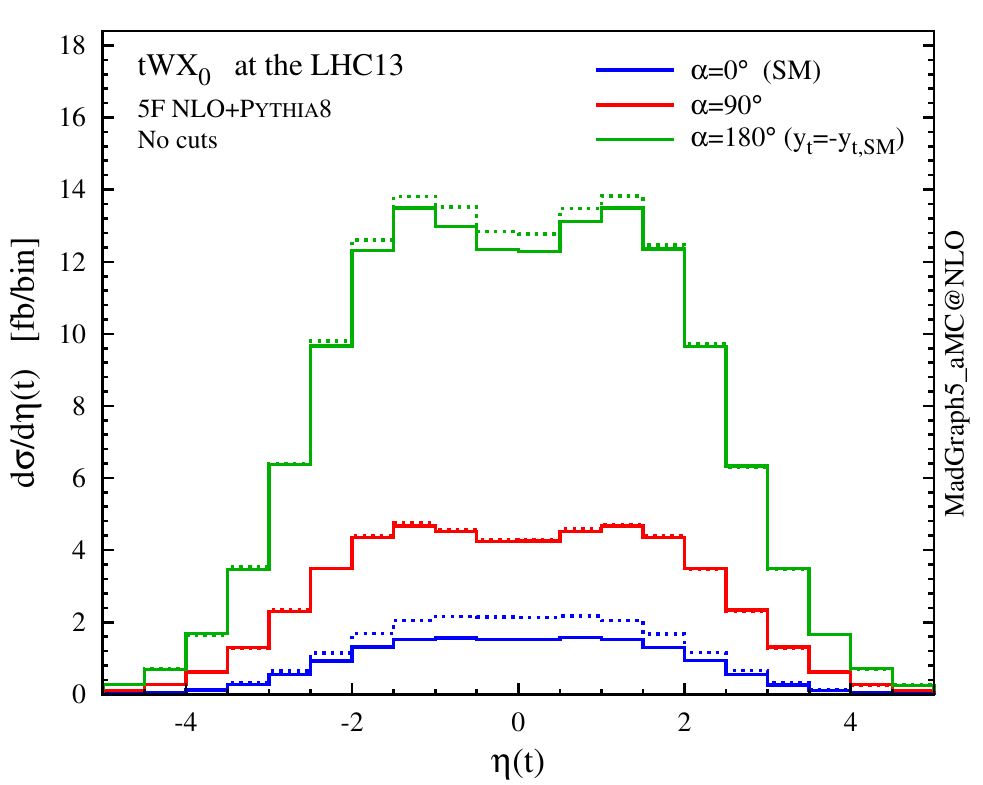}
 \includegraphics[width=0.2425\textwidth]{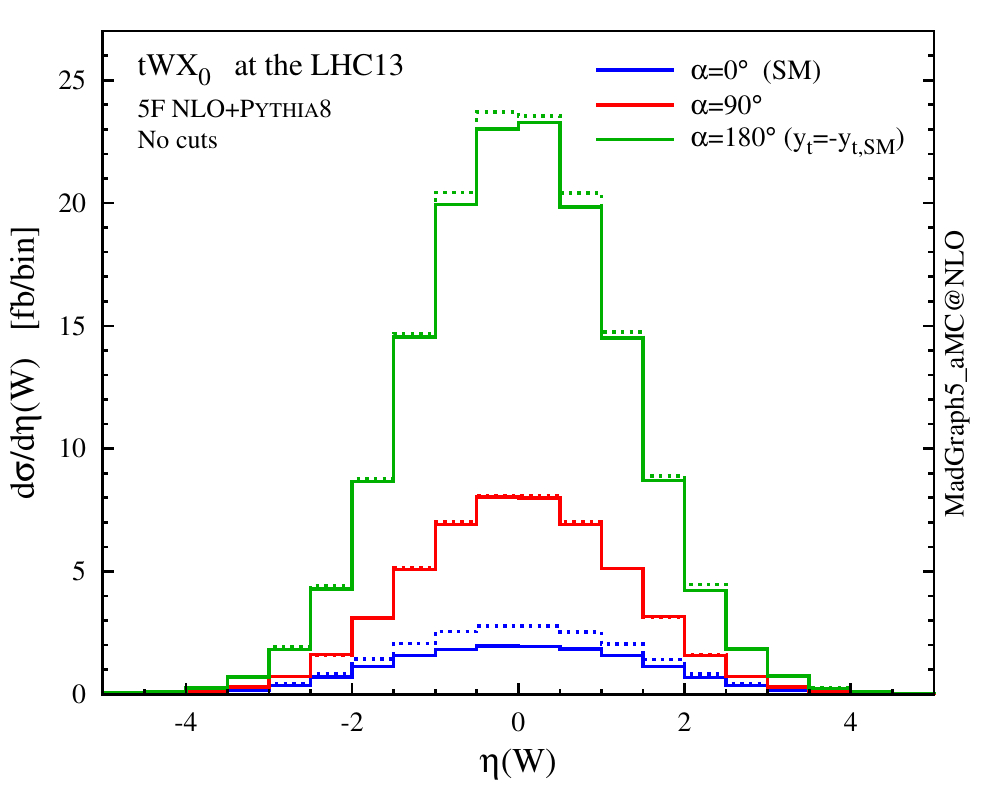}
 \includegraphics[width=0.2425\textwidth]{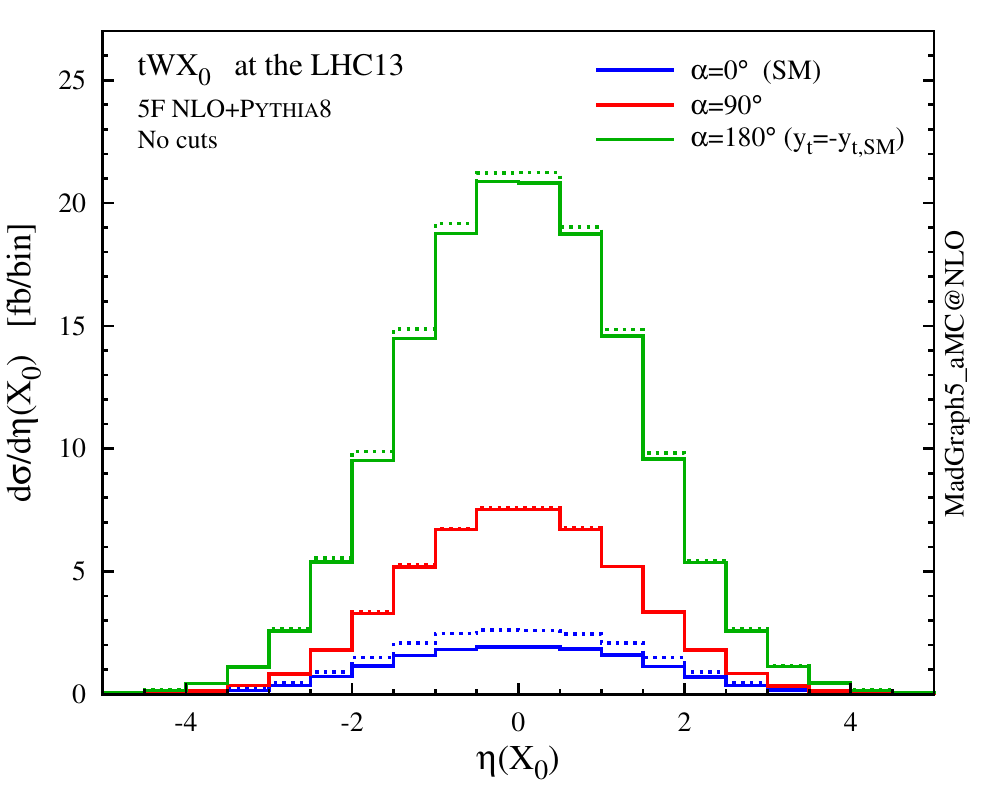} 
 \includegraphics[width=0.2425\textwidth]{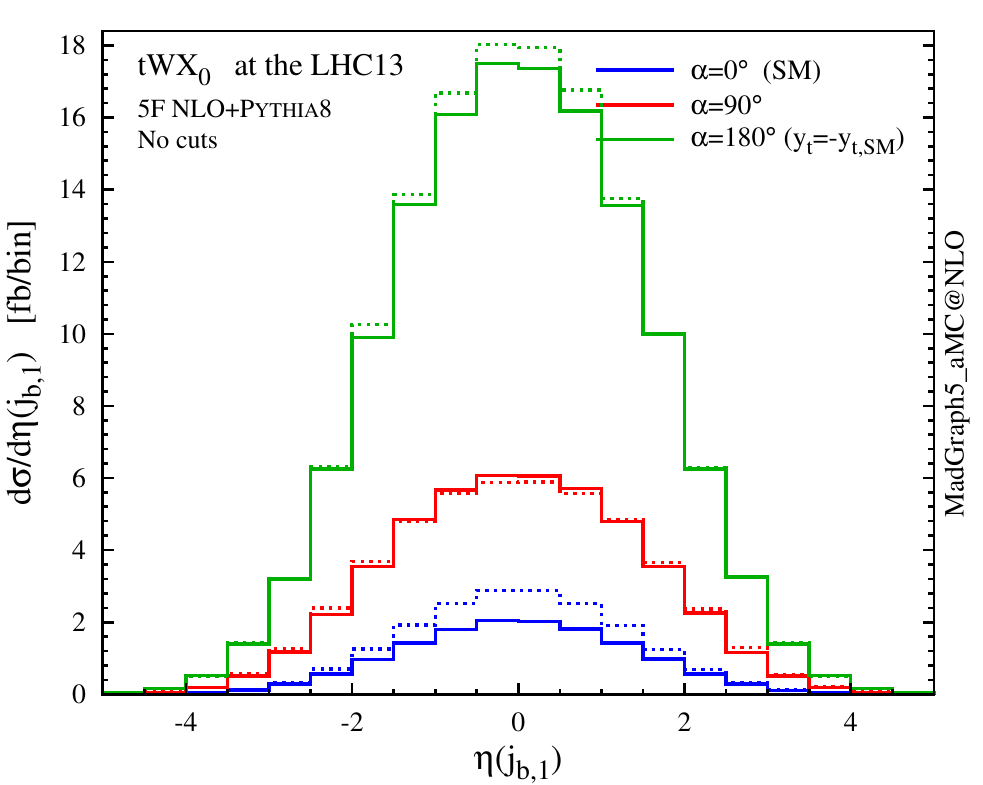}
\caption{
 $p_T$ and $\eta$ distributions for the top quark, the $W$ boson and the Higgs boson at NLO+PS accuracy in $tWH$ production at the 13-TeV LHC with different values of the CP-mixing angles between the Higgs boson and the top quark, where $\kappa_{\sss Htt}$ and $\kappa_{\sss Att}$ are set to reproduce the SM gluon-fusion cross section for every value of $\alpha$.
 The results are obtained employing DR2 (solid) and DR1 (dashed), without any cut.} 
\label{fig:HC_dist_1}
\end{figure*} 

\begin{figure*}
\center
 \includegraphics[width=0.2425\textwidth]{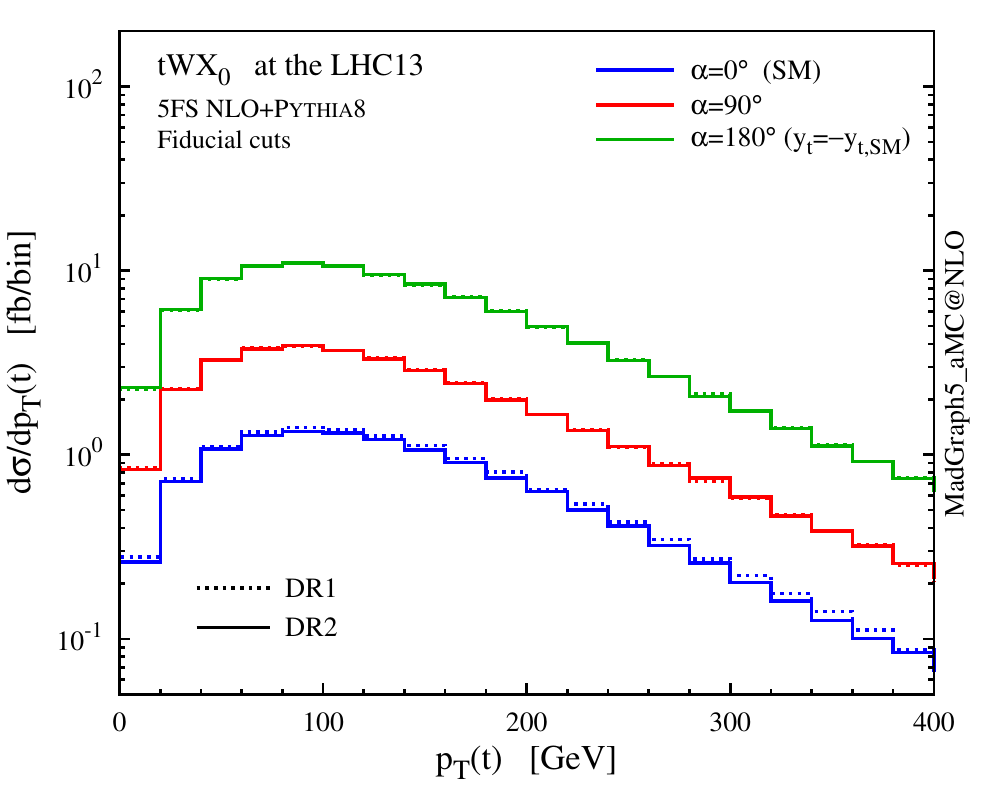}
 \includegraphics[width=0.2425\textwidth]{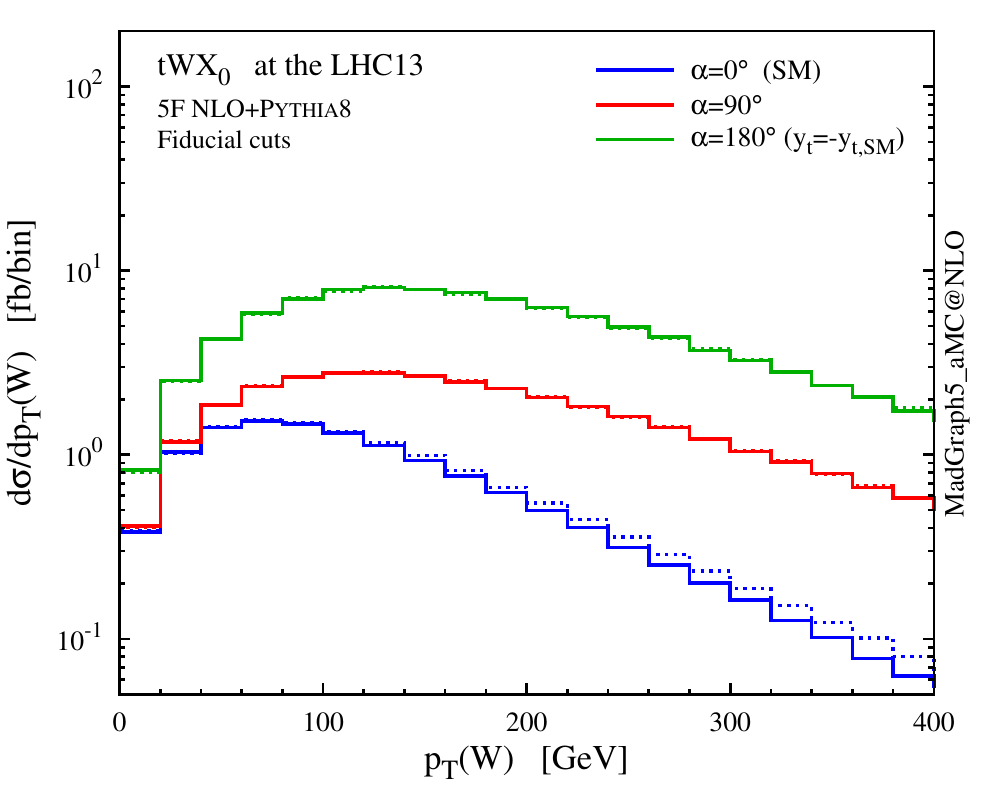} 
 \includegraphics[width=0.2425\textwidth]{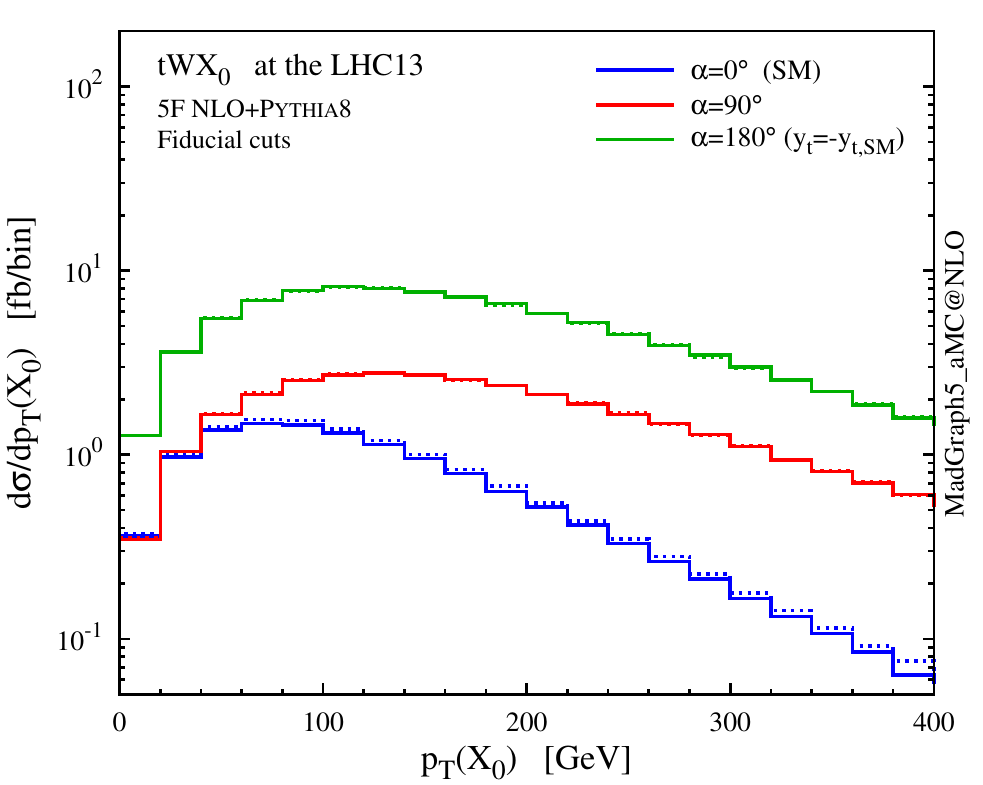}
 \includegraphics[width=0.2425\textwidth]{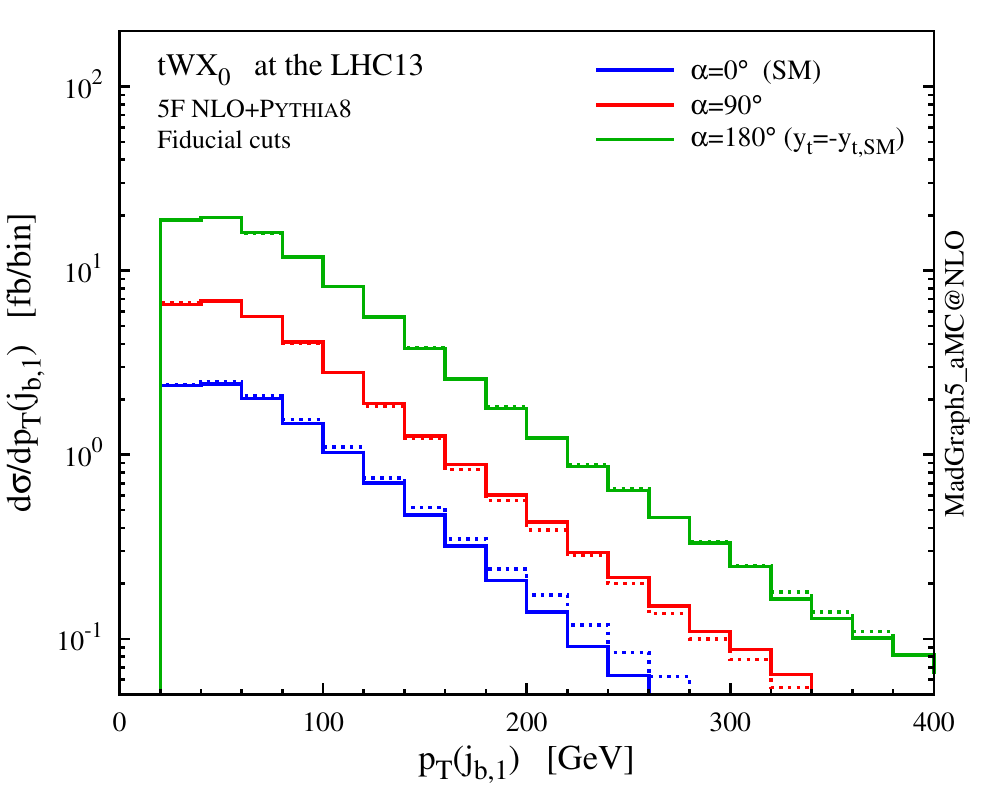}\\
 \includegraphics[width=0.2425\textwidth]{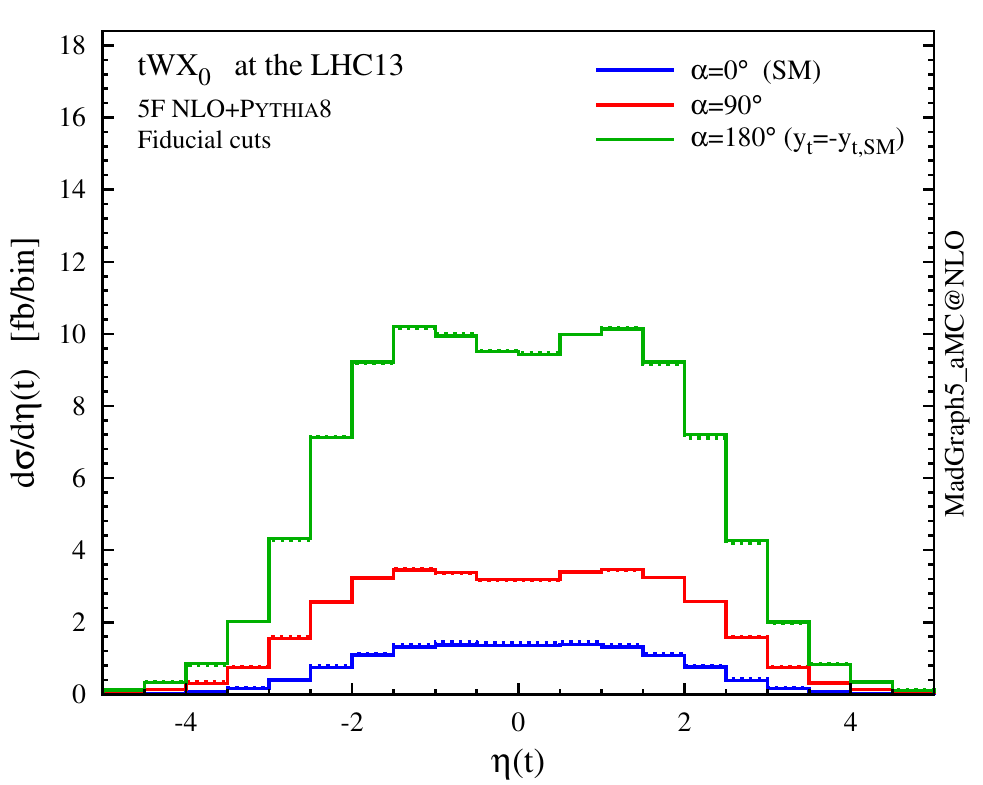}
 \includegraphics[width=0.2425\textwidth]{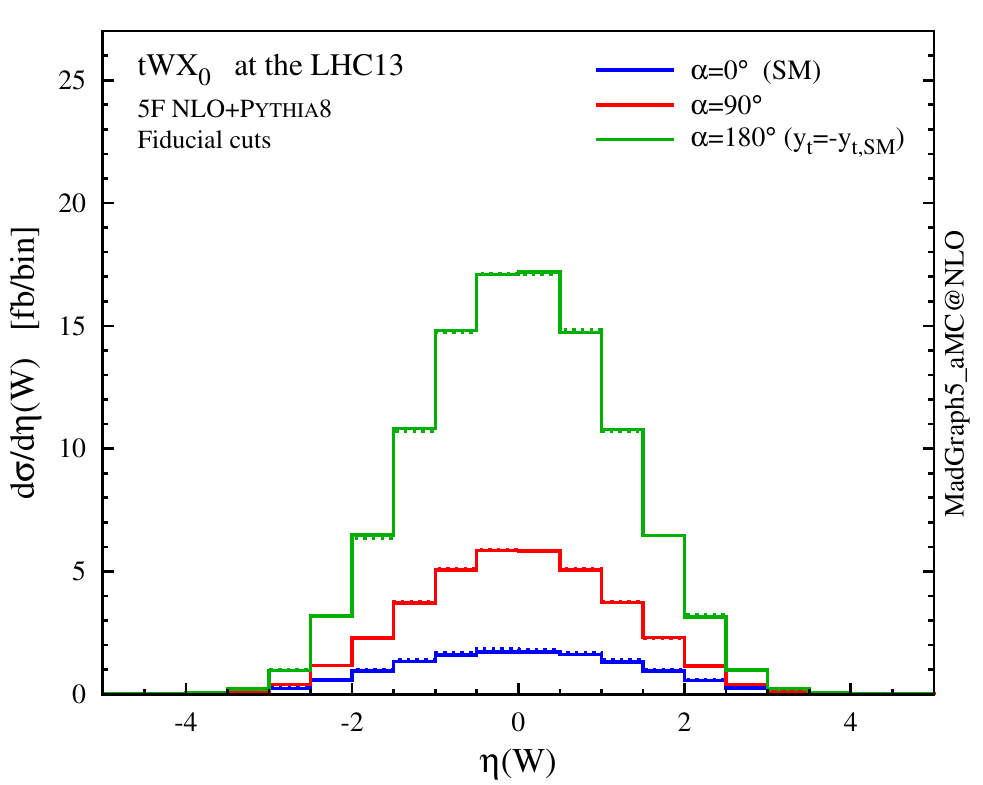}
 \includegraphics[width=0.2425\textwidth]{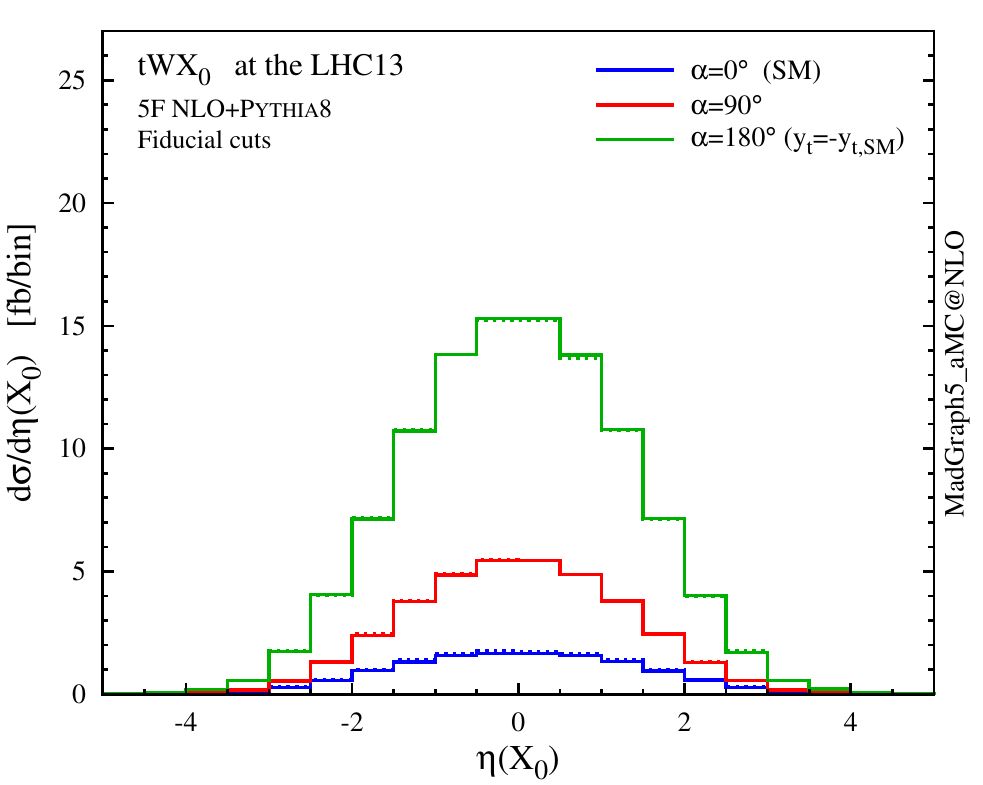} 
 \includegraphics[width=0.2425\textwidth]{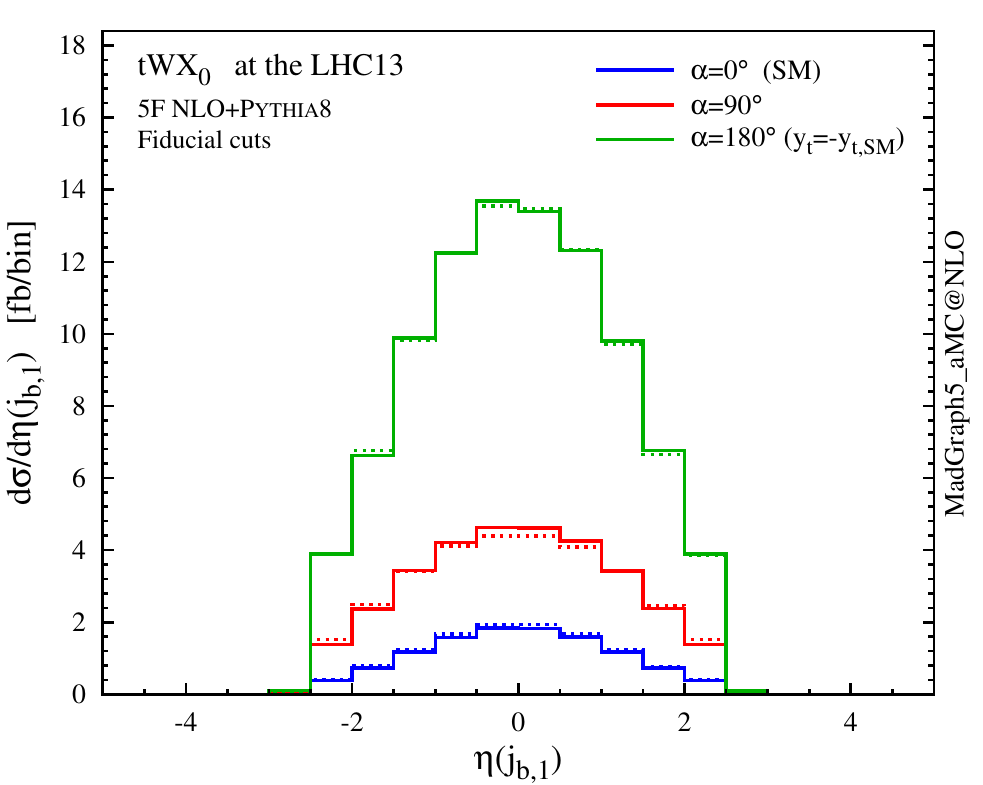}
\caption{Same as in fig.~\ref{fig:HC_dist_1}, but after applying the fiducial cuts.} 
\label{fig:HC_dist_2}
\end{figure*} 

In fig.~\ref{fig:HC_xsect} we plot the total NLO cross section for 
Higgs production in association with a top-quark pair $t \bar t X_0$ (red), 
and for the combined contribution of $t \bar t X_0$ and $tWX_0$ 
including their interference (orange), 
which is simply obtained by summing the $tWX_0$ DR2 cross section
to the $t \bar t X_0$ one.
We can immediately see that the inclusion of the
$tWX_0$ process lifts the $y_t \to - y_t$ degeneracy that is present in
$t \bar t X_0$ production.
For a flipped-sign Yukawa coupling, the interference between
single-top diagrams where the Higgs couples to the top and the ones where 
it couples to the $W$ becomes constructive, 
and the total cross section is augmented from roughly $500$~fb 
(SM, $\alpha=0^\circ$) to more than $600$~fb ($\alpha=180^\circ$). 
This enhancement can help in a combined analysis of the Higgs interactions, 
though it is less striking than the one which takes place in the
$t$-channel Higgs plus single-top process (which is also reported
in blue for comparison).
For the sake of clarity we point out that, going along the $\alpha$-axis 
in fig.~\ref{fig:HC_xsect}, the $tWX_0$ cross section includes in fact 
two different interference effects. 
On the one hand, there is the interference between single-top amplitudes 
with Higgs-to-fermion and Higgs-to-gauge-boson interactions, similar to the $tH$ process.
This is already present at LO, 
and it drives the growth of the cross section from the SM case 
(maximally destructive interference) to the case of a 
reversed-sign top Yukawa (maximally constructive).
On the other hand, employing DR2 for the computation of the $tWX_0$ 
NLO cross section means that also the interference with $t\bar{t}H$ 
is included. This is an effect present only at NLO, and its size depends 
as well on the CP-mixing angle $\alpha$ (due to the different ratio between 
$t \bar t H$ and $tWH$ amplitudes).

In fig.~\ref{fig:HC_dist_1} we compare some differential distributions 
for the SM hypothesis (blue), the purely CP-odd scenario (red)
and the flipped-sign CP-even case (green), before any cuts.
We can see that the interference between the doubly resonant
$t \bar t H$ and the singly resonant $tWH$ amplitudes is largest for the SM case.
For the case of flipped Yukawa coupling the interference gives a minor 
contribution, while for the CP-odd case it is very tiny
because the doubly resonant contribution is at its minimum.
The $W$ and Higgs transverse momentum distributions become harder 
when the mixing angle is larger.
Once the fiducial cuts are applied (fig.~\ref{fig:HC_dist_2}), 
the difference between DR1 and DR2 decreases as expected.

In conclusion, we find that the $tWH$ process can help to lift the $y_t \to - y_t$
degeneracy for $t\bar tH$ and put constraint on BSM Yukawa interactions of the Higgs boson 
in a combined analysis, on top of the most sensitive $t$-channel $tH$ production mode. 
Finally we recall that, if one also assumes a SM interaction
between the Higgs and the $W$ bosons, one can further include
the $\gamma\gamma$ decay channel data to put limits on the
CP-mixing phase $\alpha$.

\section{Summary}
\label{sec:summary}

In this work we have provided for the first time NLO accurate predictions for the $tWH$ process,
including parton-shower effects. 
In order to achieve a clear understanding of the ambiguities associated to the very definition of the process
at NLO accuracy due to its mixing with $t \bar t H$, we have revisited the currently available subtraction schemes
in the case of $tW$ production. 
We have therefore carefully analysed  $tW$ at NLO in the five-flavour scheme,
and then we have proceeded in an analogous way for $tWH$.
On the one hand, NLO corrections to these processes are crucial for a 
variety of reasons, ranging from a reliable description of the $b$ 
quark kinematics to a better modelling of backgrounds in searches for 
Higgs production in association with single top quark or a top pair.
On the other hand, they introduce the issue of interference with
$t \bar t$ or $t \bar t H$ production, which has a significant 
impact on the phenomenology of these processes.

Our first aim has been to study the pro's and the con's of the various techniques 
(which fall in the GS, DR and DS classes) that are available to subtract the resonant 
contributions appearing in the NLO corrections.
At the inclusive level these techniques can deliver rather different results, 
with differences which can often exceed the theoretical uncertainties on
the NLO cross sections estimated via scale variations. 
These differences have been traced back to whether a given technique accounts for the interference
between the $tW(H)$ and $t\bar t(H)$ processes, and to how the off-shell tails of the resonant diagrams are treated. 
They become visible at the total cross section level as well as in distributions, 
particularly those involving $b$-jet related observables.
We find the DR2 and DS2 techniques to provide a more faithful description of the underlying physics
in $tW$ and $tWH$ than that of DS1 and DR1, therefore we deem them as preferable to generate events 
for these two processes at NLO.
We stress that the aim of our work is to provide a practical and reliable technique to simulate $tW$ and $tWH$ at NLO,
when the corresponding $t \bar t$ and $t \bar t H$ process are generated separately in the on-shell approximation.
Our results have no claim of generality, and cannot be immediately extended to other SM or BSM processes.
A study of subtraction techniques should be undertaken on a process-by-process basis,
in particular for BSM physics, where different width-to-mass ratios and different amplitude structures 
(\emph{i.e.} resonance profiles) can appear.

Our second aim has been to study what happens once event selections similar
to those performed in experimental analyses are applied, and in general whether 
one can find a fiducial region where the single-top processes
$tW$ and $tWH$ can be considered well-defined \emph{per se}, and are stable under perturbative corrections.
A simple cut as requiring exactly one $b$-tagged jet in the central detector 
(which becomes three $b$ jets in the case of $tWH$ if the Higgs decays to bottom quarks) 
can greatly reduce interference effects, and thus all the process-definition systematics of $tW(H)$ at NLO.
In such a fiducial region, we find the perturbative description of $tW(H)$ to be well-behaved, 
and the inclusion of NLO corrections significantly decreases the scale dependence;
differences between the various DR and DS subtraction techniques are reduced below those due to missing perturbative
orders, making the separation of the single-top and top-pair processes meaningful. 
Given a generic set of cuts, we have provided a simple and robust recipe to estimate 
the left-over process-definition systematics, i.e. use the difference between the DR1 and DR2 predictions
(which amounts to the impact of interference effects).
In general, such approach provides a covenient way to quantify the limits in the separation of 
$t\bar t(H)$ and $tW(H)$ and the quality of fiducial regions.
In particular, this is essential for a reliable extraction of the Higgs couplings in $tWH$ production.

Finally, we have investigated the phenomenological consequences of considering a generic CP-mixed Yukawa
interaction between the Higgs boson and the top quark in $tWH$ production.
While the SM cross section is tiny, due to maximally destructive interference
between the $H$--$t$ and $H$--$W$ interactions, and direct searches for this process 
may only be feasible after the high-luminosity upgrade of the LHC,
BSM Yukawa interaction tend to increase the production rate.
For example, in the case of a reversed-sign Yukawa coupling with respect
to the SM, the $tWH$ cross section is enhanced by an order of magnitude,
similar to what happens for the dominant single-top associated mode,
i.e. the $t$-channel $tH$ production.
The large event rate predicted after the combination of these Higgs plus single
top modes will help to exclude a reversed-sign top Yukawa coupling 
already during the LHC Run II.

\section*{Acknowledgments}

We thank the LHCHXSWG and in particular the members of the $ttH/tH$ 
task force for giving us the motivation to pursue this study. 
We are grateful to Simon Fink, Stefano Frixione, Dorival Gon\c{c}alves-Netto, Michael Kr\"amer, 
David Lopez-Val, Davide Pagani, Tilman Plehn and
Francesco Tramontano for many stimulating discussions,
and to the MadGraph team (in particular Pierre Artoisenet, Rikkert Frederix,
Valentin Hirschi, Olivier Mattelaer and Paolo Torrielli) 
for their valuable help. 
KM would like to acknowledge the Mainz Institute for Theoretical Physics (MITP) 
for providing support during the completion of this work.

This work has been performed in the framework of the ERC grant 291377
``LHCtheory: Theoretical predictions and analyses of LHC physics:
advancing the precision frontier''  and of the FP7 Marie Curie Initial Training 
Network MCnetITN (PITN-GA-2012-315877). It is also supported in part by the 
Belgian Federal Science Policy Office through 
the Interuniversity Attraction Pole P7/37. 
The work of FD and FM is supported by the IISN ``MadGraph'' convention
4.4511.10 and the IISN ``Fundamental interactions'' convention 4.4517.08.
BM acknowledges the support by the DFG-funded Doctoral School
``Karlsruhe School of Elementary and Astroparticle Physics: Science and
Technology''. 
The work of KM is supported by the Theory-LHC-France initiative of the
CNRS (INP/IN2P3).
The work of MZ is supported by the
European Union's Horizon 2020 research and innovation
programme under the Marie Sklodovska-Curie grant agreement No 660171 and in part by 
the ILP LABEX (ANR-10-LABX-63), in turn supported by French state funds
managed by the ANR within the ``Investissements d'Avenir'' programme under
reference ANR-11-IDEX-0004-02.

\appendix
\section{The $\boldsymbol{tWb}$ and $\boldsymbol{tWbH}$ channels in the 4FS}
\label{app:twb_lo_4FS}

In this appendix we perform a study of the various ways to treat 
the $tWb$ channel, in particular we will discuss the performance and 
shortcomings of the diagram removal and diagram subtraction techniques, 
which are used to eliminate the $t \bar t$ resonant contribution.
Since the issue appears just in the matrix-element description, 
the study in this appendix is simply performed at the partonic level.
The $tWb$ channel is more easily addressed in the 4FS, where it appears as 
a finite and independent LO contribution, thus it can be isolated
from the other channels contributing to $tW$.
The only difference from the 5FS is that bottom mass effects are included
in the 4FS description, which act as an IR cutoff; 
the Feynman diagrams are the same ones describing the 5FS NLO real-emission 
channel, and the features and shortcomings of DR and DS are 
independent of the flavour scheme employed.
An analogous study is then repeated for the $tWbH$ channel in the 4FS.

The problem of the LO $t \bar t$ contribution in the $tW^{-} \bar b$ 
channel has first been addressed in~\cite{Tait:1999cf}, 
where it is subtracted at the cross section level
(see eq.~(4) in the reference).
This \emph{global subtraction} procedure (GS) is described in
sect.~\ref{sec:DRandDS}; an important point in the calculation is that the 
two pieces ($tW^{-} \bar b$ and $t \bar t$) are separately integrated before 
the subtraction is performed.
The GS procedure ensures that the remainder of the subtraction converges 
to a well-defined limit $\Gamma_t \to 0$, where the result is fully gauge invariant, 
and exactly all and just the LO on-shell $t \bar t$ contribution 
is subtracted.
Therefore, combining the $t \bar t$ simulation with the $tW^{-} \bar b$ 
obtained this way, one gets a well-defined total rate for producing 
the common physical final state, without double counting 
and also including interference effects;
this procedure provides a consistent way to define the $tW$ cross section.

Actually, the only way to perform 
a theoretically consistent simulation that encompasses both the top-pair 
and single-top contributions, that is gauge invariant and that includes 
interference and other finite-$\Gamma_t$ effects, is to compute 
$pp \to W^{+}bW^{-} \bar b$ in the 4FS and using a complex top-quark 
mass.
This $WbWb$ simulation will also contain the contribution
from amplitudes without any resonant top propagator $\mathcal{A}_{0t}$,
and also interference between single-top and single-antitop contributions
$\mathcal{A}_{1t}\mathcal{A}_{1 \bar t}^*$, 
which are not present in the $tWb$ simulation
\begin{align}
\label{eq:M_WbWb}
 & | \mathcal{A}_{WbWb} |^2 = | \mathcal{A}_{2t} + \mathcal{A}_{1t} + \mathcal{A}_{1 \bar t} + \mathcal{A}_{0t} |^2 
 \nn\\ & \hspace*{1em} 
 = | \mathcal{A}_{2t} |^2   \,+\,
 \Big[ \, | \mathcal{A}_{1t} |^2  
 + 2 \mathrm{Re} (\mathcal{A}_{2t}\mathcal{A}_{1t}^*) \, \Big]  
 \nn\\ & \hspace*{1em} 
 +\, \Big[\, | \mathcal{A}_{1 \bar t} |^2   \,+\, 
 2 \mathrm{Re} (\mathcal{A}_{2t}\mathcal{A}_{1 \bar t}^*) \, \Big]
 \nn\\ & \hspace*{1em} 
 +\, 2 \mathrm{Re} (\mathcal{A}_{1t}\mathcal{A}_{1 \bar t}^*) \, 
 \nn\\ & \hspace*{1em} 
 +\,  \Big[ \, | \mathcal{A}_{0t} |^2   \,+\,
 2 \mathrm{Re} \big( (\mathcal{A}_{2t}+\mathcal{A}_{1t}+\mathcal{A}_{1 \bar t}) 
 \mathcal{A}_{0t}^* \big) \, \Big] \,;
\end{align}
nonetheless, we expect the last two lines in eq.~\eqref{eq:M_WbWb}
to be negligible compared to the previous two lines,
which encompass top-pair $t \bar t$ and single-top $tWb$ production.

In the end, the reference result will be the difference between the
$WbWb$ cross section (computed in the complex-mass scheme, with a physical $\Gamma_t$) 
and the $t \bar t$ cross section (computed with on-shell top's), which 
in general guarantees a correct description of $tWb$ production.
If the non resonant contributions $\mathcal{A}_{0t}$ to $WbWb$,
the $\mathcal{A}_{1t}\mathcal{A}_{1 \bar t}^*$ interference,
and the off-shell effects related the single top kept stable in $tWb$ simulations
are small enough, this cross section will be close to the one obtained from GS.

The global subtraction schemes cannot be applied to event generation, 
where a fully \emph{local subtraction} of the top-pair contribution must be performed 
in the $2 \to 3$ phase space; this is exactly the reason why alternative techniques 
such as DR and DS have been developed and implemented in \textsc{MC@NLO} and
\textsc{POWHEG} for $tW$ production.
Nevertheless, a simple but powerful way to test the adequacy of DR and DS 
can be carried out by comparing their total cross section with the GS one,
which is the number we expect to be returned from a consistent local subtraction scheme.
We perform this comparison in table~\ref{tab:tWb_xsDRDS_4FS}, 
where cross sections are computed with the static scale $\mu_0^s$, also showing
the cross section ratio $R$ defined as
\begin{align}
\label{eq:ratioR}
 R = \frac{\sigma_{tWb}}{ \sigma_{WbWb}-\sigma_{t \bar t}} \,.
\end{align}

\begin{table}
\center  
\begin{tabular}{llc}
 \hline
 \rule{0pt}{3ex}   
 process \hspace*{2em}  
 & $\sigma_{\rm LO}$~{\small [pb]}   
 & $R$   
 \\[0.7ex] 
 \hline
 \rule{0pt}{3ex}  $WbWb$ \scriptsize{(complex $t$ mass)}	& 640.3(2)			&  -   \\[0.3ex]
 \rule{0pt}{3ex}  $t \bar t$ \scriptsize{($t$ stable)}		& 609.0(1)			&  -   \\[0.3ex]
 \rule{0pt}{3ex}  $WbWb$ $-$ $t \bar t$						& \enskip31.3(2)	&  1   \\[0.7ex]
 \hline
 \rule{0pt}{3ex}  $tWb$ GS 
                  & \enskip30.9(3)     & 0.99(1)   \\[0.7ex]
 \hline
 \rule{0pt}{3ex}  $tWb$ DR1  
                  & 40.79(1)     & 1.30(1)   \\[0.3ex]
 \rule{0pt}{3ex}  $tWb$ DR2 
                  & 31.11(1)     & 0.99(1)   \\[0.3ex]
 \rule{0pt}{3ex}  $WbWb$ $-$ $| \mathcal{A}_{2t} |^2$
                  & 31.81(1)     & 1.01(1)   \\[0.7ex]
 \hline
 \rule{0pt}{3ex}  $tWb$ DS1  
                  & 38.31(3)    & 1.22(1)   \\[0.3ex]
 \rule{0pt}{3ex}  $tWb$ DS2     
                  & 31.56(2)    & 1.01(1)   \\[0.7ex]
 \hline
\end{tabular}
\caption{LO cross sections in the 4FS at the 13-TeV LHC for the processes 
 $pp \to W^{+} b W^{-} \bar b$ (complex mass scheme), 
 $pp \to t \bar t$ ($t$ stable), 
 and singly resonant $pp \to tW^{-} \bar b$ plus $pp \to \bar t W^{+}b$ 
 computed using the GS, DR and DS prescriptions. 
 For these $tWb$ results we also report the ratio $R$
 defined in eq.~\eqref{eq:ratioR}. 
 All numbers are computed using the static scale $\mu_0^s = (m_t+m_W)/2$,
 and the numerical uncertainty affecting the last digit is reported
 in parentheses.}
\label{tab:tWb_xsDRDS_4FS}
\end{table}

\begin{figure}
\center
\includegraphics[width=1.0\columnwidth]{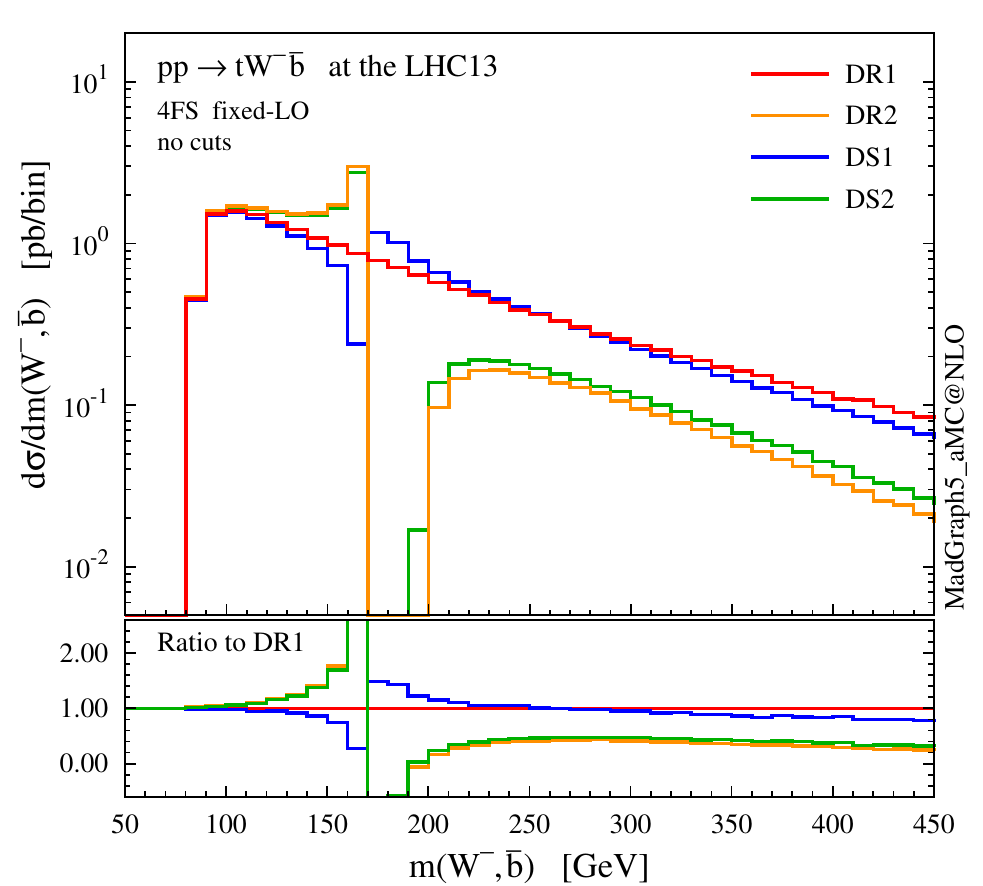}
\caption{Invariant mass $m(W^- , \bar b)$ in the $pp\to t W^- \bar b$ process, computed with DR and DS.}
\label{fig:tWb_4FS_distr}
\end{figure}

From the results in table~\ref{tab:tWb_xsDRDS_4FS}
we first notice that the $WbWb-t \bar t$ cross section 
(computed with a physical $\Gamma_t$)
is in good agreement with the $tWb$ one computed with the GS prescription 
(which is independent on the actual value of $\Gamma_t$),
thus either can be considered as the reference value.
This also confirms that non resonant contributions from ${A}_{0t}$ and
$\mathcal{A}_{1t}\mathcal{A}_{1 \bar t}^*$ interference are small,
and justifies the 5FS treatment where one top is always on-shell.

Among the two diagram removal techniques, the DR1 modelling does not
capture the $\mathcal{A}_{2t}\mathcal{A}_{1t}^*$ interference,
which amounts to more than 9~pb
(this was evident already in table~\ref{tab:tW_NLO_xsect_1}).
On the other hand, there is excellent agreement between the
DR2 cross section and the desired one from $WbWb-t \bar t$,
thus any possible violation of gauge invariance in the DR2 total rate
must be negligible.%
\footnote{We recall that in our simulations we have included only transverse
polarizations of initial-state gluons, and we have employed a covariant gauge for gluon propagators.
A non-covariant gauge (axial) was shown to lead to differences at the level
of permille in the case of $tW$ production~\cite{Frixione:2008yi}.
}
When we compute $|\mathcal{A}_{WbWb}|^2 - |\mathcal{A}_{2t}|^2$
(namely $WbWb - |\mathcal{A}_{2t}|^2$ in table~\ref{tab:tWb_xsDRDS_4FS}),
we can see that the difference with $tWb$ DR2 is a modest $2\perc$;
this provides a further confirmation that effects related to ${A}_{0t}$,
$\mathcal{A}_{1t}\mathcal{A}_{1 \bar t}^*$ interference, and off-shell $t$
are small; the subtraction of $|\mathcal{A}_{2t}|^2$
in a covariant gauge turns out to be almost equivalent to an on-shell $t \bar t$ subtraction
(compare $WbWb-t \bar t$ and $WbWb - |\mathcal{A}_{2t}|^2$).

Moving to diagram subtraction, we can see that DS2 is in rather good agreement
with GS and DR2, while DS1 clearly overestimates the total rate, 
which tends to be much closer to DR1.

The situation can be understood also at the differential level by looking at
the $m_{Wb}$ distribution in fig.~\ref{fig:tWb_4FS_distr}.
The missing of interference in DR1 leads to an underestimate of
the rate in the low-mass region $m_{Wb}<m_t$, and to an overestimate 
in the tail $m_{Wb}>m_t$; at the LHC energy, the latter region dominates,
leading to a net overestimate of the total rate.%
\footnote{We have verified that the net sum of interference effects in the total
rate is positive at collider energies below $\sim 2$~TeV, while becomes more 
and more negative at higher energies, where the phase space for $m_{Wb}>m_t$
is larger.
}
DR2 and DS2 nicely reproduce the peak-dip interference pattern, with small
differences between the two curves; since DS2 is gauge invariant, this fact 
can be interpreted as that gauge effects in DR2, when employing a covariant gauge,
are small also at the level of differential shapes.
Finally, while DS1 includes interference effects as well, it also introduces
a significant distortion in the profile of the subtraction term $\mathcal{C}_{2t}$,
as already shown in fig.~\ref{fig:mwb}; the net effect is an unreliable
$m_{Wb}$ profile, with an inverted dip-peak structure and a too large tail.


We now move on to studying the $tWbH$ channel
in $tWH$ production at NLO, which overlaps with LO $t \bar t H$. 
We follow a procedure completely analogous to the one employed for $tWb$, 
therefore we do not repeat all the details in the following discussion.

Our reference total rate is the difference between the $WbWbH$ cross section, 
computed in the complex top-quark mass scheme, and the $t \bar t H$ cross section 
computed in the approximation of stable final-state top quarks.
Once again we find GS to be in very good agreement with this reference value,
so both results can be taken as a reference for comparison with DR and DS, see table~\ref{tab:tWH_xsDRDS_4FS}.

\begin{table}
\center  
\begin{tabular}{llc}
 \hline
 \rule{0pt}{3ex}   
 process \hspace*{2em}  
 & $\sigma_{\rm LO}$~{\small [fb]}   
 & $R$   
 \\[0.7ex] 
 \hline
 \rule{0pt}{3ex}  $WbWbH$ \scriptsize{(complex $t$ mass)}	& 468.5(1)      &  -   \\[0.3ex]
 \rule{0pt}{3ex}  $t \bar t H$ \scriptsize{($t$ stable)}	& 463.0(1)      &  -   \\[0.3ex]
 \rule{0pt}{3ex}  $WbWbH$ $-$ $t \bar t H$					& \quad5.5(1)	&  1   \\[0.7ex]
 \hline
 \rule{0pt}{3ex}  $tWbH$ GS 
                  & \quad5.7(2)		& 1.04(3)   \\[0.7ex]
 \hline
 \rule{0pt}{3ex}  $tWbH$ DR1  
                  & 12.35(1)		& 2.27(5)   \\[0.3ex]
 \rule{0pt}{3ex}  $tWbH$ DR2 
                  & \enskip5.49(1)	& 1.01(2)   \\[0.3ex]
 \rule{0pt}{3ex}  $WbWbH$ $-$ $| \mathcal{A}_{2t} |^2$
                  & \enskip5.59(2)	& 1.02(2)   \\[0.7ex]
 \hline
 \rule{0pt}{3ex}  $tWbH$ DS1  
                  & 11.17(2)		&  2.05(4)  \\[0.3ex]
 \rule{0pt}{3ex}  $tWbH$ DS2           
                  & \enskip4.80(2)  &  0.88(2)  \\[0.7ex]
 \hline
\end{tabular}
\caption{LO cross sections in the 4FS at the LHC with $\sqrt{s}=13$~TeV
 for the processes 
 $pp \to W^{+} b W^{-} \bar b H$ (complex mass scheme), 
 $pp \to t \bar t H$ ($t$ stable), 
 and singly resonant $pp \to tW^{-} \bar b H$ plus $pp \to \bar t W^{+}b H$ 
 computed using the GS, DR and DS prescriptions. 
 For these $tWbH$ results we also report the ratio $R$,
 which is analogous to the one defined in eq.~\eqref{eq:ratioR}. 
 All numbers are computed using the static scale $\mu_0^s = (m_t+m_W+m_H)/2$,
 and the numerical uncertainty affecting the last digit is reported
 in parentheses.}
\label{tab:tWH_xsDRDS_4FS}
\end{table}

\begin{figure}
\center
\includegraphics[width=1.0\columnwidth]{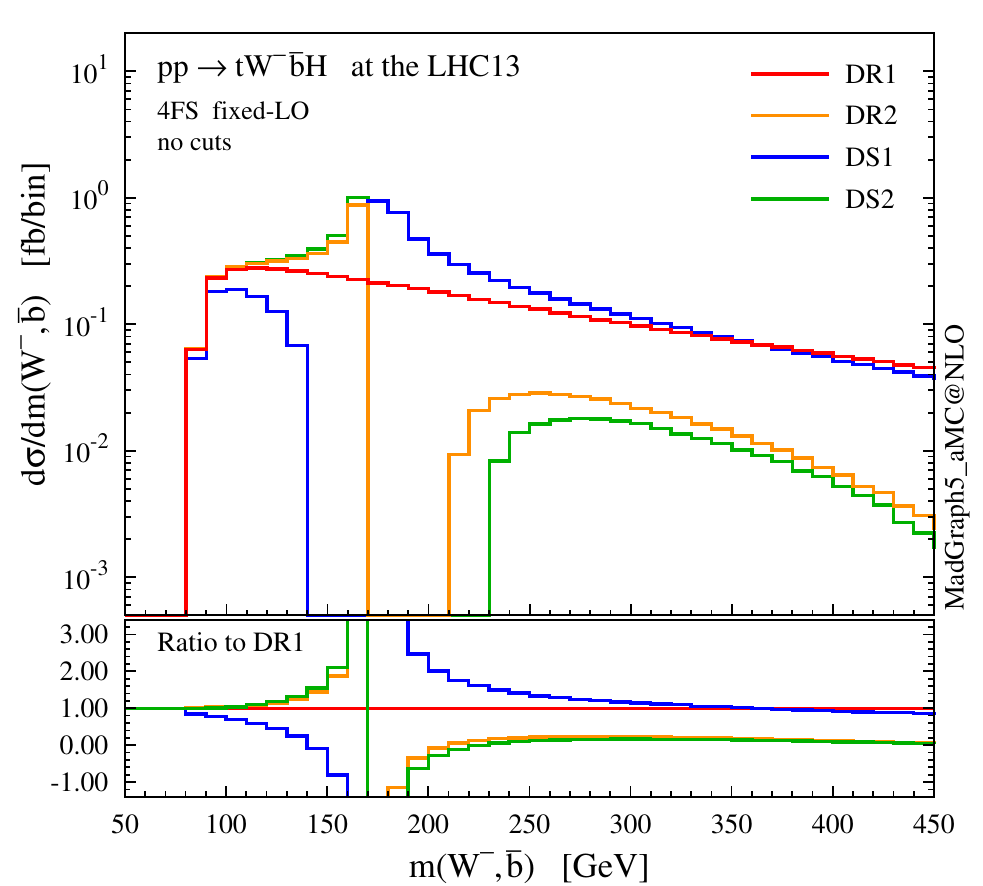}
\caption{Invariant mass $m(W^- , \bar b)$ in the $pp \to t W^- \bar b H$ process, computed with DR and DS.}
\label{fig:tWbH_4FS_distr}
\end{figure}

We can see that the ratio between top-pair and single-top amplitudes is even higher 
than for $t \bar t$ versus $tW$, and this exacerbates the same problems
we have observed in that case. 
Interference effects are very large and neglecting them results in an 
error of $O(100\perc)$ in DR1, where the cross section is more than twice
that from GS.
Once again, we find DR2 results to be in excellent agreement within
the numerical accuracy.
The impact of non resonant amplitudes and of interference between single-top 
and single-antitop contributions is very small, less than $2\perc$ of the 
DR2 rate in this channel.
The rate obtained from DS1 is overestimated by more than a factor two, 
while DS2 looks again in better agreement with GS and DR2,
although there is a residual difference of about 0.7~fb (slightly larger than
the 0.3~fb in the 5FS scheme).

In fig.~\ref{fig:tWbH_4FS_distr} we show the $m_{Wb}$ differential distribution.
A similar pattern of the one for $tWb$ is repeated: interference effects are 
large and positive in the $m_{Wb}<m_t$ region, while negative for $m_{Wb}>m_t$,
where DR1 clearly overestimates the event rate.
The interference pattern is nicely reproduced by the DR2 and DS2 shapes, although
there are some minor differences between the two methods; instead,
DS1 fails to return a physical shape, due to the visibly distorted profile 
of the subtraction term $\mathcal{C}_{2t}$, see fig.~\ref{fig:mwb}.

We would like to stress one final remark: the fact that gauge dependence is
apparently not an issue in the DR2 procedure should be regarded as a peculiarity 
of the $tWb$ and $tWbH$ channels, and not as a general result.
We cannot exclude that gauge dependence could become a significant issue at higher
perturbative orders (NNLO $tW(H)$), or in other processes with a more complex
colour flow, or using a different (i.e. non-covariant) gauge.

\providecommand{\href}[2]{#2}\begingroup\raggedright\endgroup

\end{document}